\documentclass[11pt]{article}
\usepackage{amssymb, amsmath, bbm}
\usepackage{mathrsfs}
\usepackage{dsfont}
\usepackage{authblk}

\usepackage{tikz}

\usepackage[titletoc, title]{appendix}
\usepackage{color}

\allowdisplaybreaks \numberwithin{equation}{section}
\renewcommand\arraystretch{1.25}
\newtheorem{thm}{Theorem}[section]
\newtheorem{prp}[thm]{Proposition}
\newtheorem{lem}[thm]{Lemma}
\newtheorem{dfn}[thm]{Definition}
\newenvironment{defn}{\begin{dfn} \rm }{\end{dfn}}

\newtheorem{example}[thm]{Example}
\newenvironment{exa}{\begin{example} \rm }{ \end{example}}

\newtheorem{con}{Conjecture}[section]

\newtheorem{rmk}[thm]{Remark}
\newenvironment{prf}{\noindent {\it Proof:} \ }{\hfill $\Box$}
\newenvironment{prfof}[1]{\noindent {\it Proof of #1.} \ }{\hfill $\Box$}

\newcommand\od{\mathrm{d}}
\newcommand\ad{\mathrm{ad}}

\newcommand{\nn}{\nonumber}
\newcommand\pd{\partial}
 \newcommand{\Ld}{\Lambda}
\newcommand{\al}{\alpha}
\newcommand{\gm}{\gamma} \newcommand{\Gm}{\Gamma}
\newcommand{\sg}{\sigma}
 \newcommand{\Om}{\Omega}
\newcommand{\ep}{\epsilon}   \newcommand{\vp}{\varphi}
\newcommand{\dt}{\delta}

\newcommand{\g}{\mathfrak{g}}
\newcommand{\bsigma}{\bar{\sigma}}
\newcommand{\ang}[1]{\langle #1\rangle}

\newcommand{\im}{\mathrm{Im}}
\newcommand{\Ker}{\mathrm{Ker}}

\newcommand\sL{\mathscr{L}} 
\newcommand\fg{\mathfrak{g}} \newcommand\mH{\mathcal{H}}  \newcommand\mV{\mathcal{V}}  \newcommand\mR{\mathcal{R}}
\newcommand\Z{\mathbb{Z}}  \newcommand\Zop{\mathbb{Z^{\mathrm{odd}}_+}}
\newcommand\C{\mathbb{C}}
\newcommand\R{\mathbb{R}}

\newcommand{\p}{\partial}

\newcommand{\bt}{\mathbf{t}}  
  \newcommand{\bh}{\mathbf{h}}
\newcommand{\bu}{\mathbf{u}}
\newcommand{\rs}{\mathbf{s}}

\def\bea{\begin{eqnarray}}
\def\eea{\end{eqnarray}}

\allowdisplaybreaks

\begin{document}
\title{Drinfeld-Sokolov Hierarchies and Diagram Automorphisms of Affine Kac-Moody Algebras
}

\author[1]{Si-Qi Liu}
\author[2]{ Chao-Zhong Wu}
\author[1]{Youjin Zhang}
\author[1]{Xu Zhou}
\affil[1]{Department of Mathematical Sciences, Tsinghua University, Beijing
100084, P.\,R. China \authorcr
Email: liusq@tsinghua.edu.cn; youjin@tsinghua.edu.cn}

\affil[2]{School of Mathematics, Sun Yat-Sen University,
   Guangzhou 510275, P.\,R. China \authorcr
   Email: wuchaozhong@mail.sysu.edu.cn}
\date{}
\maketitle

\begin{abstract}
For a diagram automorphism of an affine Kac-Moody algebra such that the folded diagram is still an affine Dynkin diagram, we show that the associated Drinfeld-Sokolov hierarchy also admits an induced automorphism. Then we show how to obtain the Drinfeld-Sokolov hierarchy associated to the affine Kac-Moody algebra that corresponds to the folded Dynkin diagram from the invariant sub-hierarchy of the original Drinfeld-Sokolov hierarchy.
\end{abstract}

\tableofcontents

\section{Introduction}

Starting from an affine Kac-Moody algebra with a marked vertex of its
Dynkin diagram, Drinfeld and Sokolov constructed  a
hierarchy of integrable Hamiltonian evolutionary PDEs of Korteweg –de Vries (KdV) type in \cite{DS}. It is called the
Drinfeld-Sokolov hierarchy associated to the affine Kac-Moody algebra, which plays important roles in different research fields of mathematical physics. For instance, if the Drinfeld-Sokolov hierarchy is associated to an untwisted simply laced affine Kac-Moody algebra $\fg$ with the zeroth vertex marked, then the tau function of its solution
selected by the string equation is proved to be the total descendant potential of the FJRW
theory of ADE singularities \cite{FJR, Ko, Witten}. For the untwisted affine Kac-Moody algebra of BCFG types, Ruan and the first- and the third-named authors of the present paper showed in \cite{LRZ}  that the tau function of a particular solution of the associated Drinfeld-Sokolov hierarchy coincides with the total descendant potential of the so-called $\Gm$-sector of the FJRW theory for the corresponding ambient ADE singularities.

In the above mentioned work of \cite{LRZ}, the following $\Gamma$-reduction theorem plays an important role. Let $\fg$ be an affine Kac-Moody algebra of type $A_{2n-1}^{(1)}$, $D_{n+1}^{(1)}$, $D_4^{(1)}$
and $E_6^{(1)}$ given in Table~\ref{table-diag1},
then the automorphism $\bar\sg$ of its Dynkin diagram induces a diagram
automorphism $\sg$ on $\fg$, and the $\sg$-invariant subalgebra
$\fg^\sg$ is just an affine Kac-Moody algebra of type $C_n^{(1)}$,
$B_{n}^{(1)}$, $G_2^{(1)}$ and $F_{4}^{(1)}$ respectively
\cite{Kac}. The automorphism $\sg$ generates a finite group $\Gm$,
which yields an action on the Drinfeld-Sokolov hierarchy associated to $\fg$.
\begin{thm}[Theorem 1.4 of \cite{LRZ}]\label{thm-LRZ}
The $\Gamma$-invariant flows of an ADE Drinfeld-Sokolov hierarchy define the corresponding $B_n, C_n, F_4, G_2$ Drinfeld-Sokolov hierarchy. Furthermore, the
restriction of the ADE tau function to the $\Gamma$-invariant subspace of the big phase space provides a tau function of the corresponding $B_n, C_n, F_4, G_2$ Drinfeld-Sokolov hierarchy.
\end{thm}

\begin{table}[h]
\caption{Dynkin diagrams with automorphism: $X_{\ell}^{(1)} \rightsquigarrow X_{\bar{\ell}}^{(1)}$
}\label{table-diag1}
\begin{center}
\begin{tabular}{l c}
(a1) &
\raisebox{-5ex}{
\begin{tikzpicture}[scale=1, line width=0.75pt]
\node at (3,1.35){$A_{2n-1}^{(1)}$};
\foreach \i in {1,2,4,5}{
	\foreach \j in {-0.5,0.5}{
		\draw[fill](\i,\j)circle(2pt);
	}
        \draw [<->, >=stealth, dashed, thin] (\i,-0.35) -- (\i,0.35);
}

\node at (1-1.732/2-0.25,0){$0$};  \node at (1-0.1,-0.5-0.25){$1$};
\node at (2,-0.5-0.25){$2$};  \node at (4-0.2,-0.5-0.25){$n-2$};  \node at (5+0.3,-0.5-0.25){$n-1$};
\node at (5+1.732/2+0.25,0){$n$};
 \node at (5+0.3,0.5+0.25){$n+1$};  \node at (4-0.2,0.5+0.25){$n+2$};
\node at (2+0.2,0.5+0.25){$2n-2$};
\node at (1-0.3,0.5+0.25){$2n-1$};

\coordinate (V0) at (1-1.732/2,0);
\coordinate (Vn) at (5+1.732/2,0);

\draw (2.5,-0.5) -- (1,-0.5) -- (V0) -- (1,0.5) -- (2.5,0.5);
\draw (3.5,-0.5) -- (5,-0.5) -- (Vn) -- (5,0.5) -- (3.5,0.5);
\foreach \j in {-0.5,0.5}{
		\draw[dotted, thick] (2.5,\j) -- (3.5,\j);
	}
\draw[fill=white] (V0) circle(2pt);  \draw[fill](Vn)circle(2pt);
\node at (2.3,0){$\bar\sigma$};

\node at (7,0){$\rightsquigarrow$};

\foreach \i in {9,10,12,13,14}{
		\draw[fill](\i,0)circle(2pt);
}
\coordinate (V0) at (8,0);

 \node at (8-0.1,0.25){$0$};  \node at (9,0.25){$1$};  \node at (10,0.25){$2$};
\node at (12-0.3,0.25){$n-2$};  \node at (13+0.1,0.25){$n-1$};  \node at (14+0.1,0.25){$n$};

\foreach \i in {1,-1} { \draw (8,0)+(0,\i*0.05) -- (9,\i*0.05); }
\draw (8.5,0.2) -- (8.5+0.1,0) -- (8.5,-0.2);
\draw (9,0) -- (10.5,0);
\draw[dotted, thick] (10.5,0) -- (11.5,0);

\foreach \i in {1,-1} { \draw (13,\i*0.05) -- (14,\i*0.05); }
\draw (13.5,0.2) -- (13.5-0.1,0) -- (13.5,-0.2);
\draw (11.5,0) -- (13,0);

\draw[fill=white] (V0) circle(2pt);
\node at (11,1.35){$C_{n}^{(1)}$};
\end{tikzpicture}
}
\\
\\
(d1) &
\raisebox{-5ex}{
\begin{tikzpicture}[scale=1, line width=0.75pt]

\node at (3,1){$D_{n+1}^{(1)}$};
\foreach \i in {1-1.732/2, 5+1.732/2}{
	\foreach \j in {-0.5,0.5}{
		\draw[fill](\i,\j)circle(2pt);
	}
}
\foreach \i in {1,2,4,5}{
		\draw[fill](\i,0)circle(2pt);
}

\node at (1-1.732/2-0.1,0.5+0.25){$0$};  \node at (1-1.732/2-0.1,-0.5-0.25){$1$};
\node at (1+0.1,0.25){$2$};  \node at (2,0.25){$3$};
\node at (4-0.25,-0.25){$n-2$};  \node at (5-0.2,0.35){$n-1$};
\node at (5+1.732/2,0.5+0.25){$n+1$};  \node at (5+1.732/2+0.1,-0.5-0.25){$n$};

\coordinate (V0) at (1-1.732/2,0.5);
\draw (V0) -- (1,0) -- (1-1.732/2,-0.5)  (1,0) -- (2.5,0);
\draw[fill=white] (V0) circle(2pt);

\draw[dotted, thick] (2.5,0) -- (3.5,0);

\draw (3.5,0) -- (5,0)  (5+1.732/2,0.5) -- (5,0) -- (5+1.732/2,-0.5);

\draw [<->, >=stealth, dashed, thin](5+1.732/2+0.1,-0.5+0.1) .. controls (6.1,0)  .. (5+1.732/2+0.1,0.5-0.1);

\node at (6.3,0){$\bar\sigma$};

\node at (7,0){$\rightsquigarrow$};

\foreach \j in {-0.5,0.5}{
		\draw[fill](9-1.732/2,\j)circle(2pt);
}
\foreach \i in {9,10}{
		\draw[fill](\i,0)circle(2pt);
}
\node at (9-1.732/2-0.1,0.5+0.25){$0$};   \node at (9-1.732/2-0.1,-0.5-0.25){$1$};
\node at (9+0.1,0.25){$2$};  \node at (10,0.25){$3$};
\node at (12-0.3,0.25){$n-2$};  \node at (13+0.1,0.25){$n-1$};  \node at (14+0.1,0.25){$n$};

\coordinate (V0) at (9-1.732/2,0.5);
\draw (V0) -- (9,0) -- (9-1.732/2,-0.5)  (9,0) -- (10.5,0);
\draw[fill=white] (V0) circle(2pt);

\draw[dotted, thick] (10.5,0) -- (11.5,0);

\foreach \i in {12,13,14}{
		\draw[fill](\i,0)circle(2pt);
}
\draw (11.5,0) -- (13,0);
\foreach \i in {1,-1} { \draw (13,\i*0.05) -- (14,\i*0.05); }
\draw (13.5,0.2) -- (13.5+0.1,0) -- (13.5,-0.2);
\node at (11,1){$B_{n}^{(1)}$};
\end{tikzpicture}
}
\\
\\
(d2) &
\raisebox{-6.5ex}{
\begin{tikzpicture}[scale=1, line width=0.75pt]
\node at (3.6,1.25){$D_{4}^{(1)}$};
	\foreach \j in {-0.5,0.5}{
		\draw[fill](4.5,\j*1.732)circle(2pt);
	}
\foreach \i in {3,4,5}{
		\draw[fill](\i,0)circle(2pt);
}
\node at (3-0.25,0){$0$};  \node at (4-0.1,0.25){$2$};  \node at (5+0.25,0){$3$};
\node at (4.5-0.1, -1.732/2-0.25){$1$};   \node at (4.5-0.1, 1.732/2+0.25){$4$};

\coordinate (V0) at (3,0);
\coordinate (V1) at (4.5, -1.732/2);
\coordinate (V2) at (4,0);
\coordinate (V3) at (5,0);
\coordinate (V4) at (4.5, 1.732/2);
\draw (V0) -- (V3)  (V1)--(V2)--(V4);
\draw[fill=white] (V0) circle(2pt);

\draw [->, >=stealth, dashed, thin] (4.5-0.05, 1.732/2-0.2) .. controls (4.2,0.3) and (4.2,0-0.3)  .. (4.5-0.05, -1.732/2+0.2);
\draw[->, >=stealth, dashed, thin] (4.5+0.15, -1.732/2+0.05) to [out=15,in=-105] (5,-0.15);
\draw[<-, >=stealth, dashed, thin] (4.5+0.15, 1.732/2-0.05) to [out=-15,in=105] (5,0.15);
\node at (5.2,0.5){$\bar\sigma$};

\node at (7,0){$\rightsquigarrow$};

\foreach \i in {10,11}{
		\draw[fill](\i,0)circle(2pt);
}

\coordinate (V0) at (9,0);
\coordinate (V1) at (10,0);
\coordinate (V2) at (11,0);

\node at (9-0.1, 0.25){$0$};  \node at (10, 0.25){$1$};  \node at (11+0.1, 0.25){$2$};

\draw (V0) -- (V1);
\foreach \i in {1,-1} { \draw (10,\i*0.06) -- (11,\i*0.06); } \draw[thin] (10,0) -- (11,0);
\draw[fill=white] (V0) circle(2pt);
\draw (10.5,0.2) -- (10.5+0.1,0) -- (10.5,-0.2);
\node at (10,1.25){$G_{2}^{(1)}$};
\end{tikzpicture}
}
\\
(e1) &
\raisebox{-5ex}{
\begin{tikzpicture}[scale=1, line width=0.75pt]
\node at (3.6,1.25){$E_{6}^{(1)}$};
\foreach \i in {4+1.732/2,5+1.732/2}{
	\foreach \j in {-0.5,0.5}{
		\draw[fill](\i,\j)circle(2pt);
	}

\draw[<->, >=stealth, dashed, thin] (\i+0.1, 0.5-0.1) to [out=-45,in=45] (\i+0.1,-0.5+0.1);
}
\foreach \i in {2,3,4}{
		\draw[fill](\i,0)circle(2pt);
}

\coordinate (V0) at (2,0);
\draw (V0) -- (4,0)  (5+1.732/2,0.5)--(4+1.732/2,0.5)--(4,0)--(4+1.732/2,-0.5)--(5+1.732/2,-0.5);
\draw[fill=white] (V0) circle(2pt);

\node at (6.4,0){$\bar\sigma$};

\node at (2-0.1,0.25){$0$};
\node at (3-0.1,0.25){$6$}; \node at (4-0.1,0.25){$3$};
\node at (4+1.732/2+0.1,0.5+0.25){$4$};  \node at (5+1.732/2+0.1,0.5+0.25){$5$};
\node at (4+1.732/2+0.1,-0.5-0.25){$2$};  \node at (5+1.732/2+0.1,-0.5-0.25){$1$};

\node at (7.5,0){$\rightsquigarrow$};

\foreach \i in {10,11,12,13}{
		\draw[fill](\i,0)circle(2pt);
}

\coordinate (V0) at (9,0);
\coordinate (V1) at (10,0);
\coordinate (V2) at (11,0);
\coordinate (V3) at (12,0);
\coordinate (V4) at (13,0);

\node at (9-0.1, 0.25){$0$};  \node at (10, 0.25){$1$};
\node at (11, 0.25){$2$};  \node at (12, 0.25){$3$};  \node at (13+0.1, 0.25){$4$};
\draw (V0) -- (V2)  (V3)--(V4);
\foreach \i in {1,-1} { \draw (11,\i*0.05) -- (12,\i*0.05); }
\draw[fill=white] (V0) circle(2pt);
\draw (11.5,0.2) -- (11.5+0.1,0) -- (11.5,-0.2);
\node at (11,1.25){$F_{4}^{(1)}$};
\end{tikzpicture}
}
\end{tabular}
\end{center}
\end{table}

A twisted affine Kac-Moody algebra can be realized as the invariant subalgebra of a twisted automorphism of an untwisted simply laced affine Kac-Moody algebra. So it is natural to ask: is there a $\Gamma$-reduction theorem for the Drinfeld-Sokolov hierarchy associated to the twisted affine Kac-Moody algebra? Unfortunately, since the twisted automorphism does not yield an action on the associated Drinfeld-Sokolov hierarchy, there is no such a theorem along this line. Instead, to obtain an analogue of the $\Gamma$-reduction theorem for the cases of twisted affine Kac-Moody algebras, we have to study the (untwisted) diagram automorphisms on affine Kac-Moody algebras and their invariant subalgebras.

The Dynkin diagrams of affine type on which there exist automorphisms $\bar\sg$
such that the corresponding folded diagrams are still  Dynkin diagrams of affine type are listed in Tables~\ref{table-diag1}--\ref{table-diag3}, with the vertices of the Dynkin diagrams labeled as in \cite{Kac}  (the blank node always stands for the zeroth vertex). We observe, in particular, that the Dynkin diagrams of all twisted affine Kac-Moody algebras are contained in Tables~\ref{table-diag2} and \ref{table-diag3}. Hence,  to answer the above question, we only need to clarify the relationship between the Drinfeld-Sokolov hierarchies of types  $X_{\ell'}^{(r)}$ and $X_{\bar{\ell}'}^{(\bar r)}$ listed in Tables~\ref{table-diag2} and \ref{table-diag3}.
We remark that, in Table \ref{table-diag2}, for Case (a2) with $n=1$, the reduced Lie algebra is of type $A_2^{(2)}$, whose Dynkin diagram is illustrated in Case (d5) of Table~\ref{table-diag3}. This reduced Lie algebra also appears in Table \ref{table-diag3}, Case (c2) with $n=1$, whose Dynkin diagram is obtained from the one of Case (d5)  with the two vertices exchanged.

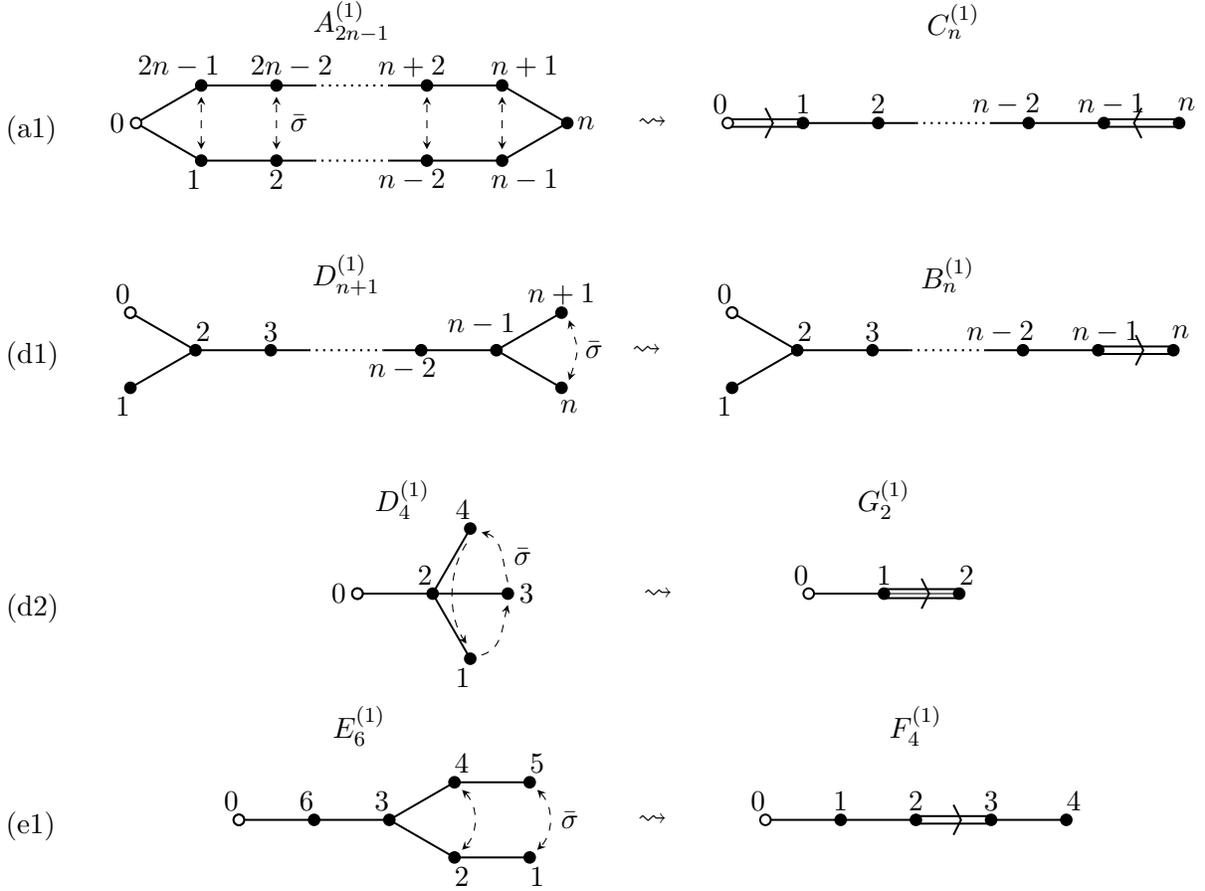
\begin{table}[]
\caption{Dynkin diagrams with automorphism: $X_{\ell'}^{(r)} \rightsquigarrow X_{\bar{\ell}'}^{(\bar r)}$
}\label{table-diag2}
\begin{center}
\begin{tabular}{lc}
(a2)&  \hspace{-12pt}
\raisebox{-5ex}{
\begin{tikzpicture}[scale=1, line width=0.75pt]
\node at (3,1.35){$A_{2n}^{(1)}$};
\foreach \i in {0,1,2,4,5}{
	\foreach \j in {-0.5,0.5}{
		\draw[fill](\i,\j)circle(2pt);
	}
}
\node at (0-0.1, -0.5-0.25){$0$};  \node at (1, -0.5-0.25){$1$};  \node at (2, -0.5-0.25){$2$};
\node at (4-0.2, -0.5-0.25){$n-2$};  \node at (5+0.2, -0.5-0.25){$n-1$};  \node at (5+1.732/2+0.25,0){$n$};
\node at (0-0.3, 0.5+0.25){$2n$};  \node at (1-0.2, 0.5+0.25){$2n-1$};  \node at (2+0.2, 0.5+0.25){$2n-2$};
\node at (4-0.2, 0.5+0.25){$n+2$};  \node at (5+0.2, 0.5+0.25){$n+1$};

\coordinate (V0) at (0,-0.5);
\coordinate (Vn) at (5+1.732/2,0);

\draw (2.5,0.5) -- (0,0.5) -- (0,-0.5) -- (2.5,-0.5);
\draw (3.5,-0.5) -- (5,-0.5) -- (Vn) -- (5,0.5) -- (3.5,0.5);
\foreach \j in {-0.5,0.5}{
		\draw[dotted, thick] (2.5,\j) -- (3.5,\j);
	}
\draw[fill=white] (V0) circle(2pt);  \draw[fill](Vn)circle(2pt);
\foreach \i in {0,1,2,4,5}{
\draw[<->, >=stealth, dashed, thin] (\i-0.1, 0.5-0.1) to [out=-135,in=135] (\i-0.1,-0.5+0.1);
}
\node at (2.1,0){$\bar\sigma$};

\node at (7,0){$\rightsquigarrow$};

\foreach \i in {9,10,12,13,14}{
		\draw[fill](\i,0)circle(2pt);

\node at (8-0.1, 0.25){$0$}; \node at (9, 0.25){$1$};  \node at (10, 0.25){$2$};
\node at (12-0.3, 0.25){$n-2$}; \node at (13+0, 0.25){$n-1$};  \node at (14+0.1, 0.25){$n$};

}
\coordinate (V0) at (8,0);
\foreach \i in {1,-1} { \draw (8,0)+(0,\i*0.05) -- (9,\i*0.05); }
\draw (8.5,0.2) -- (8.5-0.1,0) -- (8.5,-0.2);
\draw (9,0) -- (10.5,0);
\draw[dotted, thick] (10.5,0) -- (11.5,0);

\foreach \i in {1,-1} { \draw (13,\i*0.05) -- (14,\i*0.05); }
\draw (13.5,0.2) -- (13.5-0.1,0) -- (13.5,-0.2);
\draw (11.5,0) -- (13,0);

\draw[fill=white] (V0) circle(2pt);
\node at (11,1.35){$A_{2n}^{(2)}$};
\end{tikzpicture}
}
\\
\\
(a3)
 &
\raisebox{-5ex}{
\begin{tikzpicture}[scale=1, line width=0.75pt]
\node at (3,1.25){$A_{2n+1}^{(1)}$};
\foreach \i in {0,1,2,4,5,6}{
	\foreach \j in {-0.5,0.5}{
		\draw[fill](\i,\j)circle(2pt);
	}
}

\coordinate (V0) at (0,-0.5);

\draw (2.5,0.5) -- (0,0.5) -- (0,-0.5) -- (2.5,-0.5);
\draw (3.5,-0.5) -- (6,-0.5) -- (6,0.5) -- (3.5,0.5);
\foreach \j in {-0.5,0.5}{
		\draw[dotted, thick] (2.5,\j) -- (3.5,\j);
	}
\draw[fill=white] (V0) circle(2pt);
\foreach \i in {0,1,2,4,5,6}{
\draw[<->, >=stealth, dashed, thin] (\i-0.1, 0.5-0.1) to [out=-135,in=135] (\i-0.1,-0.5+0.1);
}
\node at (2.1,0){$\bar\sigma$};

\node at (0-0.1, -0.5-0.25){$0$};  \node at (1, -0.5-0.25){$1$};  \node at (2, -0.5-0.25){$2$};
\node at (4-0.2, -0.5-0.25){$n-2$};  \node at (5, -0.5-0.25){$n-1$};  \node at (6+0.1, -0.5-0.25){$n$};
\node at (0-0.1, 0.5+0.25){$2n+1$};  \node at (1, 0.5+0.25){$2n$};  \node at (2+0.2, 0.5+0.25){$2n-1$};
\node at (4-0.2, 0.5+0.25){$n+3$};  \node at (5, 0.5+0.25){$n+2$};   \node at (6+0.1, 0.5+0.25){$n+1$};

\node at (7,0){$\rightsquigarrow$};

\foreach \i in {9,10,12,13,14}{
		\draw[fill](\i,0)circle(2pt);
}

\coordinate (V0) at (8,0);
\foreach \i in {1,-1} { \draw (8,0)+(0,\i*0.05) -- (9,\i*0.05); }
\draw (8.5,0.2) -- (8.5-0.1,0) -- (8.5,-0.2);
\draw (9,0) -- (10.5,0);
\draw[dotted, thick] (10.5,0) -- (11.5,0);

\foreach \i in {1,-1} { \draw (13,\i*0.05) -- (14,\i*0.05); }
\draw (13.5,0.2) -- (13.5+0.1,0) -- (13.5,-0.2);
\draw (11.5,0) -- (13,0);

\draw[fill=white] (V0) circle(2pt);
\node at (11,1.25){$D_{n+1}^{(2)}$};

\node at (8-0.1, 0.25){$0$}; \node at (9, 0.25){$1$};  \node at (10, 0.25){$2$};
\node at (12-0.3, 0.25){$n-2$}; \node at (13+0, 0.25){$n-1$};  \node at (14+0.1, 0.25){$n$};
\end{tikzpicture}
}
\\
\\
(b)
 &  \hspace{-10pt}
\raisebox{-5ex}{
\begin{tikzpicture}[scale=1, line width=0.75pt]
\node at (3,1){$B_{n+1}^{(1)}$};
\foreach \j in {-0.5,0.5}{
		\draw[fill](1-1.732/2,\j)circle(2pt);
}
\foreach \i in {1,2}{
		\draw[fill](\i,0)circle(2pt);
}

\coordinate (V0) at (1-1.732/2,0.5);
\draw (V0) -- (1,0) -- (1-1.732/2,-0.5)  (1,0) -- (2.5,0);
\draw[fill=white] (V0) circle(2pt);

\draw[dotted, thick] (2.5,0) -- (3.5,0);

\foreach \i in {4,5,6}{
		\draw[fill](\i,0)circle(2pt);
}
\draw (3.5,0) -- (5,0);
\foreach \i in {1,-1} { \draw (5,\i*0.05) -- (6,\i*0.05); }
\draw (5.5,0.2) -- (5.5+0.1,0) -- (5.5,-0.2);

\draw[<->, >=stealth, dashed, thin] (1-1.732/2-0.1, 0.5-0.1) to [out=-135,in=135] (1-1.732/2-0.1,-0.5+0.1);
\node at (-0.35,0){$\bar\sigma$};

\node at (1-1.732/2-0.1,0.5+0.25){$0$};  \node at (1-1.732/2-0.1,-0.5-0.25){$1$};
\node at (1+0.1,0.25){$2$};  \node at (2,0.25){$3$};
\node at (4-0.1,0.25){$n-1$};  \node at (5,0.25){$n$};  \node at (6+0.1,0.25){$n+1$};

\node at (7,0){$\rightsquigarrow$};

\foreach \i in {9,10,12,13,14}{
		\draw[fill](\i,0)circle(2pt);
}
\coordinate (V0) at (8,0);
\foreach \i in {1,-1} { \draw (8,0)+(0,\i*0.05) -- (9,\i*0.05); }
\draw (8.5,0.2) -- (8.5-0.1,0) -- (8.5,-0.2);
\draw (9,0) -- (10.5,0);
\draw[dotted, thick] (10.5,0) -- (11.5,0);

\foreach \i in {1,-1} { \draw (13,\i*0.05) -- (14,\i*0.05); }
\draw (13.5,0.2) -- (13.5+0.1,0) -- (13.5,-0.2);
\draw (11.5,0) -- (13,0);

\draw[fill=white] (V0) circle(2pt);
\node at (11,1){$D_{n+1}^{(2)}$};

\node at (8-0.1, 0.25){$0$}; \node at (9, 0.25){$1$};  \node at (10, 0.25){$2$};
\node at (12-0.3, 0.25){$n-2$}; \node at (13+0, 0.25){$n-1$};  \node at (14+0.1, 0.25){$n$};
\end{tikzpicture}
}
\\
(c1)
 &
\raisebox{-5ex}{
\begin{tikzpicture}[scale=1, line width=0.75pt]
\node at (3,1.25){$C_{2n}^{(1)}$};
\foreach \i in {0,1,2,4,5}{
	\foreach \j in {-0.5,0.5}{
		\draw[fill](\i,\j)circle(2pt);
	}
}

\coordinate (V0) at (0,-0.5);
\coordinate (Vn) at (5+1.732/2,0);

\draw (2.5,0.5) -- (1,0.5)  (1,-0.5) -- (2.5,-0.5);
\draw (3.5,-0.5) -- (5,-0.5) -- (Vn) -- (5,0.5) -- (3.5,0.5);
\foreach \j in {0.5,-0.5} {
\foreach \i in {1,-1} { \draw (0,\j)+(0,\i*0.05) -- (1,\j+\i*0.05); }
\draw (0.5,\j+0.2) -- (0.5+0.1,\j) -- (0.5,\j-0.2);
}
\foreach \j in {-0.5,0.5}{
		\draw[dotted, thick] (2.5,\j) -- (3.5,\j);
	}
\draw[fill=white] (V0) circle(2pt);  \draw[fill](Vn)circle(2pt);
\foreach \i in {0,1,2,4,5}{
\draw[<->, >=stealth, dashed, thin] (\i-0.1, 0.5-0.1) to [out=-135,in=135] (\i-0.1,-0.5+0.1);
}
\node at (2.1,0){$\bar\sigma$};

\node at (0-0.1, -0.5-0.25){$0$};  \node at (1, -0.5-0.25){$1$};  \node at (2, -0.5-0.25){$2$};
\node at (4-0.2, -0.5-0.25){$n-2$};  \node at (5+0.2, -0.5-0.25){$n-1$};  \node at (5+1.732/2+0.25,0){$n$};
\node at (0-0.3, 0.5+0.25){$2n$};  \node at (1-0.2, 0.5+0.25){$2n-1$};  \node at (2+0.2, 0.5+0.25){$2n-2$};
\node at (4-0.2, 0.5+0.25){$n+2$};  \node at (5+0.2, 0.5+0.25){$n+1$};

\node at (7,0){$\rightsquigarrow$};

\foreach \i in {9,10,12,13,14}{
		\draw[fill](\i,0)circle(2pt);
}
\coordinate (V0) at (8,0);
\foreach \i in {1,-1} { \draw (8,0)+(0,\i*0.05) -- (9,\i*0.05); }
\draw (8.5,0.2) -- (8.5+0.1,0) -- (8.5,-0.2);
\draw (9,0) -- (10.5,0);
\draw[dotted, thick] (10.5,0) -- (11.5,0);

\foreach \i in {1,-1} { \draw (13,\i*0.05) -- (14,\i*0.05); }
\draw (13.5,0.2) -- (13.5-0.1,0) -- (13.5,-0.2);
\draw (11.5,0) -- (13,0);

\draw[fill=white] (V0) circle(2pt);
\node at (11,1.25){$C_{n}^{(1)}$};

\node at (8-0.1, 0.25){$0$}; \node at (9, 0.25){$1$};  \node at (10, 0.25){$2$};
\node at (12-0.3, 0.25){$n-2$}; \node at (13+0, 0.25){$n-1$};  \node at (14+0.1, 0.25){$n$};
\end{tikzpicture}
}
\\
\\
(d3)
&    \hspace{-14pt}
\raisebox{-7ex}{
\begin{tikzpicture}[scale=1, line width=0.75pt]
\node at (3,1.25){$D_{2n+1}^{(1)}$};
\foreach \i in {1,2,4,5,6}{
	\foreach \j in {-0.5,0.5}{
		\draw[fill](\i,\j)circle(2pt);
	}
}

\draw (2.5,0.5) -- (1,0.5)  (1,-0.5) -- (2.5,-0.5);
\draw (3.5,-0.5) -- (6,-0.5) -- (6,0.5) -- (3.5,0.5);
\foreach \j in {0.5,-0.5} {
    \draw[fill] (1,\j)+(-0.966, -2*\j*0.259) circle(2pt);
    \draw[fill] (1,\j)+(-1.414/2, \j*1.414) circle(2pt);
    \draw (1,\j)+(-0.966, -2*\j*0.259) -- (1,\j);
    \draw (1,\j)+(-1.414/2, \j*1.414) -- (1,\j);
}
\foreach \j in {-0.5,0.5}{
		\draw[dotted, thick] (2.5,\j) -- (3.5,\j);
	}

\coordinate (V0) at (1-0.966,-0.5+0.259);
\draw[fill=white] (V0) circle(2pt);
\foreach \i in {1,2,4,5,6}{
\draw[<->, >=stealth, dashed, thin] (\i-0.1, 0.5-0.1) to [out=-135,in=135] (\i-0.1,-0.5+0.1);
}
\draw[<->, >=stealth, dashed, thin] (1-0.966-0.15, -0.5+0.259) to [out=-160,in=160] (1-0.966-0.15, 0.5-0.259);
\draw[<->, >=stealth, dashed, thin] (1-1.414/2-0.15, -0.5-1.414/2) to [out=180,in=-180] (1-1.414/2-0.15, 0.5+1.414/2);
\node at (2.1,0){$\bar\sigma$};

\node at (0-0.1, -0.5){$0$};  \node at (0.5, -1.3){$1$};   \node at (1+0.1, -0.5-0.25){$2$};
\node at (2, -0.5-0.25){$3$};
\node at (4-0.2, -0.5-0.25){$n-2$};  \node at (5, -0.5-0.25){$n-1$};  \node at (6+0.1, -0.5-0.25){$n$};
\node at (0-0.1, 0.5){$2n+1$};  \node at (0.5, 1.45){$2n$};  \node at (1+0.25, 0.5+0.35){$2n-1$};
\node at (2+0.35, 0.5+0.25){$2n-2$};
\node at (4-0.2, 0.5+0.25){$n+3$};  \node at (5, 0.5+0.25){$n+2$};   \node at (6+0.1, 0.5+0.25){$n+1$};

\node at (7,0){$\rightsquigarrow$};

\foreach \j in {-0.5,0.5}{
		\draw[fill](9-1.732/2,\j)circle(2pt);
}
\foreach \i in {9,10}{
		\draw[fill](\i,0)circle(2pt);
}

\coordinate (V0) at (9-1.732/2,0.5);
\draw (V0) -- (9,0) -- (9-1.732/2,-0.5)  (9,0) -- (10.5,0);
\draw[fill=white] (V0) circle(2pt);

\draw[dotted, thick] (10.5,0) -- (11.5,0);

\foreach \i in {12,13,14}{
		\draw[fill](\i,0)circle(2pt);
}
\draw (11.5,0) -- (13,0);
\foreach \i in {1,-1} { \draw (13,\i*0.05) -- (14,\i*0.05); }
\draw (13.5,0.2) -- (13.5+0.1,0) -- (13.5,-0.2);
\node at (11,1.25){$B_{n}^{(1)}$};

\node at (9-1.732/2-0.1,0.5+0.25){$0$};  \node at (9-1.732/2-0.1,-0.5-0.25){$1$};
\node at (9+0.1,0.25){$2$};  \node at (10,0.25){$3$};
\node at (12-0.2,0.25){$n-2$};  \node at (13,0.25){$n-1$};  \node at (14+0.1,0.25){$n$};
\end{tikzpicture}

}
\\
\\
(d6)
&  \hspace{-8pt}
\raisebox{-5ex}{
\begin{tikzpicture}[scale=1, line width=0.75pt]
\node at (3,1.25){$D_{(2n+1)+1}^{(2)}$};
\foreach \i in {0,1,2,4,5,6}{
	\foreach \j in {-0.5,0.5}{
		\draw[fill](\i,\j)circle(2pt);
	}
}

\coordinate (V0) at (0,-0.5);

\draw (2.5,0.5) -- (1,0.5)  (1,-0.5) -- (2.5,-0.5);
\draw (3.5,-0.5) -- (6,-0.5) -- (6,0.5) -- (3.5,0.5);
\foreach \j in {0.5,-0.5} {
\foreach \i in {1,-1} { \draw (0,\j)+(0,\i*0.05) -- (1,\j+\i*0.05); }
\draw (0.5,\j+0.2) -- (0.5-0.1,\j) -- (0.5,\j-0.2);
}
\foreach \j in {-0.5,0.5}{
		\draw[dotted, thick] (2.5,\j) -- (3.5,\j);
	}
\draw[fill=white] (V0) circle(2pt);
\foreach \i in {0,1,2,4,5,6}{
\draw[<->, >=stealth, dashed, thin] (\i-0.1, 0.5-0.1) to [out=-135,in=135] (\i-0.1,-0.5+0.1);
}
\node at (2.1,0){$\bar\sigma$};

\node at (0-0.1, -0.5-0.25){$0$};  \node at (1, -0.5-0.25){$1$};  \node at (2, -0.5-0.25){$2$};
\node at (4-0.2, -0.5-0.25){$n-2$};  \node at (5, -0.5-0.25){$n-1$};  \node at (6+0.1, -0.5-0.25){$n$};
\node at (0-0.1, 0.5+0.25){$2n+1$};  \node at (1, 0.5+0.25){$2n$};  \node at (2+0.2, 0.5+0.25){$2n-1$};
\node at (4-0.2, 0.5+0.25){$n+3$};  \node at (5, 0.5+0.25){$n+2$};   \node at (6+0.1, 0.5+0.25){$n+1$};

\node at (7,0){$\rightsquigarrow$};

\foreach \i in {9,10,12,13,14}{
		\draw[fill](\i,0)circle(2pt);
}
\coordinate (V0) at (8,0);
\foreach \i in {1,-1} { \draw (8,0)+(0,\i*0.05) -- (9,\i*0.05); }
\draw (8.5,0.2) -- (8.5-0.1,0) -- (8.5,-0.2);
\draw (9,0) -- (10.5,0);
\draw[dotted, thick] (10.5,0) -- (11.5,0);

\foreach \i in {1,-1} { \draw (13,\i*0.05) -- (14,\i*0.05); }
\draw (13.5,0.2) -- (13.5+0.1,0) -- (13.5,-0.2);
\draw (11.5,0) -- (13,0);

\draw[fill=white] (V0) circle(2pt);
\node at (11,1.25){$D_{n+1}^{(2)}$};

\node at (8-0.1, 0.25){$0$}; \node at (9, 0.25){$1$};  \node at (10, 0.25){$2$};
\node at (12-0.3, 0.25){$n-2$}; \node at (13+0, 0.25){$n-1$};  \node at (14+0.1, 0.25){$n$};
\end{tikzpicture}
}
\\
\end{tabular}
\end{center}
\end{table}

\begin{table}[]
\caption{Dynkin diagrams with automorphism: $X_{\ell'}^{(r)} \rightsquigarrow X_{\bar{\ell}'}^{(\bar r)}$
}\label{table-diag3}
\begin{center}
\begin{tabular}{lc}
(a4)
&
\raisebox{-5ex}{
\begin{tikzpicture}[scale=1, line width=0.75pt]
\node at (3,1){$A_{2(n+1)-1}^{(2)}$};
\foreach \j in {-0.5,0.5}{
		\draw[fill](1-1.732/2,\j)circle(2pt);
}
\foreach \i in {1,2}{
		\draw[fill](\i,0)circle(2pt);
}
\node at (1-1.732/2-0.1,0.5+0.25){$0$};   \node at (1-1.732/2-0.1,-0.5-0.25){$1$};
\node at (1+0.1,0.25){$2$};  \node at (2,0.25){$3$};
\node at (4-0.1,0.25){$n-1$};  \node at (5,0.25){$n$};  \node at (6+0.1,0.25){$n+1$};

\coordinate (V0) at (1-1.732/2,0.5);
\draw (V0) -- (1,0) -- (1-1.732/2,-0.5)  (1,0) -- (2.5,0);
\draw[fill=white] (V0) circle(2pt);

\draw[dotted, thick] (2.5,0) -- (3.5,0);

\foreach \i in {4,5,6}{
		\draw[fill](\i,0)circle(2pt);
}
\draw (3.5,0) -- (5,0);
\foreach \i in {1,-1} { \draw (5,\i*0.05) -- (6,\i*0.05); }
\draw (5.5,0.2) -- (5.5-0.1,0) -- (5.5,-0.2);

\draw[<->, >=stealth, dashed, thin] (1-1.732/2-0.1, 0.5-0.1) to [out=-135,in=135] (1-1.732/2-0.1,-0.5+0.1);
\node at (-0.35,0){$\bar\sigma$};

\node at (7,0){$\rightsquigarrow$};

\foreach \i in {9,10,12,13,14}{
		\draw[fill](\i,0)circle(2pt);
}
\coordinate (V0) at (8,0);
\foreach \i in {1,-1} { \draw (8,0)+(0,\i*0.05) -- (9,\i*0.05); }
\draw (8.5,0.2) -- (8.5-0.1,0) -- (8.5,-0.2);
\draw (9,0) -- (10.5,0);
\draw[dotted, thick] (10.5,0) -- (11.5,0);

\foreach \i in {1,-1} { \draw (13,\i*0.05) -- (14,\i*0.05); }
\draw (13.5,0.2) -- (13.5-0.1,0) -- (13.5,-0.2);
\draw (11.5,0) -- (13,0);

\draw[fill=white] (V0) circle(2pt);
\node at (11,1){$A_{2n}^{(2)}$};

\node at (8-0.1, 0.25){$0$}; \node at (9, 0.25){$1$};  \node at (10, 0.25){$2$};
\node at (12-0.3, 0.25){$n-2$}; \node at (13, 0.25){$n-1$};  \node at (14+0.1, 0.25){$n$};
\end{tikzpicture}
}
\\
(c2)
&
\raisebox{-5ex}{
\begin{tikzpicture}[scale=1, line width=0.75pt]
\node at (3,1.25){$C_{2n+1}^{(1)}$ };
\foreach \i in {0,1,2,4,5,6}{
	\foreach \j in {-0.5,0.5}{
		\draw[fill](\i,\j)circle(2pt);
	}
}

\coordinate (V0) at (0,-0.5);

\draw (2.5,0.5) -- (1,0.5)  (1,-0.5) -- (2.5,-0.5);
\draw (3.5,-0.5) -- (6,-0.5) -- (6,0.5) -- (3.5,0.5);
\foreach \j in {0.5,-0.5} {
\foreach \i in {1,-1} { \draw (0,\j)+(0,\i*0.05) -- (1,\j+\i*0.05); }
\draw (0.5,\j+0.2) -- (0.5+0.1,\j) -- (0.5,\j-0.2);
}
\foreach \j in {-0.5,0.5}{
		\draw[dotted, thick] (2.5,\j) -- (3.5,\j);
	}
\draw[fill=white] (V0) circle(2pt);
\foreach \i in {0,1,2,4,5,6}{
\draw[<->, >=stealth, dashed, thin] (\i-0.1, 0.5-0.1) to [out=-135,in=135] (\i-0.1,-0.5+0.1);
}
\node at (2.1,0){$\bar\sigma$};

\node at (0-0.1, -0.5-0.25){$0$};  \node at (1, -0.5-0.25){$1$};  \node at (2, -0.5-0.25){$2$};
\node at (4-0.2, -0.5-0.25){$n-2$};  \node at (5, -0.5-0.25){$n-1$};  \node at (6+0.1, -0.5-0.25){$n$};
\node at (0-0.1, 0.5+0.25){$2n+1$};  \node at (1, 0.5+0.25){$2n$};  \node at (2+0.2, 0.5+0.25){$2n-1$};
\node at (4-0.2, 0.5+0.25){$n+3$};  \node at (5, 0.5+0.25){$n+2$};   \node at (6+0.1, 0.5+0.25){$n+1$};

\node at (7,0){$\rightsquigarrow$};

\foreach \i in {8,9,10,12,13,14}{
		\draw[fill](\i,0)circle(2pt);
}
\coordinate (V0) at (14,0);
\foreach \i in {1,-1} { \draw (8,0)+(0,\i*0.05) -- (9,\i*0.05); }
\draw (8.5,0.2) -- (8.5+0.1,0) -- (8.5,-0.2);
\draw (9,0) -- (10.5,0);
\draw[dotted, thick] (10.5,0) -- (11.5,0);

\foreach \i in {1,-1} { \draw (13,\i*0.05) -- (14,\i*0.05); }
\draw (13.5,0.2) -- (13.5+0.1,0) -- (13.5,-0.2);
\draw (11.5,0) -- (13,0);

\draw[fill=white] (V0) circle(2pt);
\node at (11,1.25){$A_{2n}^{(2)}$};

\node at (14+0.1, 0.25){$0$}; \node at (13, 0.25){$1$};  \node at (12, 0.25){$2$};
\node at (10+0.3, 0.25){$n-2$}; \node at (9, 0.25){$n-1$};  \node at (8-0.1, 0.25){$n$};
\end{tikzpicture}
}
\\
\\
(d4)
&
\raisebox{-7ex}{
\begin{tikzpicture}[scale=1, line width=0.75pt]
\node at (3,1.25){$D_{2n}^{(1)}$};
\foreach \i in {1,2,4,5}{
	\foreach \j in {-0.5,0.5}{
		\draw[fill](\i,\j)circle(2pt);
	}
}
\draw[fill](5+1.732/2,0)circle(2pt);

\draw (2.5,0.5) -- (1,0.5)  (1,-0.5) -- (2.5,-0.5);
\draw (3.5,-0.5) -- (5,-0.5) -- (5+1.732/2,0) -- (5,0.5) -- (3.5,0.5);
\foreach \j in {0.5,-0.5} {
    \draw[fill] (1,\j)+(-0.966, -2*\j*0.259) circle(2pt);
    \draw[fill] (1,\j)+(-1.414/2, \j*1.414) circle(2pt);
    \draw (1,\j)+(-0.966, -2*\j*0.259) -- (1,\j);
    \draw (1,\j)+(-1.414/2, \j*1.414) -- (1,\j);
}
\foreach \j in {-0.5,0.5}{
		\draw[dotted, thick] (2.5,\j) -- (3.5,\j);
	}

\coordinate (V0) at (1-0.966,-0.5+0.259);
\draw[fill=white] (V0) circle(2pt);
\foreach \i in {1,2,4,5}{
\draw[<->, >=stealth, dashed, thin] (\i-0.1, 0.5-0.1) to [out=-135,in=135] (\i-0.1,-0.5+0.1);
}
\draw[<->, >=stealth, dashed, thin] (1-0.966-0.15, -0.5+0.259) to [out=-165,in=165] (1-0.966-0.15, 0.5-0.259);
\draw[<->, >=stealth, dashed, thin] (1-1.414/2-0.15, -0.5-1.414/2) to [out=180,in=-180] (1-1.414/2-0.15, 0.5+1.414/2);
\node at (2.1,0){$\bar\sigma$};

\node at (0-0.1, -0.5){$0$};  \node at (0.5, -1.3){$1$};   \node at (1+0.1, -0.5-0.25){$2$};
\node at (2, -0.5-0.25){$3$};
\node at (4-0.2, -0.5-0.25){$n-2$};  \node at (5, -0.5-0.25){$n-1$};  \node at (5+1.732/2+0.25,0){$n$};
\node at (0-0.1, 0.5){$2n$};  \node at (0.4, 1.45){$2n-1$};  \node at (1+0.25, 0.5+0.35){$2n-2$};
\node at (2+0.35, 0.5+0.25){$2n-3$};
\node at (4-0.2, 0.5+0.25){$n+2$};  \node at (5, 0.5+0.25){$n+1$};

\node at (7,0){$\rightsquigarrow$};

\foreach \j in {-0.5,0.5}{
		\draw[fill](9-1.732/2,\j)circle(2pt);
}
\foreach \i in {9,10}{
		\draw[fill](\i,0)circle(2pt);
}

\coordinate (V0) at (9-1.732/2,0.5);
\draw (V0) -- (9,0) -- (9-1.732/2,-0.5)  (9,0) -- (10.5,0);
\draw[fill=white] (V0) circle(2pt);

\draw[dotted, thick] (10.5,0) -- (11.5,0);

\foreach \i in {12,13,14}{
		\draw[fill](\i,0)circle(2pt);
}
\draw (11.5,0) -- (13,0);
\foreach \i in {1,-1} { \draw (13,\i*0.05) -- (14,\i*0.05); }
\draw (13.5,0.2) -- (13.5-0.1,0) -- (13.5,-0.2);
\node at (11,1.25){$A_{2n-1}^{(2)}$};

\node at (9-1.732/2-0.1,0.5+0.25){$0$};  \node at (9-1.732/2-0.1,-0.5-0.25){$1$};
\node at (9+0.1,0.25){$2$};  \node at (10,0.25){$3$};
\node at (12-0.2,0.25){$n-2$};  \node at (13,0.25){$n-1$};  \node at (14+0.1,0.25){$n$};
\end{tikzpicture}
}
\\
(d5)
&
\raisebox{-5ex}{
\begin{tikzpicture}[scale=1, line width=0.75pt]
\node at (3.6,1.25){$D_{4}^{(1)}$};
\coordinate (V0) at (5-0.383,0.924);  \node at (5-0.383+0.25,0.924+0.1){$0$};
\coordinate (V1) at (5-0.924, 0.383);  \node at (5-0.924+0.1, 0.383+0.25){$1$};
\coordinate (V2) at (5,0);  \node at (5+0.25, 0){$2$};
\coordinate (V3) at (5-0.924, -0.383);   \node at (5-0.924-0.1, -0.383+0.25){$3$};
\coordinate (V4) at (5-0.383,-0.924);    \node at (5-0.383+0.25,-0.924){$4$};
\draw[fill](V0)circle(2pt);  \draw[fill](V1)circle(2pt);
\draw[fill](V2)circle(2pt);  \draw[fill](V3)circle(2pt);
\draw[fill](V4)circle(2pt);
\draw (V0) -- (V2)--(V4)  (V1)--(V2)--(V3);
\draw[fill=white] (V0) circle(2pt);

\draw[->, >=stealth, dashed, thin] (5-0.383,0.924-0.15) to [out=-90,in=0] (5-0.924+0.15, 0.383);
\draw[->, >=stealth, dashed, thin] (5-0.924+0.1, 0.383-0.1) to [out=-45,in=45] (5-0.924+0.1, -0.383+0.1);
\draw[<-, >=stealth, dashed, thin] (5-0.383,-0.924+0.15) to [out=90,in=0] (5-0.92+0.15 4, -0.383);
\draw[->, >=stealth, dashed, thin] (5-0.383-0.15,-0.924) .. controls (5-0.383-1.3,-0.924) and (5-0.383-1.3,0.924) ..  (5-0.383-0.15,0.924) ;
\node at (3.3,0){$\bar\sigma$};

\node at (7,0){$\rightsquigarrow$};

\foreach \i in {10}{
		\draw[fill](\i,0)circle(2pt);
}
\node at (9-0.1,0.25){$0$};  \node at (10+0.1,0.25){$1$};

\coordinate (V0) at (9,0);
\coordinate (V1) at (10,0);
\foreach \i in {1.5,0.5,-0.5,-1.5} { \draw[thin] (9,\i*0.05) -- (10,\i*0.05); }
\draw[fill=white] (V0) circle(2pt);
\draw (9.5,0.2) -- (9.5-0.1,0) -- (9.5,-0.2);

\node at (9.7,1.25){$A_{2}^{(2)}$};
\end{tikzpicture}
}
\\
(d7)
&
\raisebox{-5ex}{
\begin{tikzpicture}[scale=1, line width=0.75pt]
\node at (3,1.25){$D_{2n+1}^{(2)}$};
\foreach \i in {0,1,2,4,5}{
	\foreach \j in {-0.5,0.5}{
		\draw[fill](\i,\j)circle(2pt);
	}
}

\coordinate (V0) at (0,-0.5);
\coordinate (Vn) at (5+1.732/2,0);

\draw (2.5,0.5) -- (1,0.5)  (1,-0.5) -- (2.5,-0.5);
\draw (3.5,-0.5) -- (5,-0.5) -- (Vn) -- (5,0.5) -- (3.5,0.5);
\foreach \j in {0.5,-0.5} {
\foreach \i in {1,-1} { \draw (0,\j)+(0,\i*0.05) -- (1,\j+\i*0.05); }
\draw (0.5,\j+0.2) -- (0.5-0.1,\j) -- (0.5,\j-0.2);
}
\foreach \j in {-0.5,0.5}{
		\draw[dotted, thick] (2.5,\j) -- (3.5,\j);
	}
\draw[fill=white] (V0) circle(2pt);  \draw[fill](Vn)circle(2pt);
\foreach \i in {0,1,2,4,5}{
\draw[<->, >=stealth, dashed, thin] (\i-0.1, 0.5-0.1) to [out=-135,in=135] (\i-0.1,-0.5+0.1);
}
\node at (2.1,0){$\bar\sigma$};

\node at (0-0.1, -0.5-0.25){$0$};  \node at (1, -0.5-0.25){$1$};  \node at (2, -0.5-0.25){$2$};
\node at (4-0.2, -0.5-0.25){$n-2$};  \node at (5+0.2, -0.5-0.25){$n-1$};  \node at (5+1.732/2+0.25,0){$n$};
\node at (0-0.3, 0.5+0.25){$2n$};  \node at (1-0.2, 0.5+0.25){$2n-1$};  \node at (2+0.2, 0.5+0.25){$2n-2$};
\node at (4-0.2, 0.5+0.25){$n+2$};  \node at (5+0.2, 0.5+0.25){$n+1$};

\node at (7,0){$\rightsquigarrow$};

\foreach \i in {9,10,12,13,14}{
		\draw[fill](\i,0)circle(2pt);
}
\coordinate (V0) at (8,0);
\foreach \i in {1,-1} { \draw (8,0)+(0,\i*0.05) -- (9,\i*0.05); }
\draw (8.5,0.2) -- (8.5-0.1,0) -- (8.5,-0.2);
\draw (9,0) -- (10.5,0);
\draw[dotted, thick] (10.5,0) -- (11.5,0);

\foreach \i in {1,-1} { \draw (13,\i*0.05) -- (14,\i*0.05); }
\draw (13.5,0.2) -- (13.5-0.1,0) -- (13.5,-0.2);
\draw (11.5,0) -- (13,0);

\draw[fill=white] (V0) circle(2pt);
\node at (11,1.25){$A_{2n}^{(2)}$};

\node at (8-0.1, 0.25){$0$}; \node at (9, 0.25){$1$};  \node at (10, 0.25){$2$};
\node at (12-0.3, 0.25){$n-2$}; \node at (13, 0.25){$n-1$};  \node at (14+0.1, 0.25){$n$};
\end{tikzpicture}
}
\\
\\
(e2)
&
\raisebox{-7ex}{
\begin{tikzpicture}[scale=1, line width=0.75pt]
	\foreach \j in {-0.5,0.5}{
		\draw[fill](3.5,\j*1.732)circle(2pt);
\draw[fill](4.5,\j*1.732)circle(2pt);
	}
\foreach \i in {3,4,5}{
		\draw[fill](\i,0)circle(2pt);
}

\node at (2.75,1.25){$E_{6}^{(1)}$};
\node at (3.5-0.1,1.732/2+0.25){$0$};  \node at (4.5+0.1,1.732/2+0.25){$6$};
\node at (3-0.25,0){$1$};  \node at (4-0.15,0.25){$2$};  \node at (5+0.25,0){$3$};
\node at (3.5-0.1,-1.732/2-0.25){$5$};  \node at (4.5+0.1,-1.732/2-0.25){$4$};

\coordinate (V0) at (3.5,1.732/2);
\coordinate (V6) at (4.5, 1.732/2);
\coordinate (V1) at (3,0);
\coordinate (V2) at (4,0);
\coordinate (V3) at (5,0);
\coordinate (V4) at (4.5, -1.732/2);
\coordinate (V5) at (3.5, -1.732/2);
\draw (V0) -- (V6) -- (V3)  (V1)--(V2)--(V3)--(V4)--(V5);
\draw[fill=white] (V0) circle(2pt);

\foreach \i in {3.5,4.5}{
\draw [->, >=stealth, dashed, thin] (\i+0.05, -1.732/2+0.2) .. controls (\i+0.3,-0.3) and (\i+0.3,0.3)  .. (\i+0.05, 1.732/2-0.2);
\draw[->, >=stealth, dashed, thin] (\i-0.15, 1.732/2-0.05) to [out=-165,in=75] (\i-0.5,0.15);
\draw[<-, >=stealth, dashed, thin] (\i-0.15, -1.732/2+0.05) to [out=165,in=-75] (\i-0.5,-0.15);
}
\node at (2.8,0.5){$\bar\sigma$};

\node at (7,0){$\rightsquigarrow$};

\foreach \i in {10,11}{
		\draw[fill](\i,0)circle(2pt);
}
\node at (9-0.1,0.25){$0$};   \node at (10,0.25){$1$};  \node at (11+0.1,0.25){$2$};

\coordinate (V0) at (9,0);
\coordinate (V1) at (10,0);
\coordinate (V2) at (11,0);
\draw (V0) -- (V1);
\foreach \i in {1,-1} { \draw (10,\i*0.06) -- (11,\i*0.06); }  \draw[thin] (10,0) -- (11,0);
\draw[fill=white] (V0) circle(2pt);
\draw (10.5,0.2) -- (10.5-0.1,0) -- (10.5,-0.2);
\node at (10.3,1.25){$D_{4}^{(3)}$};
\end{tikzpicture}
}
\\
(e3)
&
\raisebox{-5ex}{
\begin{tikzpicture}[scale=1, line width=0.75pt]
\node at (3.4,1.25){$E_{7}^{(1)}$};
\foreach \i in {2,3,4}{
	\foreach \j in {-0.5,0.5}{
		\draw[fill](\i,\j)circle(2pt);
	}
\draw[<->, >=stealth, dashed, thin] (\i-0.1, 0.5-0.1) to [out=-135,in=135] (\i-0.1,-0.5+0.1);
}
\foreach \i in {4,5}{
		\draw[fill](\i+1.732/2,0)circle(2pt);
}
\node at (2-0.1,0.5+0.25){$0$};  \node at (3,0.5+0.25){$1$};   \node at (4+0.1,0.5+0.25){$2$};
\node at (2-0.1,-0.5-0.25){$6$};  \node at (3,-0.5-0.25){$5$};   \node at (4+0.1,-0.5-0.25){$4$};
\node at (4+1.732/2+0.1,0.25){$3$};  \node at (5+1.732/2+0.1,0.25){$7$};

\coordinate (V0) at (2,0.5);
\draw (V0) -- (4,0.5) --  (4+1.732/2,0)--(4,-0.5)--(2,-0.5) (4+1.732/2,0)--(5+1.732/2,0);
\draw[fill=white] (V0) circle(2pt);

\node at (1.5,0){$\bar\sigma$};

\node at (7.5,0){$\rightsquigarrow$};

\foreach \i in {10,11,12,13}{
		\draw[fill](\i,0)circle(2pt);
}
\node at (9-0.1,0.25){$0$}; \node at (10,0.25){$1$}; \node at (11,0.25){$2$};
\node at (12,0.25){$3$};  \node at (13+0.1,0.25){$4$};

\coordinate (V0) at (9,0);
\coordinate (V1) at (10,0);
\coordinate (V2) at (11,0);
\coordinate (V3) at (12,0);
\coordinate (V4) at (13,0);
\draw (V0) -- (V2)  (V3)--(V4);
\foreach \i in {1,-1} { \draw (11,\i*0.05) -- (12,\i*0.05); }
\draw[fill=white] (V0) circle(2pt);
\draw (11.5,0.2) -- (11.5-0.1,0) -- (11.5,-0.2);
\node at (11.3,1.25){$E_{6}^{(2)}$};
\end{tikzpicture}
}
\end{tabular}
\end{center}
\end{table}

Let us consider an affine Kac-Moody algebra $\fg$ of type $X_{\ell'}^{(r)}$ given in Tables~\ref{table-diag1}--\ref{table-diag3}, on which there is a diagram automorphism $\sg$ induced by $\bar\sg$ \cite{Kac}.  Denote by $\fg^\sg$ the subalgebra of $\fg$ that consists of all $\sg$-invariant elements. Note that the subalgebra $\fg^{\sg}$ may not be the affine Kac-Moody algebra of type $X_{\bar{\ell}'}^{(\bar r)}$ associated to the folded Dynkin diagram (as we already pointed out, $\fg^\sg$ is indeed such a subalgebra for each case listed in Table~\ref{table-diag1}). Instead, $\fg^{\sg}$ may be the direct sum of such a subalgebra and a certain linear subspace. So what we need is to figure out the difference between these subalgebras. For this purpose, let us decompose $\fg$ according to its principal gradation as  $\fg=\bigoplus_{k\in\Z}\fg^k$, then its $\sigma$-invariant subalgebra is also decomposed as $\fg^\sg=\bigoplus_{k\in\Z}(\fg^\sg)^k$. Let $\bar\fg$ be a graded subalgebra of $\fg$ generated by $(\fg^\sg)^{-1},\, (\fg^\sg)^0, \,
(\fg^\sg)^1$ . Then we have $\bar\fg\subset\fg^\sg$ and the following theorem.

\begin{thm}\label{thm-algred}
For any case listed in
Tables~\ref{table-diag1}--\ref{table-diag3},  the following assertions hold true:
\begin{enumerate} \renewcommand{\labelenumi}{\rm{(\roman{enumi})}}
\item $\bar\fg$ is an affine Kac-Moody
algebra associated to the folded Dynkin diagram
$X_{\bar{\ell}'}^{(\bar r)}$;
\item  $\fg^\sg=\bar\fg$ for any case listed in Tables~\ref{table-diag1} and \ref{table-diag2};
\item $\fg^\sg=\bar\fg\oplus\hat{\mathcal{H}}$ for any case listed in Table~\ref{table-diag3}, where
$\hat{\mathcal{H}}$ is a nontrivial subspace of the principal Heisenberg
subalgebra $\mathcal{H}$ of $\fg$ (see Theorem~\ref{thm-ggbar} below).
\end{enumerate}
\end{thm}
Although this theorem looks very basic in the theory of Lie algebra, we could not find it in the literature (except those cases in Table~\ref{table-diag1}, see, e.g. \cite{Kac}), so we will give a proof of it in the present paper.

Based on Theorem~\ref{thm-algred}, we will study the
relationship between the Drinfeld-Sokolov hierarchies associated to
$\fg$ and to $\bar\fg$.
In the original construction of the Drinfeld-Sokolov hierarchy \cite{DS}, two gradations of $\fg$ play important roles (see \eqref{sset}--\eqref{fgks} below): one is the principal gradation $\mathds{1}=\left(1, \dots, 1\right)$, and the other one
\[
\rs_m=\left(s_0, \dots, s_\ell\right), \quad \mbox{with }s_i=\delta_{i m}
\]
is given by the $m$-th marked vertex.
Observe that the gradation $\rs_m$ is not preserved by the automorphism $\bar\sg$ for all cases in Tables~\ref{table-diag1}--\ref{table-diag3}. So in order to obtain an analogue of the $\Gamma$-reduction theorem for these cases, we will start from a certain generalization of the Drinfeld-Sokolov hierarchies proposed by de Groot et al. in \cite{dGHM} (cf.\,\cite{FHM}). To construct such a generalized version of the Drinfeld-Sokolov hierarchy we need to fix, instead of the gradation $\rs_m$, an arbitrary gradation $\rs=(s_0,s_1,\dots,s_\ell)$ (which belongs to the set $S$ defined in \eqref{sset}) of $\fg$ satisfying the condition $\rs\le \mathds{1}$ and being consistent with the given diagram automorphism $\sigma$ of $\fg$, namely, $s_i\le1$, and $s_{\bar\sg(i)}=s_i$ for all $i$. Accordingly, the gradation $\rs$ induces a gradation $\bar\rs$ of $\bar\fg$.
Denote by $\mH$, $\bar\mH$ the Heisenberg subalgebras corresponding to the principal gradations of $\fg$ and $\bar\fg$ respectively, and $\mH^\sigma=\fg^\sigma\cap\mH$.
Then by appropriately choosing a basis of $\mH$ we can represent these subalgebras as
\[
\mH=\bigoplus_{j\in J}\C\Ld_j\oplus\C c,\quad \bar\mH=\bigoplus_{j\in \bar{J}}\C\Ld_j\oplus\C c,\quad  \mH^\sigma=\bigoplus_{j\in J^\sigma}\C\Ld_j\oplus\C c.
\]
Here $c$ is the center of the subalgebras, $\Lambda_j\in\fg^j$, $J, \bar{J}$ are the sets of exponents of $\fg$ and $\bar\fg$ respectively,
$J^\sigma=\{j\in J\,|\, \sigma(\Lambda_j)=\Lambda_j\}$, and they satisfy the relation
\[ \bar{J}\subseteq J^\sigma\subseteq J.\]
Then associated to the affine Kac-Moody algebra $\fg$ together with the gradations $\rs$ and $\mathds{1}$, we have the Drinfeld-Sokolov hierarchy consisting of the flows
${\p}/{\p t_j},\, j\in J_+$, where $J_+$ is the set of positive exponents of $\fg$.
Similarly, for the affine Kac-Moody algebra $\bar\fg$, we also have the Drinfeld-Sokolov hierarchy with the flows
${\p}/{\p \bar{t}_j},\, j\in \bar{J}_+$, that are associated to the gradations $\bar\rs$ and $\mathds{1}$ of $\bar{\fg}$.
The flows ${\p}/{\p t_j}$ and ${\p}/{\p \bar{t}_j}$ are defined respectively on the jet space $\mathcal{R}$ and $\bar{\mathcal{R}}$ of certain finite dimensional subspaces $\mathcal{V}\subset\fg$ and $\bar{\mathcal{V}}=\mathcal{V}\cap \bar\fg$ that will be defined in \eqref{BV} and \eqref{zh-11}.

The main result of the present paper is given by the following theorem.
\begin{thm}\label{thm-main}
Let $\fg=X_{\ell'}^{(r)}$ be an affine Kac-Moody algebra whose Dynkin diagram is given in
Tables~\ref{table-diag1}--\ref{table-diag3}, on which there is a
diagram automorphism $\sg$ induced by $\bar\sg$.
Fix a gradation $\rs\le\mathds{1}$ on $\fg$
being consistent with $\sg$, and let $\bar\rs$ be the gradation on $\bar\fg$ induced by $\rs$.
Denote by $\Gm$ the finite group generated by $\sg$, then this group acts on the Drinfeld-Sokolov hierarchy associated to $(\fg, \rs, \mathds{1})$. Furthermore, we have the following assertions:
\begin{enumerate} \renewcommand{\labelenumi}{\rm{(\roman{enumi})}}
\item  The flows $\pd/\pd t_j$ are $\Gm$-invariant if and only if $j\in J^\sg_+$.
\item   When $j\in \bar{J}_+$, the restrictions of $\pd/\pd t_j$ on $\bar{\mathcal{R}}$ are proportional to the Drinfeld-Sokolov hierarchy $\p/\p\bar{t}_j, j\in\bar{J}_+$, associated to $(\bar\fg, \bar\rs, \mathds{1})$.
\item  When $j\in J^\sg_+\backslash\bar{J}_+$, the restrictions of $\pd/\pd t_j$ on $\bar{\mathcal{R}}$ are zeroes.
\item  For any case in Tables \ref{table-diag1} and \ref{table-diag2}, suppose that $\tau^\rs$ is a
$\Gm$-invariant tau function (see Definition~\ref{def-Gmtau} below) of the Drinfeld-Sokolov hierarchy associated to $(\fg, \rs, \mathds{1})$, then
\begin{equation}\label{tautaubar0}
\log\bar{\tau}^{\bar\rs}=\frac{1}{\mu}\left.\log\tau^\rs
\right|_{t_j=\sqrt{\mu}\,\bar{t}_j, \, j\in\bar{J}_+; ~ t_j=0, \,
j\in J_+\setminus\bar{J}_+}
\end{equation}
gives a tau function
$\bar{\tau}^{\bar\rs}$ of the Drinfeld-Sokolov hierarchy associated to
$(\bar\fg, \bar\rs, \mathds{1})$. Here the notion of tau function is given by Definition \ref{zh-10}, and $\mu$ is a constant listed in Appendix~\ref{zh-2}.
\end{enumerate}
\end{thm}

This theorem generalizes Theorem~\ref{thm-LRZ}. Especially, we conclude that the Drinfeld-Sokolov hierarchies of twisted type can be obtained by reduction of those hierarchies of ADE type. Here we remark that the notion of tau function given by Definition \ref{zh-10} generalizes the ones for the Drinfeld-Sokolov hierarchies that are introduced in \cite{EF, Mi, Wu}, see Remark \ref{rmk-tau} below.
We also emphasize that the fourth assertion of Theorem \ref{thm-main} does not hold true for cases listed in Table \ref{table-diag3} in general, see Remark~\ref{rmk-tauTab3} below for a counterexample.

The paper is organized as follows. In Section~2 we recall some basic properties
of affine Kac-Moody algebras. In Section~3 we give the definitions of
Drinfeld-Sokolov hierarchies and their tau functions used in the present paper.
In Section~4 we prove Theorems~\ref{thm-algred} and \ref{thm-main}.
In the final section, we give some remarks on the application of Theorem~\ref{thm-main} to the so-called topological tau functions.

\section{Preliminary knowledge on affine Kac-Moody algebras}\label{sec-g}

In this section, we recall the definition and some basic properties of affine
Kac-Moody algebras mainly following \cite{DS, Kac}, as a preparation for what follows.

\subsection{Affine Kac-Moody algebra and its principal Heisenberg subalgebra}\label{zh-15}

Let $A=(a_{i j})_{0\le i, j\le \ell}$ be a generalized Cartan
matrix of affine type $X_{\ell'}^{(r)}$ with $r=1,2,3$ (in
particular, $\ell=\ell'$ when $r=1$), and $\{k_0, k_1, \dots, k_\ell\}$,
$\{k_0^\vee, k_1^\vee, \dots, k_\ell^\vee\}$ be the sets of Kac labels and dual Kac labels
respectively (see \cite{dGHM} for their definition). It is known that $k_i^\vee a_{i j}k_j=k_j^\vee a_{j i}
k_i$ for $i, j=0,1, \dots, \ell$.

Denote by $\fg(A)$ the complex affine Kac-Moody algebra associated to
$A$, with $\mathfrak{h}$ being a fixed Cartan subalgebra. Let  $\Pi=\{\al_0,
\al_1, \dots, \al_\ell\}$ and  $\Pi^\vee=\{\al_0^\vee, \al_1^\vee,
\dots, \al_\ell^\vee\}$ be the sets of  simple roots and simple
coroots corresponding to $\mathfrak{h}$ respectively.  Then  $\fg(A)$
has the following root space decomposition:
\begin{equation}\label{rootdec}
\fg(A)=\mathfrak{h}\oplus\left(\bigoplus_{\al\in\Delta}\fg_\al
\right),
\end{equation}
where $\Delta$ is the root system of $\fg(A)$.
In  $\fg(A)$ there is  a set of Chevalley generators $\{e_i\in\fg_{\al_i}, f_i\in\fg_{-\al_i} \mid i=0, 1, \dots,
\ell\}$, which satisfy the following Serre relations:
\begin{align}\label{}
[e_i, f_j]=\dt_{i j}\al^\vee_i,& \quad [\al^\vee_i, \al^\vee_j]=0, \\
[\al^\vee_i, e_j]=a_{i j}e_j, & \quad [\al^\vee_i, f_j]=-a_{i j}f_j,
\\
(\ad_{e_i})^{1-a_{i j}}e_j=0, & \quad (\ad_{f_i})^{1-a_{i j}}f_j=0.
\label{adee}
\end{align}
Here $\dt_{ij}$ is the Kronecker delta function, $i, j=0, 1, 2, \dots,
\ell$, and $i\ne j$ in the equations of \eqref{adee}.

The Cartan subalgebra of $\fg(A)$ is given by
\[
\mathfrak{h}=\C\al_0^\vee\oplus\C\al_1^\vee\oplus\dots\oplus\C\al_\ell^\vee\oplus\C
d,
\]
where $d$ is a scaling element satisfying
\begin{equation}\label{}
[d, e_i]=e_i, \quad [d, f_i]=-f_i, \quad i=0,1,\dots, \ell.
\end{equation}
Note that
\begin{equation}
c=\sum_{i=0}^\ell k_i^\vee \al^\vee_i\in \mathfrak{h}
\end{equation}
is the  canonical central element of $\fg(A)$.
On $\mathfrak{h}$ there is a nondegenerate symmetric bilinear form
such that
\begin{align}\label{}
(\al_i^\vee\mid\al_j^\vee)=a_{i j}\frac{k_j}{k_j^\vee}, \quad
(d\mid\al_j^\vee)=\frac{k_j}{k_j^\vee}, \quad (d\mid d)=0,
\end{align}
where $i,j=0,1, \dots, \ell$. Clearly,
\begin{equation}\label{Coxeter}
(d\mid c)=\sum_{i=0}^\ell k_i=:h,
\end{equation}
which is called the Coxeter number of $\fg(A)$. This bilinear form can be
uniquely extended to the normalized invariant symmetric bilinear
form $(\cdot\mid\cdot)$ on $\fg(A)$. We remark that, when restricted to the simple Lie algebra generated by $\{e_i, f_i\mid i=1,2,\dots\ell\}$, this bilinear form is a multiple of the Cartan-Killing form on the simple Lie algebra, see \cite{Kac} for more details.

Let $\fg=[\fg(A),\fg(A)]$ be the derived algebra of $\fg(A)$, then we have
$\fg(A)=\fg\oplus\C d$. In what follows we will mainly
use the derived algebra $\fg$ rather than $\fg(A)$. Note that $\fg$ is generated by the Chevalley generators, and we will also call $\fg$
the affine Kac-Moody algebra associated to $A$.

The adjoint action of $d$ induces on $\fg$ the principal gradation
\begin{equation}\label{prin}
\fg=\bigoplus_{k\in\Z}\fg^k, \quad \fg^k=\left\{X\in\fg\mid [d,X]=k
X\right\}.
\end{equation}
Fix the following cyclic element
\[
\Ld=\sum_{i=0}^\ell  e_i\in\fg^1
\]
and consider its adjoin action on $\fg$, then we obtain
\begin{equation}\label{dec}
\fg=\im\,\ad_{\Ld}+\mathcal{H}, \quad
\im\,\ad_{\Ld}\cap\mathcal{H}=\C c,
\end{equation}
where $\mathcal{H}=\{X\in\fg\mid \ad_\Lambda X\in \C c\}$ is called the principal Heisenberg subalgebra of $\fg$. More precisely, the  principal Heisenberg subalgebra can be represented as
\begin{equation}\label{zh-1}
\mathcal{H}=\bigoplus_{j\in J}\C\Ld_j\oplus\C c,
\end{equation}
where
\begin{equation}\label{mirh}
J=\{m_1, m_2, \dots, m_{\ell'}\}+r h\Z, \quad 1= m_1< m_2\le m_3\le \dots \le m_{\ell'-1}
< m_{\ell'}=r h-1,
\end{equation}
is the set of exponents of $\fg$ and $\Ld_j\in\fg^j$, and the basis elements can be chosen to satisfy the following relations:
\begin{align}
[\Ld_i, \Ld_j]=i \dt_{i,-j} c, \quad i, j\in J. \label{Ldij}
\end{align}
In particular, we have
\[
\Ld_1=\nu\Ld
\]
for a certain constant $\nu$. Note that these generators also have the following properties:
\begin{equation}\label{blfLd}
(\Ld_i\mid\Ld_j)=\frac{1}{i}([d,\Ld_i]\mid\Ld_j)=\frac{1}{i}(d\mid[\Ld_i,\Ld_j])=\dt_{i,-j}h.
\end{equation}

Denote
\begin{equation}\label{zh-4}
\mathcal{I}^k=\left(\im\,\ad_\Ld\right)\cap \fg^k\ \textrm{for}\ k\ne 0,\quad \mathcal{I}^0=\ad_\Ld\left(\mathcal{I}^{-1}\right),\quad
\mathcal{I}=\bigoplus_{k\in\Z}\mathcal{I}^k.
\end{equation}
From the relations given in \eqref{dec} and the definition of $\mathcal{H}$  it follows that $c\notin\mathcal{I}^0$, hence
\begin{equation}\label{zh-7}
\fg=\mathcal{I}\oplus \mathcal{H}.
\end{equation}
This implies that $\mathcal{I}\cap\Ker\,\ad_\Ld=\{0\}$, from which it follows that the maps
\begin{equation}\label{zh-6}
\ad_\Ld: \mathcal{I}^k\longrightarrow \mathcal{I}^{k+1}, \quad
k\in\Z
\end{equation}
are injective. On the other hand, the space $\mathcal{I}^k$ are finite-dimentional ($\dim\mathcal{I}^k=\ell$, see Proposition~3.8 in \cite{Kac78}), so the maps \eqref{zh-6} are bijections.

\subsection{Gradations on an affine Kac-Moody algebra}
Let us denote
\begin{equation}\label{sset}
{S}=\{\rs=(s_0, s_1, \dots,
s_\ell)\in\Z^{\ell+1}\mid s_i\ge0, s_0+s_1+\cdots+s_\ell>0\}.
\end{equation}
For each $\rs=(s_0, s_1, \dots, s_\ell)\in S$, there is an element $d^{\rs}$ in the Cartan subalgebra
$\mathfrak{h}$ of $\fg(A)$ fixed via the nondegenerate bilinear form $(\cdot\mid\cdot)$ by the relations
\begin{equation}\label{dsal}
(d^{\rs}\mid \al_i^\vee)=\frac{k_i}{k_i^{\vee}}s_i, \quad
i=0,1,\dots,\ell; \quad (d^{\rs}\mid d^{\rs})=0.
\end{equation}
It follows from the invariance of $(\cdot\mid\cdot)$ that $(d^\rs\mid X)=0$ when $X|_{\fg^0}=0$ with respect to the decomposition \eqref{prin}. So, for $X\in\fg$ with $X|_{\fg^0}=\sum_{i=0}^\ell x_i\al_i^\vee$  we have
\begin{equation}\label{zh-5}
(d^\rs\mid X)=\sum_{i=0}^\ell x_i\left(d^\rs\mid\al_i^\vee\right)=\sum_{i=0}^\ell\frac{  x_i k_i s_i}{ k_i^\vee}.
\end{equation}
In particular,
\begin{align}\label{dsc}
&(d^{\rs}\mid c)=k_0 s_0+k_1 s_1+\dots+k_\ell
s_\ell=:h^\rs.
\end{align}
The fact $h^\rs>0$ also implies that $\fg\oplus\C d^\rs=\fg\oplus\C d=\fg(A)$.
In particular, if we denote $\mathds{1}:=(1,1,\dots,1)$, then we have $d^{\mathds{1}}=d$, and $h^\mathds{1}=h$ is just the Coxeter number of $\fg(A)$.

The element $d^{\rs}$ satisfies the commutation relations
\[
[d^{\rs}, e_i]=s_i e_i, \quad [d^{\rs}, f_i]=-s_i f_i, \quad
i=0,1,\dots,\ell,
\]
and it induces a gradation
\begin{equation}\label{fgks}
\fg=\bigoplus_{k\in\Z}\fg_{k [\rs]}, \quad \fg_{k [\rs]}=\{X\in\fg\mid [d^\rs, X]=k X\}.
\end{equation}
For instance, when $\rs=\mathds{1}$ we arrive at the principal gradation \eqref{prin} of $\fg$.
We also have the following relations for any $X_k\in \fg_{k[\rs]}$, $X_l\in \fg_{l [\rs]}$:
\begin{equation}\label{brcen}
(d^\rs\mid [X_k, X_l] )=\dt_{k,-l} k (X_k\mid X_l).
\end{equation}

Let us recall briefly how to realize $\fg$ starting from a simple Lie algebra, see \S\,8.6
of \cite{Kac} for details.
Let $\mathcal{G}$ be a simple Lie algebra of type $X_{\ell'}$,
on which there is a diagram automorphism of order $r$. The integer
vector $\rs\in S$ induces on $\mathcal{G}$ a $\Z/r h^\rs\Z$-gradation
$\mathcal{G}=\bigoplus_{k=0}^{r h^\rs-1}\mathcal{G}_k$. Then $\fg$ with gradation $\rs$ can be
realized as
\begin{equation}\label{gAs}
\fg=\bigoplus_{k\in\Z}\left( z^k\otimes\mathcal{G}_{k\, \mathrm{mod}\, r
h^{\rs}}\right)\oplus \C\,c.
\end{equation}
More precisely, for any elements $X(k), Y(k)\in
 z^k\otimes\mathcal{G}_{k\, \mathrm{mod}\, r h^\rs}$ and parameters $\xi, \eta\in \C$, the Lie bracket is defined by
\begin{align}\label{XYbr}
&[X(k)+\xi\cdot c, Y(l)+\eta \cdot c]=[X, Y](k+l) +\dt_{k,-l}\frac{k}{r
h^\rs}(X\mid Y)_\mathcal{G}\cdot c,
\end{align}
where $(\,\cdot\mid\cdot\,)_\mathcal{G}$ is the standard
nondegenerate invariant symmetric bilinear form on $\mathcal{G}$.  The normalized invariant bilinear form on $\fg$ is then given by
\begin{equation}\label{}
(X(k)+\xi \cdot c\mid Y(l)+\eta\cdot c)=\dt_{k,-l}\frac{1}{r}(X\mid Y)_\mathcal{G}.
\end{equation}
In fact, one can choose  elements $E_i$, $F_i$ and $H_i$ of
$\mathcal{G}$ with $i=0, 1, \dots, \ell$ which yield the Weyl generators
of $\fg$ as  follows (see \S\,8.3 of \cite{Kac}):
\begin{equation}\label{weyls}
e_i=E_i(s_i), \quad  f_i=F_i(-s_i), \quad
\al_i^{\vee}=H_i(0)+\frac{k_i s_i}{k^\vee_i h^\rs}\cdot c.
\end{equation}

\section{The Drinfeld-Sokolov hierarchies}\label{sec-3}

In this section, we recall the definition of the Drinfeld-Sokolov hierarchies of
both untwisted and twisted types, and then introduce the notion of their tau
functions. To this end, let us consider two gradations on $\fg$: the principal gradation $\mathds{1}$, and a gradation  $\rs=(s_0,s_1,s_2,\dots,s_\ell)\in S$ with
$s_i\le1$ ($\rs\le\mathds{1}$ for short). For convenience we will write
\[
\fg^k=\fg_{k [\mathds{1}]}, \quad \fg_k=\fg_{k [\rs]},
\]
and use the notations $\fg_{\ge l}=\bigoplus_{k\ge
l}\fg_k$, $\fg^{<l}=\bigoplus_{k< l}\fg^k$ etc.

\subsection{The dressing lemma and the pre-Drinfeld-Sokolov hierarchy}

Let us introduce a Borel subalgebra
of $\fg$ as follows:
\begin{equation}\label{Borel}
\mathcal{B}=\left\{X\in\fg_{0}\cap\fg^{\le 0}\mid (d^\rs\mid
X)=0\right\},
\end{equation}
and consider operators of the form
\begin{equation}\label{sL}
\sL=\frac{\od}{\od x}+\Ld+Q,\quad Q=Q(x)\in C^{\infty}(\mathbb{R}, \mathcal{B}).
\end{equation}
The following lemma plays a crucial role in the what follows, and we will call it the dressing lemma (cf. \cite{DS,dGHM,Wu}).

\begin{lem}\label{thm-dr0}
Given an operator $\sL$ of the form \eqref{sL}, then there exists a unique
function $U\in C^\infty(\R, \fg^{<0})$ satisfying the following two
conditions:
\begin{align}
(i)\quad   &e^{-\ad_U}\sL=\frac{\od}{\od x}+\Ld+H,  \quad H\in
C^\infty(\R, \mathcal{H}\cap\fg^{<0}), \label{UL}\\
(ii)\quad &\left(d^\rs\mid e^{\ad_U}\Ld_j\right)=0\quad
\textrm{for any positive exponent}\ j\in J_{+}.\label{ULdc}
\end{align}
Moreover, both $U$ and $H$ are differential polynomials in the components of $Q$ with respect to a basis of $\mathcal{B}$ (differential polynomials in $Q$ for short below) with zero constant terms.
\end{lem}
\begin{prf}
The existence of $U\in C^\infty(\R, \fg^{<0})$  and  $H\in C^\infty(\R, \mathcal{H}\cap\fg^{<0})$  satisfying the condition \eqref{UL} was proved by Drinfeld and Sokolov in \cite{DS} based on the decomposition property
\eqref{dec} of $\fg$. The condition \eqref{ULdc} is imposed to ensure the uniqueness of $U$ and $H$. In the case when $\rs=(1,0,0,\dots, 0)$ the conclusion is proved in \cite{Wu}. For the general case the proof is similar to that of \cite{Wu}, which we give below for the convenience of the readers.

Let us write the equation \eqref{UL} in the form
\begin{equation}
e^{\ad_{\sum_{k\le-1}U_k} }\left(\frac{\od}{\od x}+\Ld+\sum_{k\le-1}H_k\right)=\frac{\od}{\od x}+\Ld +
\sum_{k\le0}Q_k,
\end{equation}
where $U_k, H_k, Q_k$ take values in $\fg^k$ (note that $Q_k=0$ when $k\le-h'$, where $h'$ is the smallest positive integer satisfying the condition $\mathcal{B}\cap\fg^{-h'}=\{0\}$). By comparing the
homogeneous terms of both sides of the above equation, we obtain
\begin{align}\label{Um1}
[U_{-1}, \Ld]&=Q_0, \\
[U_{k},\Ld]+H_{k+1}+W_k&=Q_k, \quad k=-2,-3,\dots. \label{HUk}
\end{align}
Here $W_k\in C^\infty(\mathbb{R}, \fg^{k+1})$ depends on
$U_i$ and  $H_{i+1}$ with $i>k$ and their $x$-derivatives.

From the decomposition \eqref{dec} of $\fg$ and the properties \eqref{Ldij} of
the elements of the Heisenberg subalgebra \eqref{zh-1}, it follows the existence of
$U_{-1}\in C^{\infty}(\mathbb{R}, \fg^{-1})$ satisfying \eqref{Um1}. The property of the Heisenberg subalgebra also implies that the map
$\ad_{\Ld}: \fg^{-1}\to\fg^0$ is injective, so the solution $U_{-1}$ of the
equation \eqref{Um1} is unique. Moreover,
the condition \eqref{ULdc} with $j=1$ is satisfied since
\[
(d^\rs\mid [U_{-1}, \Ld])=(d^\rs\mid Q_0)=0.
\]
For $k\le-2$, suppose that $U_i$ and $H_{i+1}$ with $i>k$ are given, then $W_k$ is known. By using the decomposition
\eqref{dec} of $\fg$ we know the existence of solution $U_k\in C^\infty(\mathbb{R}, \fg^{k})$, $H_{k+1}\in C^\infty(\mathbb{R}, \fg^{k+1})$ of the equation \eqref{HUk}, and we also know the uniqueness of $H_{k+1}$ since $k<-1$.
In order to prove the uniqueness of $U_k$,
let us expand $U_k=U_k|_{\mathcal{I}}+U_k|_{\mathcal{H}}$ with respect to the decomposition \eqref{zh-7}, and consider the following two cases:
\begin{itemize}
\item  If $k\not\in J$, then $U_k|_{\mathcal{H}}=0$, and from the property of the Heisenberg subalgebra $\mathcal{H}$ we know that the map $\ad_{\Ld}: \fg^k\to \fg^{k+1}$ is injective, so
$U_k=U_k|_{\mathcal{I}}$ is uniquely determined by the equation \eqref{HUk}.
\item  If $k\in J$, then $U_k|_{\mathcal{I}}$ is uniquely determined by the equation \eqref{HUk} for the same reason as above, so we are left to determine $U_k|_{\mathcal{H}}$. In the case when the linear space $\mathcal{H}\cap\fg^k$ is $1$-dimensional, $U_k|_{\mathcal{H}}=a_k\Ld_k$ for some scalar function $a_k$. From \eqref{zh-5} we know that the condition \eqref{ULdc} takes the form
\[
\left(d^\rs\mid [a_k\Ld_k, \Ld_{-k}]\right)=\left(d^\rs\mid Z_k\right),
\]
where
\begin{equation}\label{bkU}
Z_k:=-[ U_k|_{\mathcal{I}}, \Lambda_{-k} ]-\sum_{m=2}^{-k}\frac{1}{m!}\sum_{\substack{k+1\le k_i\le -1 \\ k_1+k_2+\dots+k_m=k}} \ad_{U_{k_1}} \ad_{U_{k_2}}\dots \ad_{U_{k_m}} \Ld_{-k}.
\end{equation}
Hence by using \eqref{Ldij} and \eqref{dsc} we obtain
\[
a_k=\frac{\left(d^\rs\mid Z_k\right)}{(d^\rs\mid k c)}=\frac{\left(d^\rs\mid Z_k\right)}{k h^\rs}.
\]
In the case when the linear space $\mathcal{H}\cap\fg^k$ is $2$-dimensional (only when $k\in \{n+2n\Z\}$ for the affine Kac-Moody algebras of type $D_{2(n+1)}^{(1)}$), $U_k|_{\mathcal{H}}=a_k\Ld_k+a_{k'}\Ld_{k'}$ with (recall \eqref{blfLd})
\[
(\Ld_k\mid \Ld_{k})=(\Ld_{k'}\mid \Ld_{k'})=h, \quad (\Ld_k\mid \Ld_{k'})=0.
\]
Then the condition \eqref{ULdc} takes the form
\[
\left(d^\rs\mid [a_k\Ld_k+a_{k'}\Ld_{k'}, \Ld_{-k}]\right)=\left(d^\rs\mid Z_k\right), \quad \left(d^\rs\mid [a_k\Ld_k+a_{k'}\Ld_{k'}, \Ld_{-k'}]\right)=\left(d^\rs\mid Z_{k'}\right),
\]
where $Z_k$ and $Z_{k'}$ have similar expressions as that of \eqref{bkU}, so the coefficients $a_k$ and $a_{k'}$ are uniquely fixed.
\end{itemize}
The lemma is proved.
\end{prf}

\begin{rmk}
Due to the uniqueness result of the dressing lemma, we will
use $U(Q)$ and $H(Q)$ to denote the functions $U$ and $H$ that are determined by the dressing lemma to emphasize their dependence on $Q$ when it is needed.
\end{rmk}

Let the function $U$ be determined by Lemma~\ref{thm-dr0}, then from the properties \eqref{zh-5}, \eqref{brcen} we know that the
following family of evolutionary equations are well defined:
\begin{equation}\label{Lt2}
\frac{\pd \sL}{\pd t_j}=[-(e^{\ad_U}\Ld_j)_{\ge 0}, \sL], \quad j\in
J_{+}.
\end{equation}
Here and in what follows, we denote by $X_{\ge 0}$ the projection of $X\in\fg$ to
$\fg_{\ge 0}$ w.r.t. the gradation \eqref{fgks} given by $\rs$.

Denote $\sL'={\od}/{\od x}+\Ld+H$ and $V={\pd U}/{\pd t_j}$, then from the identity $\sL=e^{\ad_U}\sL'$ we obtain
\begin{align}
 \frac{\pd\sL}{\pd t_j}&=\sum_{k\ge 0} \frac1{(k+1)!} \sum_{l=0}^k \ad_U^l \ad_{V}\ad_U^{k-l} \sL'+e^{\ad_U} \frac{\pd H}{\pd t_j}\notag\\
&=\sum_{k\ge 0} \frac1{(k+1)!} \sum_{l=0}^k \sum_{m=0}^l \binom{l}{m}\ad_{\ad_U^m V}\ad_U^{l-m+k-l} \sL'+e^{\ad_U} \frac{\pd H}{\pd t_j}\notag\\
&=\sum_{k\ge 0} \frac1{(k+1)!} \sum_{m=0}^k \binom{k+1}{m+1}\ad_{\ad_U^m V}\ad_U^{k-m} \sL'+e^{\ad_U} \frac{\pd H}{\pd t_j}\notag\\
&=\sum_{m\ge 0}\sum_{k\ge m} \frac1{(k-m)! (m+1)!} \ad_{\ad_U^m V}\ad_U^{k-m} \sL'+e^{\ad_U} \frac{\pd H}{\pd t_j}\notag\\
&=\sum_{m\ge 0} \frac1{(m+1)!} \ad_{\ad_U^m V}e^{\ad_U} \sL'+e^{\ad_U} \frac{\pd H}{\pd t_j}\notag\\
&=\left[\nabla_{t_j, U} U, \sL\right]+e^{\ad_U} \frac{\pd H}{\pd t_j}.\label{sLtj}
\end{align}
Here we use the notation
\begin{equation}\label{nablaUt}
\nabla_{t_j, U} U=\sum_{m\ge0}\frac1{(m+1)!}(\ad_U)^m\frac{\pd
U}{\pd t_j}.
\end{equation}
We remark that the identity \eqref{sLtj} was stated, without proof, in \cite{TT}.

Denote
\begin{equation}\label{zh-18}
G(\Ld_j)=e^{-\ad_U}\left(\nabla_{t_j, U} U-(e^{\ad_U}\Ld_j)_{<0}\right),
\quad j\in J_+,
\end{equation}
then we have the following lemma.

\begin{lem}\label{thm-Gi}
The functions $G(\Ld_j)$ take value in $\mathcal{H}\cap\fg^{<0}$, and
satisfy the equations
\begin{align}
&\frac{\pd G(\Ld_j)}{\pd x}=\frac{\pd H}{\pd t_j}, \quad [\Ld,
G(\Ld_j)]=[\Ld_j, H]. \label{Gj}\\
&\frac{\pd}{\pd
t_j}e^{\ad_U}\Ld_i=\left[\left(e^{\ad_U}\Ld_j\right)_{<0},
e^{\ad_U}\Ld_i\right]+[G(\Ld_j), \Ld_i], \quad i,j\in J_+.\label{eUt}
\end{align}
\end{lem}
\begin{prf}
From the equations \eqref{Lt2} and \eqref{sLtj} it follows that
\begin{equation}\label{eq1}
[\nabla_{t_j, U} U-(e^{\ad_U}\Ld_j)_{<0}, \sL]+e^{\ad_U}\frac{\pd
H}{\pd t_j}+[e^{\ad_U}\Ld_j,\sL]=0.
\end{equation}
Due to the decomposition \eqref{zh-7}, the function $G(\Ld_j)$ can be uniquely represented as
\[
G(\Ld_j)=A+B, \ \textrm{where}\  A\in \mathcal{I}\cap\fg^{<0},\ B\in \mathcal{H}\cap\fg^{<0}.
\]
Then the equation \eqref{eq1} can be rewritten as
\begin{equation}\label{HU}
 \left[A+B, \frac{\od}{\od x}+\Ld+H\right]+\frac{\pd H}{\pd t_j}+
 [\Ld_j, H]=0.
\end{equation}
We write $A=\sum_{i<0}A_i$ and $B=\sum_{i<0}B_i$ according to the principal gradation $\fg^{<0}=\bigoplus_{i<0}\fg^i$.
The degree zero part of \eqref{HU} reads
\[
[A_{-1}+B_{-1}, \Ld] + [\Ld_j, H]=0.
\]
Note that $[A_{-1},\Ld]\in \mathcal{I}^0$, and $[B_{-1}, \Ld] + [\Ld_j, H]\in\mH$, so from \eqref{zh-7} and the bijection \eqref{zh-6}  it follows that
\begin{equation}\label{zh-8}
A_{-1}=0, \quad [B_{-1}, \Ld] + [\Ld_j, H]=0.
\end{equation}
The negative degree part of \eqref{HU} yields the following two equations:
\begin{align}
&-\frac{\pd A}{\pd x}+[A,\Ld+H]=0, \label{tGj} \\
\label{HG}
&-\frac{\pd B}{\pd x}+\frac{\pd H}{\pd t_j}=0.
\end{align}
By considering the homogeneous terms in the equation \eqref{tGj} with
respect to the principal gradation and by using the bijection \eqref{zh-6},
we arrive at $A=0$ and the fact that $G(\Ld_j)\in \mathcal{H}\cap\fg^{<0}$. Then the two equations given in \eqref{Gj} follow from the equation \eqref{HG} and the second equation of \eqref{zh-8}.

Due to the fact that $[G(\Ld_j),\Ld_i]$ is a
multiple of the canonical center $c$, we have
\begin{align*}
\frac{\pd}{\pd t_j}e^{\ad_U}\Ld_i=&\left[\nabla_{t_j,U}U,
e^{\ad_U}\Ld_i\right]
=\left[\left(e^{\ad_U}\Ld_j\right)_{<0},
e^{\ad_U}\Ld_i\right]+e^{\ad_U}[G(\Ld_j),\Ld_i] \\
=&\left[\left(e^{\ad_U}\Ld_j\right)_{<0},
e^{\ad_U}\Ld_i\right]+[G(\Ld_j),\Ld_i],
\end{align*}
thus $G(\Ld_j)$ satisfies the equation \eqref{eUt}. The lemma is proved.
\end{prf}

Now we arrive at the following proposition.
\begin{prp}
The flows $\pd/\pd t_j$ defined in \eqref{Lt2} commute
with each other, and they form the so-called
\emph{pre-Drinfeld-Sokolov hierarchy} associated to the triple $(\fg, \rs, \mathds{1})$.
\end{prp}
\begin{prf}
To simplify the notations let us write $\vp(\Ld_j)=e^{\ad_U}\Ld_j$. By using Lemma~\ref{thm-Gi}, we have, for $i,j\in J_+$,
\begin{align*}
\frac{\p^2\sL}{\p t_j\p t_i}-\frac{\p^2\sL}{\p t_i\p t_j} =&\frac{\p}{\p t_j}\left[ -\vp(\Ld_i)_{\ge0},\sL\right]- \frac{\p}{\p t_i}\left[ -\vp(\Ld_j)_{\ge0},\sL\right] \\
=& \left[ -[\vp(\Ld_j)_{<0}, \vp(\Ld_i)]_{\ge0}-[G(\Ld_j),\Ld_i],\sL\right]+\left[ -\vp(\Ld_i)_{\ge0},\left[ -\vp(\Ld_j)_{\ge0},\sL\right]\right] \\
&\quad -\left[ -[\vp(\Ld_i)_{<0}, \vp(\Ld_j)]_{\ge0}-[G(\Ld_i),\Ld_j],\sL\right]-\left[ -\vp(\Ld_j)_{\ge0},\left[ -\vp(\Ld_i)_{\ge0},\sL\right]\right] \\
=&\left[ [ \vp(\Ld_i)_{\ge0}, \vp(\Ld_j)_{<0}]_{\ge0}+[\vp(\Ld_i)_{\ge0}, \vp(\Ld_j)_{\ge0}]+[\vp(\Ld_i)_{<0}, \vp(\Ld_j)]_{\ge0},\sL\right]  \\
&\quad-\left[[G(\Ld_j),\Ld_i]-[G(\Ld_i),\Ld_j],\sL\right] \\
=&\left[ [ \vp(\Ld_i), \vp(\Ld_j)]_{\ge0},\sL\right]+ \left[ \frac{\p H}{\p t_j},\Ld_i\right]-\left[ \frac{\p H}{\p t_i},\Ld_j \right] \\
=&\left[ \frac{\p H}{\p t_j},\Ld_i\right]-\left[ \frac{\p H}{\p t_i},\Ld_j \right],
\end{align*}
where in the third equality we used the Jacobi identity and the fact that $[G(\Ld_i),\Ld_j]$ is a multiple of the central element $c$.
By using the definition of $\sL$ and \eqref{Borel} , we see that
\[
\left(d^\rs\mid \left[ \frac{\p H}{\p t_j},\Ld_i\right]-\left[ \frac{\p H}{\p t_i},\Ld_j \right]\right)=0,
\]
which, together with the fact that $[H, \Ld_j]$ is a scalar multiple of $c$ for any $j\in J_+$, implies that
\[
 \left[ \frac{\p H}{\p t_j},\Ld_i\right]-\left[ \frac{\p H}{\p t_i},\Ld_j \right]=0, \quad \frac{\p^2\sL}{\p t_j\p t_i}-\frac{\p^2\sL}{\p t_i\p t_j}=0.
\]
The proposition is proved.
\end{prf}

\subsection{The Drinfeld-Sokolov hierarchy and its Hamiltonian structure}

Note that the Borel subalgebra $\mathcal{B}$ contains a nilpotent subalgebra  $\mathcal{N}=\fg_{0}\cap\fg^{<0}$, which is generated by the generators $f_i$ with $s_i=0$. According to the Serre relations, we have
\begin{equation}\label{LdN}
  \ad_{\Ld}\mathcal{N}=\ad_{I}\mathcal{N}\subset\mathcal{B}, \quad I=\sum_{i\mid s_i=0}e_i.
\end{equation}
Since $\mathcal{N}\cap\mathcal{H}=\{0\}$ the map $\ad_{\Ld}:\mathcal{N}\to\mathcal{B}$ is injective, hence we have a decomposition
\begin{equation}\label{BV}
  \mathcal{B}=\ad_{\Ld}\mathcal{N}\oplus\mathcal{V}
\end{equation}
with  $\mathcal{V}$ being some $\ell$-dimensional subspace of $\mathcal{B}$. Clearly \eqref{LdN} and \eqref{BV} hold true when $\Ld$ is replaced by $\Ld_1=\nu\Ld$.

Observe that operators of the form \eqref{sL} admit the following gauge transformations:
\begin{equation}\label{gauge}
\sL\mapsto \tilde{\sL}= e^{\ad_N}\sL,  \quad N\in C^\infty(\R, \mathcal{N}).
\end{equation}
The results of the following two lemmas can be derived by using the method given in \cite{BGHM, DS}, based on the decomposition \eqref{BV} of the Borel subalgebra $\mathcal{B}$.

\begin{lem}\label{thm-gauge}
Let $\sL$ be  an operator of the form \eqref{sL}. Fix a complementary subspace $\mathcal{V}\subset \mathcal{B}$ as in \eqref{BV}, then there exists a unique function $N\in C^\infty(\R, \mathcal{N})$  such that $\sL^\mV=e^{\ad_N}\sL$ takes the form
\begin{equation}\label{QV}
\sL^\mV=\frac{\od}{\od x}+\Ld+Q^\mathcal{V}, \quad Q^\mathcal{V}\in C^\infty(\R, \mathcal{V}).
\end{equation}
Moreover, both $N$ and $Q^\mathcal{V}$ are differential polynomials in $Q$ with zero constant terms.
\end{lem}
We call $Q^\mV$ a gauge of the function $Q$, or $\sL^\mV$ a gauge of the operator $\sL$. In fact, if we take another complementary subspace $\tilde{\mV}$, then the components of $Q^{\tilde{\mV}}$ can be represented as differential polynomials in
the components of $Q^\mV$, i.e. they are related by Miura-type transformations.
In what follows, we will fix a basis $\gm_1, \gm_2, \dots, \gm_\ell$ of the subspace $\mathcal{V}$, then $Q^\mV$ takes the form
\begin{equation}\label{qcan}
Q^\mV=\sum_{i=1}^\ell u_i \gm_i.
\end{equation}
We will denote by $\mathcal{R}$ the ring of differential polynomials in $\mathbf{u}=(u_1, u_2, \dots, u_\ell)$, i.e.,
\begin{equation}\label{Rv}
\mathcal{R}=\C[\mathbf{u}, \mathbf{u}', \mathbf{u}'', \dots]
\end{equation}
where the prime means taking derivatives with respect to $x$.

\begin{lem}\label{thm-QQV}
For the operators $\sL$ and $\sL^\mV$ given above, the functions $U$ and $H$ determined in Lemma~\ref{thm-dr0} have the following properties:
\begin{equation}\label{HHV}
e^{\ad_{U(Q^\mV)} }=e^{\ad_{N} }e^{\ad_{U(Q)} }, \quad  H(Q)=H(Q^\mV).
\end{equation}
In particular, the second property implies that the function $H(Q)$ is invariant w.r.t. the gauge transformations \eqref{gauge}.
\end{lem}

By using the method of \cite{DS} (see Lemma~3.8 therein), we obtain the following result.
\begin{lem}\label{thm-Rj}
Let $\sL^\mV$ be an operator of the form \eqref{QV}. For any $j\in J_+$, there is a unique function $R(Q^\mV, \Ld_j)\in C^\infty(\R, \mathcal{N})$ satisfying the condition that
\[\left[-(e^{\ad_{U(Q^\mV)} }\Ld_j)_{\ge 0}+R(Q^\mV, \Ld_j), \sL^\mV\right]\] takes value in $\mV$. Moreover, the components of $R(Q^\mV, \Ld_j)$ are differential polynomials in $Q^\mV$ with zero constant terms.
\end{lem}

\begin{thm}\label{zh-19}
The equations \eqref{Lt2} lead to a system of evolutionary equations on the jet space of $\mV$:
\begin{align}
\frac{\pd \sL^\mV}{\pd t_j}=\left[-(e^{\ad_{U(Q^\mV)} }\Ld_j)_{\ge 0}+R(Q^\mV, \Ld_j), \sL^\mV\right], \quad j\in J_+,
\label{LVt}
\end{align}
where $R(Q^\mV, \Ld_j)\in C^\infty(\R, \mathcal{N})$ is determined as in Lemma \ref{thm-Rj}.
\end{thm}
\begin{prf}
From \eqref{Lt2} it follows that
\begin{align}
\frac{\pd \sL^\mV}{\pd t_j}=&\frac{\pd}{\pd t_j}\left(e^{\ad_N}\sL\right) =\left[\nabla_{t_j, N}N, e^{\ad_N}\sL\right]+e^{\ad_N}\left[-(e^{\ad_{U(Q)}}\Ld_j)_{\ge 0}, \sL\right] \nn\\
=&\left[-(e^{\ad_{U(Q^\mV)} }\Ld_j)_{\ge 0}+\nabla_{t_j, N}N, \sL^\mV\right].
\label{LVt2}
\end{align}
Here $\nabla_{t_j, N}N$ is defined as in \eqref{nablaUt}, and in the last equality we used the relations \eqref{HHV}. Since $\frac{\pd \sL^\mV}{\pd t_j}$ takes value in $\mV$, it follows from Lemma \ref{thm-Rj} that
$\nabla_{t_j, N}N=R(Q^\mV, \Ld_j)$.
The theorem is proved.
\end{prf}

Due to the above theorem, we can represent the equations \eqref{LVt} in the form
\begin{equation}\label{zh-9}
\frac{\pd u_i}{\pd t_j}=X_{j}^i(\mathbf{u}, \mathbf{u}', \mathbf{u}'', \dots)\in\mathcal{R},\quad i=1,\dots, \ell;\, j\in J_+.
\end{equation}
Note that $\mathcal{R}$ is the set of differential polynomials that are invariant with respect to the gauge transformations \eqref{gauge}.

\begin{defn}
The hierarchy of evolutionary PDEs \eqref{LVt} or \eqref{zh-9} is called the Drinfeld-Sokolov hierarchy associated to $(\fg, \rs, \mathds{1})$.
\end{defn}
By using the fact that $\Ld_1=\nu\Ld$, we know that the first flow $\frac{\pd}{\pd t_1}$ of the Drinfeld-Sokolov hierarchy is given by
\[\frac{\pd}{\pd t_1}=\nu \frac{\pd}{\pd x}.
\]
So we will identify $t_1$ with $x/\nu$ in what follows.

\begin{thm} [\cite{BGHM, DS}] \label{thm-dsham}
The Drinfeld-Sokolov hierarchy \eqref{zh-9} can be written, in terms of a given gauge slice $u_1,\dots, u_\ell$, as a hierarchy of
Hamiltonian systems
\begin{equation}\label{DSbh2}
\frac{\pd u_i}{\pd t_j}=\{u_i(x),\mathscr{H}_{j}\},
\quad i=1,\dots, \ell,\,  j\in J_+,
\end{equation}
with Hamiltonians
\begin{equation}\label{sHj}
\mathscr{H}_j=\int h_j\,\od x, \quad  h_j=(-\Ld_j\mid H).
\end{equation}
Here the function $H$ is given by Lemma~\ref{thm-dr0}, and  $h_j\in\mathcal{R}$ due to \eqref{HHV}.
\end{thm}

More precisely, the Hamiltonian systems \eqref{DSbh2} can be
represented  as
\begin{equation}\label{utham}
\frac{\pd u_i}{\pd{t}_j}=\{u_i(x), \mathscr{H}_j\}=\sum_{m=1}^\ell
\sum_{k\ge0}P^{i m}_k\frac{\pd^k}{\pd x^k}\frac{\dt\mathscr{H}_j}{\dt
u_m}, \quad j\in J_+,
\end{equation}
where $P^{i m}_k\in\mathcal{R}$ (note that these coefficients depend on the gradation $\rs$), and the variational derivatives of $\mathscr{H}_j$ are given by
\begin{equation}\label{dtHdtu}
\frac{\dt \mathscr{H}_j}{\dt
u_m}=\sum_{k\ge0}(-1)^k\frac{\pd^k}{\pd x^k}\frac{\pd h_j}{\pd
u_m^{(k)}}.
\end{equation}
The above theorem shows that the function $H$ is a generating function of the Hamiltonian densities $h_j$. In fact, from \eqref{blfLd} it follows that
\begin{equation}\label{Hh}
H=-\frac{1}{h}\sum_{j\in J_+}h_j\Ld_{-j}.
\end{equation}

\begin{rmk}\label{rmk-hj}
Let $m_1\le m_2\dots\le m_\ell$ be the first $\ell$ elements of $J_+$, then the Hamiltonian densities $h_{m_1}, h_{m_2}, \dots, h_{m_\ell}$ generate the ring $\mathcal{R}$ of gauge-invariant differential polynomials. So they can also be used as the unknown functions of the Drinfeld-Sokolov hierarchy, which can be written as
\begin{equation}\label{ht}
\frac{\pd h_{m_i} }{\pd t_j}=Y_{j}^i(\mathbf{h}, \mathbf{h}', \mathbf{h}'', \dots),\quad i=1,\dots, \ell;\, j\in J_+,
\end{equation}
where $\mathbf{h}=(h_{m_1}, h_{m_2}, \dots, h_{m_\ell})$ and $Y_{j}^i\in C^\infty(\mathbf{h})[[\mathbf{h}', \mathbf{h}'', \dots]]$. In the notion of \cite{DZ} such Hamiltonian densities are called the normal coordinates of the Drinfeld-Sokolov hierarchy.
\end{rmk}

\begin{rmk}\label{diff-s}
Given an operator $\sL$ of the form \eqref{sL}, and a gradation $\rs\le\mathds{1}$ on $\fg$, there are different choices of $\mV$. Note that any two systems of generators for $\mathcal{R}$ can be represented as differential polynomials in each other, hence the equations \eqref{LVt} with different $\mV$ are related by Miura-type transformations.
Furthermore, by following the approach of \cite{DS, dGHM}, we can prove that the Drinfeld-Sokolov hierarchy associated to $(\fg, \rs, \mathds{1})$ is related to the one associated to $(\fg, \mathds{1}, \mathds{1})$  by a certain Miura-type transformation. Hence the Drinfeld-Sokolov hierarchies associated to $(\fg, \rs, \mathds{1})$ with different gradations $\rs$ are also related to each by Miura-type transformations. This fact is illustrated in Examples~\ref{exa-A1tau} and \ref{exa-A22tau} below.
\end{rmk}

\subsection{The tau functions of the Drinfeld-Sokolov hierarchy}
In this subsection, we give the definition of the tau function of the
Drinfeld-Sokolov hierarchy based on that of \cite{Mi, Wu}.

Let $Q$ be a solution of the pre-Drinfeld-Sokolov hierarchy
\eqref{Lt2} associated to $(\fg, \rs, \mathds{1})$, and $U$ be
given as in Lemma~\ref{thm-dr0}. We introduce a class of
differential polynomials in the components of $Q$ as follows:
\begin{equation}\label{Omij}
\Om^\rs_{i j}=\frac{1}{h^\rs}\left(d^\rs\mid
\left[\left(e^{\ad_U}\Ld_i\right)_{\ge0},
e^{\ad_U}\Ld_j\right]\right), \quad i,j\in J_+.
\end{equation}
\begin{prp}\label{thm-Omsym}
The following identities hold true:
\begin{equation}\label{Omsym}
\Om^\rs_{i j}=\Om^\rs_{j i}, \quad \frac{\pd\Om^\rs_{i j}}{\pd
t_k}=\frac{\pd\Om^\rs_{k j}}{\pd t_i},\quad i, j, k\in J_+.
\end{equation}
Moreover, the differential polynomials $\Om_{i j}^\rs$ are invariant with respect to the gauge
transformations \eqref{gauge}, namely, $\Om_{i j}^\rs\in\mathcal{R}$.
\end{prp}
\begin{prf}
Since $(d^\rs\mid [X, Y])=0$ for any $X, Y\in\fg_0$, it follows from
$\left[e^{\ad_U}\Ld_i, e^{\ad_U}\Ld_j\right]=0$ that
\begin{align*}
\Om^\rs_{i j}=&\frac{1}{h^\rs}\left(d^\rs\mid
\left[\left(e^{\ad_U}\Ld_i\right)_{\ge0},
\left(e^{\ad_U}\Ld_j\right)_{<0}\right]\right) \\
=&\frac{1}{h^\rs}\left(d^\rs\mid
-\left[\left(e^{\ad_U}\Ld_i\right)_{<0},
\left(e^{\ad_U}\Ld_j\right)_{\ge0}\right]\right)\\
=&\frac{1}{h^\rs}\left(d^\rs\mid
\left[\left(e^{\ad_U}\Ld_j\right)_{\ge0},
\left(e^{\ad_U}\Ld_i\right)_{<0}\right]\right) \\
=&\Om^\rs_{j i}.
\end{align*}
To simplify notations, we denote $X=e^{\ad_U}\Ld_i$,
$Y=e^{\ad_U}\Ld_j$ and $Z=e^{\ad_U}\Ld_k$. Clearly they
commute with each other. By using \eqref{eUt} we obtain
\begin{align*}
\frac{\pd\Om^\rs_{i j}}{\pd t_k}=&\frac{1}{h^\rs}\left(d^\rs\mid
\left[ \left[ Z_{<0}, X\right]_{\ge0}, Y\right]+\left[ X_{\ge0},
\left[Z_{<0}, Y\right]\right] \right)\\
=&\frac{1}{h^\rs}\left(d^\rs\mid -\left[ \left[ Z_{\ge0},
X\right]_{\ge0}, Y\right]-\left[ X_{\ge0},
\left[Z_{\ge0}, Y\right]\right] \right)\\
=&\frac{1}{h^\rs}\left(d^\rs\mid \left[ \left[  X,
Z_{\ge0}\right]_{\ge0}, Y\right]-\left[\left[ X_{\ge0},
Z_{\ge0}\right], Y\right]-\left[ Z_{\ge0},
\left[X_{\ge0}, Y\right]\right] \right)\\
=&\frac{1}{h^\rs}\left(d^\rs\mid \left[ \left[ X_{<0},
Z_{\ge0}\right]_{\ge0}, Y\right]-\left[ Z_{\ge0}, \left[X_{\ge0},
Y\right]\right] \right)\\
=&\frac{1}{h^\rs}\left(d^\rs\mid -\left[ \left[X_{\ge0},
Z\right]_{\ge0}, Y\right]-\left[ Z_{\ge0}, \left[X_{\ge0},
Y\right]\right] \right) \\
=&\frac{\pd\Om^\rs_{k j}}{\pd t_i}.
\end{align*}

Finally, for any $N\in C^\infty(\R,\mathcal{N})$ one has $e^{\ad_N}d^\rs=d^\rs$, hence
the right
hand side of \eqref{Omij} is invariant whenever $e^{\ad_U}$ is replaced by $e^{\ad_{U(Q^\mV)} }=e^{\ad_{N} }e^{\ad_{U(Q)} }$ (recall Lemma~\ref{thm-QQV}). Thus the proposition is proved.
\end{prf}

The proposition implies that for any solution $\mathbf{u}(\bt)$ with $\bt=\{t_j\}_{j\in J_+}$ of the Drinfeld-Sokolov hierarchy, there exists a function
$\tau^\rs=\tau^\rs(\bf{t})$ such that
\begin{equation}\label{tauOm}
\frac{\pd^2\log\tau^\rs}{\pd t_i\pd t_j}=\Om^\rs_{i j}(\mathbf{u}, \mathbf{u}', \mathbf{u}'', \dots)|_{\mathbf{u}=\mathbf{u}(\bt)}, \quad i, j\in J_+.
\end{equation}
Note that $\log\tau^\rs$ is determined up to a linear function of
the time variables, and that $\log\tau^\rs$ depends on the choice of the gradation $\rs$ (see Examples~\ref{exa-A1tau} and \ref{exa-A22tau} below).

\begin{dfn}\label{zh-10}
The function $\tau^\rs$ is called the tau function of the
Drinfeld-Sokolov hierarchy \eqref{zh-9}.
\end{dfn}

The following proposition shows that the Hamiltonian densities $h_j$ given in \eqref{sHj} are tau symmetric in the sense of \cite{DZ}.
\begin{prp}
The tau function and the Hamiltonian densities for the
Drinfeld-Sokolov hierarchy \eqref{zh-9} are related by
\begin{align}\label{tauh}
\frac{\pd^2\log\tau^\rs}{\pd x\pd t_j}=\frac{j}{h}h_j, \quad j\in
J_+.
\end{align}
\end{prp}
\begin{prf}
By using \eqref{ULdc} and \eqref{eUt} we obtain
\begin{equation}\label{OmGi}
\Om^\rs_{j i}=\frac{1}{h^\rs}(d^\rs\mid [G(\Ld_j),\Ld_i]),
\end{equation}
where $G(\Ld_j)$ is defined by \eqref{zh-18}. So from  \eqref{Gj} and
$\pd/\pd t_1=\nu\,\pd/\pd x$ it follows that
\begin{align}\label{tauh1}
\frac{\pd^2\log\tau^\rs}{\pd x\pd t_j}=\frac{1}{\nu}\Om^\rs_{j
1}=\frac{1}{\nu h^\rs}(d^\rs\mid
[G(\Ld_j),\Ld_1])=\frac{1}{h^\rs}(d^\rs\mid
[G(\Ld_j),\Ld])=-\frac{1}{h^\rs}(d^\rs\mid [\Ld_j,H]).
\end{align}
Then by using \eqref{Hh} we complete the proof of the
proposition.
\end{prf}

We emphasize that the tau function $\tau^\rs$ depends essentially on the gradation $\rs\le\mathds{1}$. Let us illustrate it with the following examples.

\begin{exa}{\bf (Examples~5.1 and A.1 in \cite{Wu}) } \label{exa-A1tau}
Let $\fg$ be the affine Kac-Moody algebra of type $A_1^{(1)}$, with the elements $\Ld_j$ chosen as in \cite{DS}. For $\rs=(1,0)$, we take the gauge
$Q^\mV=-u f_1$, then the first two nontrivial equations of the
Drinfeld-Sokolov hierarchy are ($t_1=x$):
\begin{align}\label{KdVt3}
&u_{t_3}=\frac{3 u u_x}{2}+\frac{u_{x x x}}{4}, \\
&u_{t_5}=\frac{15 u^2 u_x}{8}+\frac{5}{8}u u_{x x x} +\frac{5 u_x
   u_{x x}}{4}+\frac{u_{x x x x x}}{16}, \label{KdVt5}
\end{align}
where the subscripts $t_j$ mean the partial derivatives with respect to them.
As is well known, this integrable hierarchy is just the KdV hierarchy.

For $\rs=(1,1)$, the nilpotent subalgebra $\mathcal{N}$ is trivial, so we take
$Q^\mV=Q=\frac{v}{2}(\al_1^\vee-\al_0^\vee)$. The first two nontrivial equations of the Drinfeld-Sokolov hierarchy are given by
\begin{align}\label{mKdVt3}
&v_{t_3}=-\frac{3 }{2}v^2 v_x+\frac{1}{4}v_{x x x}, \\
&v_{t_5}=\frac{15}{8} v^4
   v_x-\frac{5}{8} v^2 v_{x x x}-\frac{5}{2} v v_x
v_{x x}-\frac{5 }{8}v_x^3+\frac{1}{16}v_{x x x x x}. \label{mKdVt5}
\end{align}
This integrable hierarchy is called the modified KdV
hierarchy, which is related to the KdV hierarchy by the Miura
transformation $u=-v^2+v_x$.

By using Definition \ref{zh-10}, we know that the second-order derivatives of the logarithms of the tau functions for the above two different choices of gradations satisfy the relations:
{\renewcommand\arraystretch{2}
\setlength{\doublerulesep}{0pt}
\[
\begin{array}{ccc}
\hline\hline
\rs &  \frac{\pd^2\log\tau^\rs}{\pd {t_1}^2} &  \frac{\pd^2\log\tau^\rs}{\pd t_1 \pd t_3 } \\
\hline
 (1,0)  &  \frac{1}{2}u &  \frac{1}{8}(3 u^2+u_{x x x})  \\
(1,1) &  \quad  -\frac{1}{2}v^2 \quad  &  \frac{1}{8}( 3 v^4 -2 v v_{x x}+{v_{x}}^2)  \\
\hline\hline
\end{array}
\]
}
\end{exa}

\begin{exa}\label{exa-A22tau}
Let $\fg$ be the affine Kac-Moody algebra of type $A_2^{(2)}$, and its elements $\Ld_j$ be chosen as in Example~5.6 of \cite{Wu}. For  the Drinfeld-Sokolov hierarchies associated to different gradations $\rs$ on $\fg$, the gauge $Q^\mV$, the first
nontrivial equation and the second-order derivatives of the logarithms of the tau functions are given by
\renewcommand\arraystretch{2}
\setlength{\doublerulesep}{0pt}
\[
\begin{array}{cccc}
\hline\hline
\rs & Q^\mV  & t_5\hbox{-flow} & \frac{\pd^2\log\tau^\rs}{\pd {t_1}^2} \\
\hline
(1, 0) &  -u f_1 &  u_{t_5}=-\frac{1}{108}\left(20
u^3+30 u u''+3 u^{(4)} \right)' & \frac{1}{3}u \\
(0, 1) & -\frac{w}{4}f_0 & w_{t_5}=-\frac{1}{108}\left(20
w^3+30 w w''+\frac{45}{2}(w')^2+ 3 w^{(4)} \right)' & \frac{1}{3}w  \\
(1, 1) & \quad -\frac{v}{3\sqrt{2}}(\al_0^\vee-4 \al_1^\vee) \quad & v_{t_5}=\frac{1}{36}\left(-16 v^5 +  20
v (v')^2+ 20 v^2 v''+ 10 v' v'' -v^{(4)}\right)' & -\frac{2}{3}v^2
\\
\hline\hline
\end{array}
\]
Here we note that $x=\sqrt{2}\,t_1$ since $\nu=\sqrt{2}$, and in the present example the prime means to take derivative with respect to $t_1$.
Observe that the evolutionary equations of $u$ and of $w$ are known as the Sawada-Kotera equation \cite{SK} and the Kaup-Kupershmidt equation \cite{Ka} respectively, and the above three equations are related by the following Miura transformations:
\[
u=-2\, v^2+v', \quad w=-2\, v^2 -2\, v'.
\]
\end{exa}

\begin{rmk}\label{rmk-tau}
Formulae of the form \eqref{Omij} were introduced by Miramontes in \cite{Mi} to define tau functions for generalized Drinfeld-Sokolov hierarchies (see Equation (3.8) and (4.13) therein). The method of \cite{Mi} relies on an assumption, in terms of our notations, that the functions $U$ and $H$ given by \eqref{UL} must fulfill the following constraints (Equation (2.25) therein):
\begin{equation}\label{Mi}
U\in C^\infty\left(\R, \fg_{<0}\right), \quad H \in C^\infty\left(\R, (\mathcal{H}\cap\fg_{<0})\oplus\C c\right).
\end{equation}
When $\rs$ is chosen to be the principal gradation $\mathds{1}$, such constraints are satisfied automatically, and the tau function introduced in \cite{Mi} coincides with the one defined by \eqref{tauOm}. However, when $\rs\ne\mathds{1}$, the above constraints are no longer trivial, since $H$ may contain nontrivial component along the basis vector $\Ld_{-1}\in\mathcal{H}$, however in this case $\Ld_{-1}\notin\fg_{<0}$.
For this reason, the condition \eqref{Mi} was replaced in \cite{Wu} by the condition \eqref{ULdc} for the case when $\rs=\rs^0$ is the homogeneous gradation.

In the present paper, we use the condition \eqref{ULdc}  to fix the functions $U$ and $H$ for general $\rs\le\mathds{1}$, and define the tau function of the Drinfeld-Sokolov hierarchy by \eqref{tauOm} which generalizes the definition of the tau function given in \cite{Wu}. In term of the present definition,  $\tau^{\rs^0}$ coincides with the one defined in \cite{Wu} (which corresponds to the partition functions of cohomological field theories), and the tau function $\tau^{\mathds{1}}$ coincides with the one given by Miramontes \cite{Mi} (and also by Erinquez and Frenkel \cite{EF}).

As we pointed out in Remark \ref{diff-s}, the Drinfeld-Sokolov hierarchies corresponding
to all $\rs$ are Miura-equivalent to each other. For such a Miura-equivalence class
of Hamiltonian integrable hierarchies, we have shown in \cite{DLZ} that there are many ways to introduce
its tau functions via the so-called \emph{tau structures}. For each choice of $\rs$, our construction gives
rise to a representative of the tau structures, which is, roughly speaking, the collection of
differential polynomials $\{\Om_{ij}^{\rs}\}$. The corresponding tau functions $\tau^{\rs}$
have good behaviors in the reduction procedures, see Theorem \ref{thm-taured}.
\end{rmk}

\section{Proof of the main results}
From now on we assume that $\fg$ is one of the affine Kac-Moody algebras that are listed in Tables~\ref{table-diag1}--\ref{table-diag3} (or listed in Appendix~\ref{zh-2}). Let us proceed to study the  diagram automorphism $\sg$ on $\fg$, and the induced symmetry of the corresponding Drinfeld-Sokolov hierarchy.

\subsection{Diagram automorphisms and affine subalgebras}\label{sec-gsg}
Let $\bar{\sg}$ be a permutation of the index set $\{0, 1, 2, \dots,
\ell\}$ given in Tables~\ref{table-diag1}--\ref{table-diag3}, such that it preserves the generalized Cartan matrix $A$ of $\fg$, i.e.,
\[
a_{\bar\sg(i)\bar\sg(j)}=a_{i j}, \quad i,j=0,1,2,\dots,\ell.
\]
Then a diagram
automorphism $\sg$ on the affine Kac-Moody algebra $\fg$ is determined by
\begin{equation}\label{}
\sg(e_i)=e_{\bar\sg(i)}, \quad \sg(f_i)=f_{\bar\sg(i)}, \quad
i=0,1,2,\dots,\ell.
\end{equation}

Assume that the order of the diagram automorphism $\sg$ is $p$, then its
eigenvalues are given by
$1, \ep, \ep^2, \dots, \ep^{p-1}$
with
$\ep=\exp\left(2\pi\mathbf{i}/p\right)$. Since $\sg$ is consistent
with the principal gradation on $\fg$, we have a series of bijections
\[
\sg: \fg^k\to \fg^k,\quad  k\in\Z,
\]
and the eigenspace decompositions
\[
\fg^k=\fg^{k,1}\oplus\fg^{k,\ep}\oplus\dots\oplus\fg^{k,\ep^{p-1}} \quad \hbox{with} \quad \fg^{k,\ep^l}=\left\{X\in\fg^k\mid \sg(X)=\ep^l X\right\}.
\]
Let $\fg^\sg$ be the subalgebra of $\fg$ that consists of $\sg$-invariant elements. It
can be represented as
\begin{equation}\label{gsg}
\fg^\sg=\bigoplus_{k\in\Z}(\fg^\sg)^k, \quad
(\fg^\sg)^k=\fg^\sg\cap\fg^k=\fg^{k,1}.
\end{equation}
Since $\Ld, c\in \fg^\sg$, $\sg$ can be restricted to
$\mathcal{I}$ and $\mathcal{H}$ (see their definitions given in Section \ref{zh-15}), we can choose the
basis elements $\Lambda_j$ of $\mathcal{H}$ so that they are eigenvectors of
$\sg$.  Denote
\begin{equation}\label{Hsg}
\mathcal{I}^\sg=\fg^\sg\cap \mathcal{I},\quad \mathcal{H}^\sg=\fg^\sg\cap \mathcal{H},\quad J^\sg=\{j\in J\mid \sg(\Lambda_j)=\Lambda_j\},
\end{equation}
then we have
\begin{equation}\label{}
\mathcal{H}^\sg=\bigoplus_{j\in J^\sg}\C\Ld_j\oplus\C c,
\end{equation}
and  the gradation
$\mathcal{I}^\sg=\bigoplus_{k\in\Z}(\mathcal{I}^\sg)^k$. We also have the  bijections
\begin{equation}
\ad_\Ld: (\mathcal{I}^\sg)^k\to (\mathcal{I}^\sg)^{k+1}, \quad
k\in\Z.
\end{equation}

Following the notions of \cite{Kac68}, let $G^\sg=(\fg^\sg)^{-1}\oplus(\fg^\sg)^0\oplus(\fg^\sg)^1$ be the local part of the graded Lie algebra $\fg^\sg$. We consider the minimal graded Lie algebra whose local part is $G^\sg$, that is,
\begin{equation}\label{fgbar}
\bar\fg=\langle(\fg^\sg)^{-1},\, (\fg^\sg)^0, \, (\fg^\sg)^1\rangle.
\end{equation}
Clearly, the Lie algebra $\bar\fg$ is a subalgebra of $\fg^\sg$ which has the gradation
\begin{equation}\label{gbarprin}
\bar\fg=\bigoplus_{k\in\Z}\bar{\fg}^k, \quad
\bar{\fg}^k=\bar\fg\cap\fg^k.
\end{equation}
In particular, we have
$\bar{\fg}^k=\fg^{k,1}$ for $k=-1, 0, 1$ and $\Ld\in\bar\fg$. We will show in the following theorem that $\bar{\fg}$ is the derived algebra of an affine Kac-Moody algebra $\fg(\bar{A})$ for a certain generalized Cartan matrix $\bar{A}$. Thus $\bar{\fg}$ has the decomposition
\begin{equation}\label{gbardec}
\bar{\fg}=\bar{\mathcal{I}}\oplus\bar{\mathcal{H}},
\end{equation}
where $\bar{\mathcal{I}}=\bigoplus_{k\in\mathbb{Z}}\bar{\mathcal{I}}^k$ and $\bar{\mathcal{H}}$ are defined as in \eqref{zh-4}, \eqref{zh-1}. From the bijections
\begin{equation}
\ad_\Ld: \bar{\mathcal{I}}^k\to \bar{\mathcal{I}}^{k+1}, \quad
k\in\Z
\end{equation}
and the facts $\bar{\mathcal{I}}^k=(\mathcal{I}^\sg)^k, \ k=-1, 0, 1$, it follows that $\bar{\mathcal{I}}=\mathcal{I}^\sg$. On the other hand, we have $\bar\mH\subset\mH^\sg\subset\mH$. More precisely, we can modify, if needed,  the choice of the basis $\Lambda_j\, (j\in J)$ of the Heisenberg subalgebra $\mathcal{H}$ of $\fg$ so that
\[\bar{\mathcal{H}}=\bigoplus_{j\in\bar{J}}\C\Ld_j\oplus\C c,\]
where $\bar{J}$ is a subset of $J^\sg$. Note that the sets $J$, $J^\sg$
and $\bar{J}$ of ``exponents'' satisfy the relation $\bar{J}\subset J^\sg\subset
J$.

The following theorem is a slightly more detailed account of Theorem \ref{thm-algred}.
\begin{thm}\label{thm-ggbar}
Let $\fg$ be an affine Kac-Moody algebra of type $X_{\ell'}^{(r)}$, on which there is a diagram automorphism $\sg$ that is induced by $\bar\sg$ given in
Tables~\ref{table-diag1}--\ref{table-diag3} of Section\,1. Then the following assertions hold true:
\begin{enumerate} \renewcommand{\labelenumi}{\emph{(\roman{enumi})}}
\item The subalgebra $\bar\fg$ is an affine Kac-Moody
algebra corresponding to the folded Dynkin diagram of type
$X_{\bar{\ell}'}^{(\bar r)}$. Its Chevalley generators are given in Appendix \ref{zh-3}, and its canonical center $\bar c=\mu c$ with a constant $\mu$ given in Appendix \ref{zh-2}.
\item For any case listed in Tables~\ref{table-diag1} and \ref{table-diag2}, one has
$\fg^\sg=\bar\fg$;
\item For any case listed in Table~\ref{table-diag3}, one has
$\fg^\sg=\bar\fg\oplus\hat{\mathcal{H}}$, where
\begin{equation}\label{hatH}
\hat{\mathcal{H}}=\bigoplus_{j\in J^\sg\setminus\bar{J}}\C \Ld_j
\end{equation}
with $J^\sg\setminus\bar{J}$ nonempty (see  Appendix \ref{zh-2}).
\end{enumerate}
\end{thm}
\begin{prf}
By taking average of the elements of the $\sg$-orbits of the Chevalley
generators of $\fg$, we obtain a system of generators $E_i, F_i, H_i$, $i=0, 1,\dots, \bar{\ell}$ of $\bar\fg$ (see \eqref{chevalleygbar} of Appendix A). To prove the first assertion of the theorem, we only need to show that these generators satisfy the Serre relations
for a certain generalized Cartan matrix $\bar{A}=(A_{ij})$ (see its definition given in  \eqref{Abar}) corresponding to the folded Dynkin diagram. We present the
details of the proof of this fact in Appendix \ref{zh-3}. In particular,   from \eqref{chevalleygbar} one sees that
\[
\Ld=\sum_{ i=0}^{\bar\ell}E_i,
\]
hence it is also the cyclic element of $\bar\fg$.

In order to prove the second and the third assertions of the theorem, it is sufficient to compare the
dimensions of $\bar\fg^k$ and of $(\fg^\sg)^k$ for $k\in\Z$ due to the fact that  $\bar\fg\subset\fg^\sg$. From the realization \eqref{gAs} with $\rs=\mathds{1}$ it follows that we only need to
consider the integers $k$ which lie in an interval of $\Z$ with
length $r h$. We do this comparison case by case, and illustrate this procedure
for Case (a2) in Table~\ref{table-diag2} and Case (a4) in Table~\ref{table-diag3} (also listed in Appendix \ref{zh-2}).
To simplify the presentation we assume that the canonical center $c$ is trivial in this proof.

For Case (a2), it follows from \eqref{zh-7} and properties of the exponents of $\g=A^{(1)}_{2n}$ that
\begin{equation}\label{}
\dim\fg^k=\left\{\begin{array}{cl} 2n+1,&\quad 2n+1\nmid k,\\
2n, &\quad  2n+1\mid k.
\end{array}\right.
\end{equation}
We first note that the linear space $\fg^0$ has a basis $\al_i^{\vee}=[e_i,
f_i],\,  i=1,\dots, 2n$, from this fact it is easy to see that
\begin{equation}\label{dimgsg0}
\dim(\fg^\sg)^0=n.
\end{equation}
In the case when $k=1,2,3,\dots,2n$, the linear space $\fg^k$ is
spanned by $X_0^k, X_1^k, \dots, X_{2n}^k$ with
\[
X_i^k=[\dots[ [ e_i, e_{\overline{i+1}}], e_{\overline{i+2}}],
\dots, e_{\overline{i+k-1}}],
\]
where $\bar{j}\in\{0,1,2,\dots,2n\}$ equals $j$ modulo $2n+1$. We have
\begin{align}\label{}
\sg(X_i^k)=& [\dots[ [ e_{\overline{2n-i}}, e_{\overline{2n-i-1}}],
e_{\overline{2n-i-2}}], \dots, e_{\overline{2n-i-k+1}}] \nn \\
=&[e_{\overline{2n-i}},\dots,[  e_{\overline{2n-i-k+3}},
[e_{\overline{2n-i-k+2}}, e_{\overline{2n-i-k+1}}] ]\dots ]\nn\\
=&(-1)^{k-1} [\dots[ [ e_{\overline{2n-i-k+1}},
e_{\overline{2n-i-k+2}}],
e_{\overline{2n-i-k+3}}], \dots, e_{\overline{2n-i}}] \nn\\
=&(-1)^{k-1} X_{\overline{2n-i-k+1}}^k,
\end{align}
where the second equality holds true due to the Jacobi identity and the fact that
$[e_{i}, e_{j}]=0$ unless $|i-j|\in\{1,2n\}$. Under the action of $\sg$, $\fg^k$
is decomposed into $n$ $2$-dimensional orbit spaces
and one $1$-dimensional orbit space. The $1$-dimensional orbit space
is spanned by
{\renewcommand\arraystretch{2}
\begin{equation}\label{}
Y^k=\left\{\begin{array}{cl}X_{\overline{n-j}}^{2j+1},&\quad
k=2j+1,\\
X_{\overline{2n-j+1}}^{2j}, &\quad k=2j,
\end{array}\right.
\end{equation}
}
which satisfies the relation
\[
\sg(Y^k)=(-1)^{k-1}Y^k.
\]
Thus we obtain, for $1\le k\le 2n$,
\begin{equation}\label{}
\dim(\fg^\sg)^k=\left\{\begin{array}{cl} n+1,&\quad k \hbox{ is odd},\\
n, &\quad k \hbox{ is even}.
\end{array}\right.
\end{equation}
This together with \eqref{dimgsg0} leads to, for general $k\in\Z$, that
\begin{equation}\label{dimgsgk}
\dim(\fg^\sg)^k=\left\{\begin{array}{cl} n+1,&\quad k \hbox{ is odd, and } 2n+1\nmid k,\\
n, &\quad \hbox{other cases}.
\end{array}\right.
\end{equation}
From this fact and the property of the affine Kac-Moody algebra $\bar{\fg}$ it follows that $\dim(\fg^\sg)^k=\dim\bar{\fg}^k$ for $k\in\Z$. Thus Case (a2) is verified.

For Case (a4), the affine Kac-Moody algebra $\fg$ is of type $A_{2(n+1)-1}^{(2)}$. From the values of the exponents of $\fg$ that is given in Appendix~\ref{zh-2}, it follows that
\begin{equation}\label{}
\dim\fg^k=\left\{\begin{array}{cl} n+2,&\quad  k \hbox{ odd},\\
n+1, &\quad   k \hbox{ even}.
\end{array}\right.
\end{equation}
Note that the automorphism $\sg$ on $\fg$ is defined by (see Table~\ref{table-diag3})
\[
\sg(K_i)=\left\{\begin{array}{cl}
                  K_{1-i}, & i=0,1; \\
                  K_i, & i=2,3,\dots,n+1,
                \end{array}\right.
\]
where $K_i=e_i, \ f_i$ or $\al_i^\vee=[e_i, f_i]$.

Let us first note that
\begin{align*}
\fg^0=&\mathrm{span}\left\{\al_1^\vee-\al_0^\vee, \al_2^\vee, \al_3^\vee, \dots, \al_{n+1}^\vee \right\}, \\
\fg^1=&\mathrm{span}\left\{e_0, e_1, e_2, \dots, e_{n+1} \right\}, \\
\fg^2=& \mathrm{span}\left\{  [e_0, e_2],  [e_1, e_2], [e_2, e_3], \dots, [e_{n}, e_{n+1}] \right\}.
\end{align*}
Hence for $0\le k\le 2$, we have
$\dim(\fg^\sg)^k=\dim \fg^k-1$.
When $3\le k\le 2n$, by considering the connected  sub-diagrams of the Dynkin diagram of the affine Kac-Moody algebra of type $A_{2(n+1)-1}^{(2)}$, we can choose a basis $X_1^k, X_2^k, \dots, X_{\dim\fg^k}^k$ of $\fg^k$ such that
\[
X_1^k=[\dots[ [ e_0, e_2], e_{i_3}], \dots, e_{i_k}], \quad
X_2^k=[\dots[ [ e_1,
e_2], e_{i_3}], \dots, e_{i_k}]
\]
with certain $i_l\ge2$, and $X_i^k$ ($3\le i\le \dim\fg^k$) being either of the following forms:
\begin{equation}\label{Xkei}
[\dots [ [ [ e_0, e_2], e_1], e_{j_4}], \dots, e_{j_k}],\quad
[\dots [ [ [ e_{j_1}, e_{j_2}], e_{j_3}], e_{j_4}], \dots, e_{j_k}]
\end{equation}
with certain $j_l\ge 2$. Observe that
\[
\sg(X_i^k)=\left\{\begin{array}{cl}
                  X_{2-i}^k, & i=1,2; \\
                  X_i^k, & i=3,4,\dots,\dim\fg^k,
                \end{array}\right.
\]
hence
$\dim(\fg^\sg)^k=\dim\fg^k-1$ for $3\le k\le 2n$.
When $k=2n+1$, we can choose a basis of $\fg^{2n+1}$ consisting of elements
\[
X_1^{2n+1}=[ [\dots[ [ e_0, e_2], e_{i_3}], \dots, e_{i_{2n}}], e_0], \quad
X_2^{2n+1}=[ [\dots[ [ e_1,
e_2], e_{i_3}], \dots, e_{i_{2n}}], e_1]
\]
with certain $i_l\ge2$, and $X_i^{2n+1}$ ($3\le i\le n+2$) either of the following form:
\[
[\dots [ [ [ e_0, e_2], e_1], e_{j_4}], \dots, e_{j_{2n+1} }],\quad
[e_0, [e_1, [ [ [ e_{j_3}, e_{j_4}], e_{j_5}], \dots, e_{j_{2n+1}}] ] ]
\]
with $j_l\ge2$.
Thus, similar to the above case when $3\le k\le 2 n$, we have $\dim(\fg^\sg)^{2n+1}=\dim\fg^{2n+1}-1$.

In a similar way we can prove,  by replacing $e_i$ with $f_i$, that $\dim(\fg^\sg)^{k}=\dim\fg^{k}-1$ for $-2 n\le k\le -1$.
Thus we arrive at
\begin{equation}\label{}
\dim(\fg^\sg)^k=\dim\fg^k-1
=\left\{\begin{array}{cl}
                                     n+1, & k \hbox{ odd}, \\
                                     n, & k \hbox{ even}
                                   \end{array}\right.
\end{equation}
for all $k\in\Z$. This property is different from that for $\bar\fg=A_{2n}^{(2)}$, namely,
\begin{equation}\label{dimd-2}
\dim \bar{\fg}^k=\left\{\begin{array}{cl}
                                     n+1, & k \hbox{ odd and } (2n+1)\nmid k, \\
                                     n, & k \hbox{ even or } (2n+1)\mid k.
                                   \end{array}\right.
\end{equation}

All other cases
can be verified in a similar way. The theorem is proved.
\end{prf}

The first assertion of the above theorem implies that we can normalized the generators of the principal Heisenberg subalgebra $\bar\mH$ of $\bar\fg$ as
\begin{equation}\label{Ldjbar}
\bar\Ld_j=\sqrt{\mu}\,\Ld_j, \quad  j\in \bar{J},
\end{equation}
such that
\begin{equation}
[\bar\Ld_i, \bar\Ld_j]=\dt_{i, -j} i \bar{c}.
\end{equation}
On the other hand, we can extend the automorphism $\sg$ on $\fg$ to $\sg:\fg(A)\to\fg(A)$ by setting $\sg(d)=d$, so that the standard bilinear form $(\cdot\mid\cdot)$ on $\fg(A)$ is invariant with respect to $\sg$.
Then $\bar\fg$ is the derived algebra of
\[
\fg(\bar A):=\bar\fg\oplus\C d
\]
on which the standard bilinear form is given by
\begin{equation}\label{bilinearbar}
(\cdot\mid\cdot)^-=\kappa(\cdot\mid\cdot)|_{\fg(\bar A)}.
\end{equation}
Here the normalization constant $\kappa$ is given in Appendix~\ref{zh-2}. This constant satisfies the relations
\[
\bar{h}=(\bar{\Ld}_j\mid\bar{\Ld}_{-j})^-=\kappa\mu(\Ld_j\mid\Ld_{-j})=\kappa\mu h,
\]
where $h$ and $\bar{h}$ are the Coxeter numbers of $\fg$ and $\bar\fg$ respectively.
\begin{rmk}
We can also consider composition of diagram automorphisms, and obtain
chains of subalgebras like  $A_{4n+1}^{(1)}\supset C_{2n+1}^{(1)}
\supset A_{2 n}^{(2)}$, $D_{n+2}^{(1)}\supset B_{n+1}^{(1)}
\supset D_{n+1}^{(2)}$ etc.
\end{rmk}

\subsection{The $\Gm$-reduction theorem} \label{sec-redI}

Let us note that the principal gradation $\mathds{1}$ of $\fg$ is consistent with the diagram automorphism $\sg$. In what follows we fix a gradation
$\rs\le\mathds{1}$ of $\fg$ that is also consistent with $\sg$, i.e., it satisfies the
condition $s_{\bar\sg(i)}=s_i$.

Recall the decomposition \eqref{BV} of the Borel subalgebra $\mathcal{B}$. Since both $\mathcal{B}$ and its subspace $\ad_\Ld\mathcal{N}$ are invariant w.r.t. the action of $\sg$, we can take a $\sg$-invariant complementary subspace $\mathcal{V}\subset\mathcal{B}$.  We can also choose a basis $\gm_1,\gm_2,\dots,\gm_n$ of $\mathcal{V}$ such
that
\[
\sg(\gm_i)=\ep_i\gm_i, \quad i=1,2,\dots,\ell,
\]
where $\ep_i$ satisfy $\ep_i^p=1$, and $p$ is the order of $\sg$.
The automorphism
$\sg:\mathcal{V}\to\mathcal{V}$ induces a pullback of the coordinate functions of  $Q^\mV=\sum_{i=1}^n u_i\gm_i$, defined  in \eqref{QV}, as follows:
\begin{equation}\label{sgv}
\sg^* (u_i)=\ep_i u_i, \quad i=1, 2, \dots, \ell.
\end{equation}
This pullback action extends naturally to the ring $\mathcal{R}$ of differential polynomials in $u_1, u_2, \dots, u_\ell$.

On the other hand, as indicated in Subsection~\ref{sec-gsg}, the generators $\Ld_j\in\fg^j$ of the principal
Heisenberg subalgebra $\mH$ can be chosen such that
\[
\sg(\Ld_j)=\zeta_j\Ld_j, \quad j\in J,
\]
where $\zeta_j$ satisfy $\zeta_j^p=1$ (we remark that only the vectors $\Ld_j$ in \eqref{zh-1} with multiple exponents $j$ may need to be adjusted).  By using \eqref{blfLd}, \eqref{gAs}
and the fact that $\Ld_1=\nu\sum_{i=0}^\ell e_i$, we obtain the relations
\begin{equation}\label{}
\zeta_1=1, \quad \zeta_{-j}=\zeta_j^{-1}, \quad \zeta_{j+r h}=\zeta_j.
\end{equation}
With respect to such generators of $\mH$ the Drinfeld-Sokolov hierarchy is defined by \eqref{LVt}, whose flows  $\p/\p t_j$ are regarded as vector fields on the jet space of $\mV$ acting on the differential polynomial ring $\mathcal{R}$.

\begin{prp}\label{thm-sg}
The pullback and the pushforward maps induced by $\sg: \mathcal{V}\to\mV$ have the following properties:
\begin{enumerate} \renewcommand{\labelenumi}{\rm{(\roman{enumi})}}
\item  $\sg^*(h_{j} )=\zeta_j^{-1}h_{j}$;
\item  $\sg^*(\Om_{j k}^\rs)=\zeta_j^{-1}\zeta_k^{-1}\Om_{j k}^\rs $;
\item  $\sg^*(X_j^i)=\zeta_j^{-1}\epsilon_i X_j^i$;
\item  $\od\sg\left(\dfrac{\pd}{\pd t_j} \right)=\zeta_j \dfrac{\pd}{\pd
t_j}$.
\end{enumerate}
Here $j, k\in J_+$, the functions $h_j$, $\Om_{jk}^\rs$, $X_j^i$  and the flows $\p/\p t_j$ are defined respectively in \eqref{sHj}, \eqref{Omij}, \eqref{zh-9} and \eqref{LVt}.
\end{prp}
\begin{prf} The proof is almost the same as the one for Proposition 4.8 of \cite{LRZ}. For the convenience of the readers, let us write it down briefly here.

We first verify the validity of the second property as follows. Recall that $\Om_{j k}$ defined in \eqref{Omij} are invariant with respect to the gauge transformations \eqref{gauge} (see Proposition~\ref{thm-Omsym}). So we have
\begin{align*}
\sg^*(\Om^\rs_{j k}) =&\frac{1}{h^\rs}\left(d^\rs\mid
\left[\left(e^{\ad_{\sg ({U(Q^\mV)}) }}\Ld_j\right)_{\ge0},
\left(e^{\ad_{\sg ({U(Q^\mV)}) }}\Ld_k\right)_{<0}\right]\right) \\
=&\frac{1}{\zeta_j \zeta_k h^\rs}\left(\sg (d^\rs)\mid
\sg\left(\left[\left(e^{\ad_{{U(Q^\mV)} }}\Ld_j\right)_{\ge0},
\left(e^{\ad_{{U(Q^\mV)} }}\Ld_k\right)_{<0}\right]\right)\right)\\
=&\frac{1}{\zeta_j \zeta_k h^\rs}\left( d^\rs\mid
\left[\left(e^{\ad_{{U(Q^\mV)} }}\Ld_j\right)_{\ge0},
\left(e^{\ad_{{U(Q^\mV)} }}\Ld_k\right)_{<0}\right]\right)\\
=&\zeta_j^{-1} \zeta_k^{-1} \Om^\rs_{j k}.
\end{align*}
Here the second equality is due to $\sg( d^\rs)=d^\rs$ which
follows from the consistency of $\rs$ with $\sg$, and the relation
$\sigma({U(Q^\mV)})=U(\sigma(Q^\mV))$ which can be seen from the proof of Lemma \ref{thm-dr0}; while the third equality is due to the fact that the bilinear form is invariant with respect to $\sg$.
The first property of the proposition then follows from the second one, the formulae
given in \eqref{tauOm}, \eqref{tauh} and the fact that $\zeta_1=1$.
To prove the third property,  let us apply $\sg$ on both sides of \eqref{LVt} to obtain
\begin{equation}\label{sgLt}
\sg\left(\frac{\pd \sL^\mV }{\pd
t_j}\right)=\sg\left(\left[-\left(e^{\ad_{{U(Q^\mV)} }}\Ld_j\right)_{\ge0}+R(Q^\mV, \Ld_j), \sL^\mV \right]\right),
\quad j\in J_{+}.
\end{equation}
From these equations we have
\begin{align*}
\mathrm{l.h.s.}=&\sum_{i=1}^\ell\frac{\pd u_i}{\pd t_j}\sigma(\gm_i)=\sum_{i=1}^\ell \ep_i\frac{\pd u_i}{\pd t_j} \gm_i, , \\
\mathrm{r.h.s.}=&\left[-\left(e^{\ad_{\sg (U(Q^\mV) )}}\sg(\Ld_j)\right)_{\ge0}+ R(\sg (Q^\mV), \sg(\Ld_j)),
\sg(\sL^\mV )\right]
 \\
=&\zeta_j\left[-\left(e^{\ad_{U(\sg(Q^\mV)) }}\Ld_j\right)_{\ge0}+R(\sg (Q^\mV), \Ld_j),
\sg(\sL^\mV)\right]\\
=&\zeta_j\sum_{i=1}^\ell \sg^*\left(\frac{\pd u_i}{\pd t_j}\right)\gm_i.
\end{align*}
Here in computing the right hand side of the equation \eqref{sgLt} we used the property of linear dependence of $R(Q^\mV, \Ld_j)$ on $\Ld_j$, which can be seen from Lemma \ref{thm-Rj} and the proof of Theorem \ref{zh-19}.
By comparing both sides of \eqref{sgLt} and  by using \eqref{sgv},  we obtain
\[
\sg^*\left(\frac{\pd u_i}{\pd t_j}\right)=\frac{\ep_i}{\zeta_j}\frac{\pd u_i}{\pd t_j}, \quad
\od\sg\left(\frac{\pd}{\pd t_j}
\right)=(\sg^*)^{-1}\frac{\pd}{\pd t_j}\sg^*=\zeta_j \frac{\pd}{\pd
t_j}.
\]
The proposition is proved.
\end{prf}

The above proposition implies that, for $j\in J^\sg_+$ (i.e., $\zeta_j=1$, see \eqref{Hsg}), the flow $\pd/\pd t_j$ of the Drinfeld-Sokolov
hierarchy associated to $\fg$ can be restricted to the jet space of
\[\mV^\sg:=\mV\cap\fg^\sigma=\{X\in\mV \mid \sigma(X)=X\}.
\]
We want to compare such reduced flows with those of the Drinfeld-Sokolov hierarchy associated to the affine Kac-Moody subalgebra $\bar\fg$ of $\fg$ introduced in the previous subsection. Note that the gradations $\mathds{1}$ and $\rs$ of $\fg$ are consistent with $\sg$, so they induce two gradations $\bar{\mathds{1}}$ and $\bar\rs$ on $\bar\fg$ respectively. In fact, the adjoint actions of the corresponding derivations $d^{\bar{\mathds{1}} }$ and $d^{\bar\rs}$ on $\bar\fg$ are equal to that of $d^{\mathds{1}}$ and $d^\rs$ respectively. By abusing notations, the principal gradation $\bar{\mathds{1}}$ on $\bar\fg$ will also be written as $\mathds{1}$. Note that the Borel subalgebra of $\bar\fg$ is given by $\bar{\mathcal{B}}=\mathcal{B}\cap\bar\fg$ (see its definition \eqref{Borel}). Since $\bar\fg$ and $\fg$ have the same cyclic element $\Ld$, we have the following decomposition
\[
\bar{\mathcal{B}}=\ad_\Ld\bar{\mathcal{N}}\oplus\bar\mV,
\]
where
\begin{equation}\label{zh-11}
\bar{\mathcal{N}}:=\mathcal{N}\cap\bar\fg, \quad \bar{\mathcal{V}}:=\mV\cap\bar\fg.
\end{equation}
On the other hand, from Theorem~\ref{thm-ggbar} and properties of the exponents of $\bar\fg$ that are given in Appendix B it follows that $\hat\mH\subset\bigoplus_{|k|\ge \bar{h}}\left(  \mathfrak{g}^{\sigma} \right)^{k}$, which together with  $\fg_0\cap\bigoplus_{|k|\ge \bar{h}}\left(  \mathfrak{g}^{\sigma} \right)^{k}=\{0\}$ implies that $\fg_0\cap\hat\mH=\{0\}$. Hence, we have $\fg_0\cap\bar{\fg}=\fg_0\cap\fg^\sg$, and consequently
\[
\bar{\mV}=\mV^\sg.
\]
We denote by $\bar{\mathcal{R}}$ the ring of differential polynomials of the jet space
of $\bar{\mathcal{V}}$, i.e.
\begin{equation}\label{Rsg}
\bar{\mathcal{R}}=\C\left[u_i, u_i', u_i'', \dots \mid 1\le i\le\ell,
\sg^*(u_i)=u_i\right].
\end{equation}

With the help of the above notations,
we can construct the Drinfeld-Sokolov hierarchy associate to $(\bar\fg, \bar\rs, \mathds{1})$ as was done in Section \ref{sec-3}. More precisely, this hierarchy can be represented in the form of \eqref{LVt}, i.e.
\begin{equation}\label{Lt3}
\frac{\pd {\bar\sL}^{\bar\mV} }{\pd\bar{t}_j}=\left[-(e^{\ad_{\bar{U}( {\bar Q}^{\bar\mV} ) }
}\bar{\Ld}_j)_{\ge0}+R( {\bar Q}^{\bar\mV}, \bar\Ld_j) , {\bar\sL}^{\bar\mV} \right], \quad j\in \bar{J}_{+}.
\end{equation}
Here
\begin{equation}\label{sLbar}
{\bar\sL}^{\bar\mV} =\frac{\od}{\od x}+\Ld+\bar{Q}^{\bar\mV}, \quad \bar{Q}^{\bar\mV}\in
C^\infty(\R,\bar{\mathcal{V}}),
\end{equation}
the function $\bar{U}( {\bar Q}^{\bar\mV} )\in(\R,\bar\fg^{<0})$  is determined by Lemma~\ref{thm-dr0} with $\fg$ replaced by $\bar\fg$ and the principal gradation defined in \eqref{gbarprin}, the normalized generators $\bar\Ld_j$ are defined in \eqref{Ldjbar}, and the function $R( {\bar Q}^{\bar\mV}, \bar\Ld_j)$ takes value in $\bar{\mathcal{N}}$ (see Lemma~\ref{thm-Rj}).
We emphasize that the equations \eqref{Lt3} are well defined since the gradations $\mathds{1}$ and $\rs$ of $\fg$ are consistent with $\sg$.

The following theorem is the main result of the present section, and it will be called the $\Gm$-reduction theorem.
\begin{thm}\label{thm-qred}
Each flow $\pd/\pd t_j$ with $j\in J^\sg_+$ of the Drinfeld-Sokolov
hierarchy \eqref{LVt} associated to $(\fg, \rs, \mathds{1})$ can be restricted to
$\bar\mR$. Moreover,
\begin{enumerate} \renewcommand{\labelenumi}{\rm{(\roman{enumi})}}
\item   The reduced flows $\pd/\pd t_j$ with $j\in\bar{J}_+$
form the Drinfeld-Sokolov hierarchy associated to $(\bar\fg, \bar\rs, \mathds{1})$, and they are related to the normalized flows \eqref{Lt3}
by $\sqrt{\mu}{\pd}/{\pd t_j}={\pd}/{\pd\bar{t}_j}$;
\item  When  $j\in
J^\sg_+\setminus\bar{J}_+$ (nonempty only for the cases listed in Table \ref{table-diag3}),  the reduced flows $\pd/\pd t_j$ on $\bar\mR$ are trivial.
\end{enumerate}
\end{thm}

The theorem will be proved in the following subsection.
Before doing so, we observe that the affine Kac-Moody algebra $\bar\fg$ given in Table~\ref{table-diag1} is always of untwisted type, however, it can be of any twisted type in Tables \ref{table-diag2} and \ref{table-diag3}. In other words,
the $\Gm$-reduction theorem confirms that Drinfeld-Sokolov hierarchies associated to affine Kac-Moody algebras of both BCFG type and of twisted type can be obtained via reductions of the ones associated to affine Kac-Moody algebras of ADE type. Let us give two examples to illustrate the reduction procedure.

\begin{exa}{\bf (Case (a2) of Table~\ref{table-diag2})}\label{exa-A2sg}
Let  $\fg$ be the affine Kac-Moody algebra of type $A_2^{(1)}$, and $\rs=(1,0,1)$. The Borel and the nilpotent subsalgebras are given by
$\mathcal{B}=\mathrm{span}\{\al_0^\vee-\al_2^\vee, \al_1^\vee, f_1\}$ and $\mathcal{N}=\mathrm{span}\{f_1\}$.
Take $\mV=\mathrm{span}\{\al_0^\vee-\al_2^\vee,  f_1\}$, then we have the gauge
\[
Q^\mV=v(\al_0^\vee-\al_2^\vee)-u f_1.
\]
The first three nontrivial systems of equations of the Drinfeld-Sokolov hierarchy are given by (with $t_1=x$):
\begin{align}
 u_{t_2}=&2 v u_{x}+4 u v_{x}+v_{x x x},\quad v_{t_2}=-\frac{u_{x}}{3}-2 v v_{x}; \\
u_{t_4}=&\frac{2}{3} u_{x x x} v+2 u_{x x} v_{x}+\frac{8 u_{x} v_{x
x}}{3}-\frac{4}{3} v^3 u_{x}+4 u v
   u_{x}+\frac{8 u^2 v_{x}}{3}+2 u v_{x x x}-8 u v^2 v_{x}\nn\\
   &+\frac{v_{x x x x x}}{3}-4
   \left(v_{x}\right)^3-2 v_{x x x} v^2-12 v v_{x} v_{x x}, \\
v_{t_4}=&-\frac{u_{x x x}}{9}+\frac{2 v^2 u_{x}}{3}-\frac{2 u
u_{x}}{9}+\frac{4}{3} u v
   v_{x}-\frac{2}{3} v_{x x x} v-\frac{4 v_{x} v_{x x}}{3}+\frac{20 v^3
v_{x}}{3};
   \\
u_{t_5}=&-\frac{5 u^2 u_{x}}{9}-\frac{5 u_{x} u_{x x}}{9}-\frac{5 u
u_{x x x}}{9}-\frac{u_{x x x x x}}{9}+\frac{5}{3} u_{x x x} v^2+10 v
u_{x x} v_{x}+10 v
   u_{x} v_{x x}\nn\\
   &+\frac{25}{3} u_{x} \left(v_{x}\right)^2
   +\frac{5 v^4 u_{x}}{3}+10 u v^2
   u_{x}+\frac{40}{3} u^2 v v_{x}+\frac{10}{3} u
   v_{x x x} v+\frac{10}{3} u v_{x} v_{x x}\nn\\
   &+\frac{40}{3} u v^3 v_{x}+20 v \left(v_{x}\right)^3-
   5 v_{x x x} v_{x x}-\frac{5}{3} v_{x x x x} v_{x}+\frac{10}{3} v_{x x x} v^3
   +30 v^2 v_{x} v_{x x}, \label{A2ut5} \\
v_{t_5}=&-\frac{5 u_{x x} v_{x}}{9}-\frac{10 u_{x} v_{x
x}}{9}-\frac{10 v^3 u_{x}}{9}-\frac{10}{9} u v
   u_{x}-\frac{5 u^2 v_{x}}{9}-\frac{5}{9} u v_{x x x}-\frac{10}{3} u v^2
   v_{x}\nn\\
   &-\frac{v_{x x x x x}}{9} +\frac{5 \left(v_{x}\right)^3}{3}+\frac{5}{3} v_{x x x}
   v^2+\frac{20}{3} v v_{x} v_{x x}-\frac{35}{3} v^4 v_{x}.
\end{align}
The second-order derivatives of the logarithm of the tau function are given by
\begin{align}\label{A2Om1}
\Om_{1 1}^\rs=& \frac{u}{3}-v^2, \quad \Om_{1 2}^\rs=\frac{4 u
v}{3}+\frac{v_{x x}}{3}+\frac{4 v^3}{3}, \\
\Om_{1 4}^\rs=&\frac{4 v u_{x x}}{9}+\frac{2 u_{x} v_{x}}{9}+\frac{8
u^2 v}{9}+\frac{2 u v_{x x}}{3}-\frac{16 u
   v^3}{9}+\frac{v_{x x x x}}{9}-\frac{4}{3} v \left(v_{x}\right)^2-\frac{8
   v^5}{3}+\frac{2 v^2 v_{x x}}{3}, \\
\Om_{1 5}^\rs=&-\frac{5 u^3}{81}-\frac{5 u u_{x x}}{27}-\frac{u_{x x
x x}}{27}+\frac{5 v^2 u_{x x}}{9}+\frac{10}{3} v u_{x}
   v_{x}+\frac{25 u^2 v^2}{9}+\frac{20}{9} u v v_{x x}-\frac{5}{9} u
   \left(v_{x}\right)^2\nn\\
   &+\frac{25 u v^4}{9}+\frac{2}{9} v_{x x x x} v-\frac{4
   \left(v_{x x}\right)^2}{9}+\frac{35 v^6}{9}-\frac{7}{9} v_{x x x} v_{x}-\frac{20 v^3
   v_{x x}}{9}+\frac{5}{3} v^2 \left(v_{x}\right)^2, \\
\Om_{2 2}^\rs=&-\frac{u_{x x}}{9}-\frac{2 u^2}{9}+\frac{4 u v^2}{3}-2 v^4+\frac{2 v v_{x x}}{3}-\frac{4 \left(v_{x}\right)^2}{3}, \\
\Om_{2 4}^\rs=&-\frac{u_{x x x x}}{27}-\frac{8 u^3}{81}+\frac{2 v^2 u_{x x}}{3}+\frac{8 u^2 v^2}{3}-\frac{2 u u_{x x}}{9}+\frac{8}{3} v
   u_{x} v_{x}+\frac{8 u v^4}{9}+\frac{20}{9} u v v_{x x}\nn\\
   &-\frac{4}{3} u \left(v_{x}\right)^2 +\frac{40 v^6}{9}+\frac{2}{9} v
   v_{x x x x}-\frac{2 \left(v_{x x}\right)^2}{9}-\frac{28 v^3 v_{x x}}{9}-\frac{10}{9}v_{x} v_{x x x}+4 v^2
   \left(v_x\right)^2. \label{A2Om2}
\end{align}
From \eqref{tauh} it follows that $h_1=3\,\Om_{1 1}^\rs$ and $h_2=\frac{3}{2}\,\Om_{1 2}^\rs$, hence the map $(u, v)\mapsto (h_1, h_2)$ is a Miura-type transformation, which illustrates Remark~\ref{rmk-hj}.

On $\fg$ the diagram automorphism $\sg$ is induced by the permutation
\[
\bar{\sg}(0)=2, \quad \bar{\sg}(1)=1, \quad \bar{\sg}(2)=0.
\]
The action of $\sg$ on the basis of $\mV$ is given by
\[
\sg(f_1)=f_1, \quad \sg(\al_0^\vee-\al_2^\vee)=-(\al_0^\vee-\al_2^\vee),
\]
which induces the map
\[
\sg^*(u)=u, \quad \sg^*(v)=-v.
\]
On the other hand, by comparing $J$ and $J^\sg$, we see that the generators $\Ld_j$ satisfy
\[
\sg(\Ld_j)=(-1)^{j-1}\Ld_j, \quad j=3 k\pm 1 \hbox{ with } k\in\Z.
\]
It is easy to see that the flows $\p/\p t_j$ and the differential polynomials $\Om_{i j}^\rs$
satisfy the relations
\[
\od\sg\left(\frac{\p}{\p t_j}\right)=(-1)^{j-1}\frac{\p}{\p t_j}, \quad \sg^*(\Om_{i j}^\rs)=(-1)^{(i-1)+ (j-1)}\Om_{i j}^\rs.
\]

By setting $v\equiv0$, the flows $\pd u/\pd t_j$ with $j\in\{1,
5\}+6\Z_+$ are reduced to the Drinfeld-Sokolov hierarchy associated to $\bar\fg$
of type $A_2^{(2)}$ with gradation $\bar\rs=(1,0)$. Note that $\mu=2$, so ${\pd}/{\pd\bar{t}_j}=\sqrt{2}\,{\pd}/{\pd t_j}$.
In particular, equation \eqref{A2ut5} is reduced to the Sawada-Kotera
equation (see Example~\ref{exa-A22tau})
\begin{equation}\label{SK}
\frac{\pd u}{\pd \bar{t}_5}=-\frac{1}{108}\frac{\pd}{\pd \bar{t}_1}\left(20
u^3+30 u \frac{\pd^2 u}{\pd \bar{t}_1^2}+3 \frac{\pd^4 u}{\pd
\bar{t}_1^4}\right).
\end{equation}
\end{exa}

\begin{exa}\label{exa-C1tau} {\bf (Case (c2) of Table \ref{table-diag3})}
Let $\fg$ be the affine Kac-Moody algebra of type $C_3^{(1)}$, with a gradation
$\rs=(0,1,1,0)$.
Let us take a gauge
\[
Q^\mV=v(\al_2^\vee-\al_1^\vee)-u
(f_0+f_3)-w(f_0-f_3).
\]
With $\Ld_j$ taken as in \cite{DS}, the first nontrivial
equations of the Drinfeld-Sokolov hierarchy are given by
\begin{align}\label{}
u_{t_3}=&-6 u  v  v_{x} -3 v_{x x x}  v +3 w  v_{x x} +18 v^3 v_{x} +\frac{3}{2} v  w_{x x} -3 w  w_{x},\\
v_{t_3}=&v  u_{x} +u  v_{x} -\frac{1}{2} v_{x x x} +3 v^2 v_{x} +\frac{3 w_{x x} }{4}, \\
w_{t_3}=&-v  u_{x x} -w  u_{x} +3 u  v_{x x} -2 u  w_{x}
+\frac{3}{2} v_{x x x x} -9 v ^2 v_{x x} -6 v  w  v_{x} -18 v
   v_{x}^2-\frac{5}{4} w_{x x x}, \\
u_{t_5}=&   -\frac{5}{9} u^2 u_{x} -\frac{5}{9} u_{x}
   u_{x x}-\frac{5}{9} u  u_{x x x}-\frac{1}{9} u_{x x x x x}
-\frac{5}{6} u_{x x x} v^2-\frac{5}{2} v  u_{x x}  v_{x}
+\frac{5}{2} v
   w  u_{x x}\nn \\
   & -\frac{25}{6} v  u_{x}  v_{x x} +\frac{10}{3} w  u_{x}  v_{x} -\frac{5}{3} u_{x}  v_{x}^2 -\frac{5}{6} v  u_{x}
   w_{x} +5 v^4 u_{x} -\frac{10}{3} u  v^2 u_{x} +\frac{5}{6} w^2 u_{x} \nn \\
   &-\frac{5}{2} u  v_{x x x}  v -\frac{5}{3} u  w  v_{x x}  -5 u  v_{x}  w_{x} +20 u  v^3 v_{x} -\frac{20}{3}
   u^2 v  v_{x}  -5 u  v_{x}  v_{x x} -\frac{10}{3} u  v  w_{x x} \nn \\
   & +\frac{5}{3} u  w  w_{x} +\frac{5}{12} v_{x x x x x}
   v -\frac{5}{3} v_{x x x x}  w  -\frac{10}{3} v_{x x x}  w_{x} -\frac{5}{2} v_{x x x}  v^3-\frac{15}{4} v_{x x}  w_{x x} +10
   v^2 w  v_{x x}\nn \\
   & -\frac{5}{3} w_{x x x}  v_{x} +20 v^2 v_{x}  w_{x} +25 v  w  v_{x}^2 +10 v  w^2 v_{x} +15 v
   v_{x}^3-\frac{5}{3} v_{x x x x}  v_{x} -\frac{5}{6} v_{x x x}  v_{x x}\nn \\
   & -\frac{35}{24} v  w_{x x x x} +5 v^3 w_{x x} +5 v^2
   w  w_{x}  +\frac{5}{3} w  w_{x x x} +\frac{5}{4} w_{x}  w_{x x} .
\end{align}
Here we omit $v_{t_5}$ and $w_{t_5}$, each term of which contains $v$, $w$ or their $x$-derivatives.  The first few differential polynomials $\Om_{j k}^\rs$ have the expressions
\begin{align*}
  \Om_{1 1}^\rs=&\frac{u}{3}-v^2,  \qquad
  \Om_{1 3}^\rs= -2 u v^2+w v_{x}-v
   w_{x}-\frac{w^2}{2}, \\
  \Om_{1 5}^\rs=&-\frac{u_{x x x x}}{27}-\frac{5 u^3}{81}-\frac{5 u^2
   v^2}{3}-\frac{5 u u_{x x}}{27}-\frac{5}{9} v u_{x} v_{x}+\frac{5}{3} v w
   u_{x}+\frac{5 u v^4}{3}-\frac{5}{3} u v v_{x x}-\frac{5}{9} u w
   v_{x} \\
   &-\frac{10}{9} u v w_{x}+\frac{5 u w^2}{18}-\frac{5
   v^6}{3}-\frac{11}{18} v_{x x x x} v-\frac{5}{9} v_{x x x} w+\frac{5 v^2
   w^2}{2}-\frac{5 v_{x x} w_{x}}{9}-\frac{\left(v_{x x}\right)^2}{6}\\
   &-\frac{25
   v_{x} w_{x x}}{36}+5 v^3 v_{x x}+\frac{1}{18} v_{x x x} v_{x}+5 v^2 w
   v_{x}+\frac{20}{3} v^2 \left(v_{x}\right)^2+\frac{5}{36} v
   w_{x x x}+\frac{5 w w_{x x}}{9}-\frac{5 \left(w_{x}\right)^2}{72}.
\end{align*}

The diagram automorphism $\sg$ on $\fg$ is given by
\[
\bar{\sg}(i)=3-i, \quad i=0, 1, 2, 3.
\]
It induces the map
\[
\sg^*(u)=u, \quad \sg^*(v)=-v, \quad \sg^*(w)=-w,
\]
and satisfies, due to $J^\sg=J$, the relations
\[
\sg(\Ld_j)=\Ld_j, \quad j\in 2\Z+1.
\]
Clearly, the above flows $\p/\p t_j$ and the differential polynomials $\Om_{j k}^\rs$ satisfy
\[
\od\sg\left(\frac{\p}{\p t_j}\right)=\frac{\p}{\p t_j}, \quad \sg^*(\Om_{j k}^\rs)=\Om_{j k}^\rs,
\]
which agree with the results of Proposition~\ref{thm-sg}. Since $J^\sg=\bar{J}\cup 3\Z^{\mathrm{odd}}=J$, from Theorem~\ref{thm-qred} it follows that the flows $\p/\p t_{j}$ with $j\in 3\Zop$ must be trivial when restricted to $\bar\mR=\C[u, u_x, u_{x x}, \dots]$. We can see this fact for the flow $\p/\p t_{3}$ that is illustrated above. Moreover, the flows $\pd u/\pd t_j$ with $j\in \{1, 5\}+6\Z_+$ are reduced to the flows of the Drinfeld-Sokolov hierarchy that is associated to the affine Kac-Moody algebra $\bar\fg$
of type $A_2^{(2)}$ with gradation $\bar\rs=(1,0)$.
In particular, the
flow $\pd /\pd t_5$ is reduced to the Sawada-Kotera
equation \eqref{SK} (note that in this case $\mu=2$, so ${\pd}/{\pd\bar{t}_j}=\sqrt{2}\,{\pd}/{\pd t_j}$).
\end{exa}

\subsection{Proof of the $\Gm$-reduction theorem}

Let $\sL$ be the operator introduced in \eqref{sL} with $Q\in C^\infty(\R, \mathcal{B})$, and the functions
\begin{equation}\label{UHQ}
U(Q)\in C^\infty(\R, \fg^{<0}), \quad H(Q)\in C^\infty(\R, \mH\cap\fg^{<0})
\end{equation}
be defined  by Lemma~\ref{thm-dr0}.

Note that the Borel subalgebras $\mathcal{B}$ and $\bar{\mathcal{B}}=\mathcal{B}\cap\bar\fg$ are invariant under the action of the automorphism $\sg$, and
\[
\bar{\mathcal{B}}=\mathcal{B}\cap\fg^\sg=\{X\in\mathcal{B}\mid \sg(X)=X\}
\]
due to the fact that $\mathcal{B}\cap\hat{\mH}=\{0\}$ with $\hat\mH$ defined in Theorem \ref{thm-ggbar}. Then we have the following decomposition of subspaces:
\begin{equation}\label{BbarBhat}
\mathcal{B}=\bar{\mathcal{B}}\oplus\hat{\mathcal{B}},
\end{equation}
where $\hat{\mathcal{B}}\subset\mathcal{B}$ is spanned by the eigenvectors of $\sg$ with eigenvalues different from $1$.

Let $\bar{Q}=Q|_{\bar{\mathcal{B}}}$, then we have the following analogue of the operator \eqref{sL} associated to the affine Kac-Moody algebra $\bar\fg$:
\[
\bar\sL=\frac{\od}{\od x}+\Ld+\bar{Q}\in \C\frac{\od}{\od x}\ltimes C^\infty(\R,\bar{\mathcal{B}}).
\]
We denote the functions $U$ and $H$ that are defined by Lemma \ref{thm-dr0} as follows:
\begin{equation}\label{UHQbar}
\bar{U}(\bar{Q})\in C^\infty(\R, \bar\fg^{<0}), \quad \bar{H}(\bar{Q} )\in C^\infty(\R, \bar{\mH}\cap\bar{\fg}^{<0}).
\end{equation}
The operator $\bar\sL$ admit a group of gauge transformations given by the nilpotent subalgebra $\bar{\mathcal{N}}=\mathcal{N}\cap\bar\fg$, and a gauge slice $\bar\sL^{\bar\mV}$ of it is given by \eqref{sLbar}.

We have the following lemma.
\begin{lem}\label{thm-red}
The functions given in \eqref{UHQ} and \eqref{UHQbar} satisfy the following relations:
\begin{equation}\label{}
U(\bar{Q})=\bar{U}(\bar{Q}), \quad H(\bar{Q})=\bar{H}(\bar{Q}).
\end{equation}
\end{lem}
\begin{prf}
We regard $\bar\sL$ as an operator in $\C\od/\od x\ltimes C^\infty(\R,\mathcal{B})$, and $\bar{U}(\bar{Q})$, $\bar{H}(\bar{Q})$ as functions taking  values in $C^\infty(\R, \fg^{<0})$, $C^\infty(\R, \mH\cap\fg^{<0})$ respectively. Due to the uniqueness result of Lemma \ref{thm-dr0}, we only need to show that
\begin{equation}\label{}
\left(d^\rs\mid e^{\ad_{\bar{U}(\bar{Q})}}{\Ld}_j\right)=0, \quad j\in J_+.
\end{equation}
The case when $j\in\bar{J}_+$ follows immediately from the definition of $\bar{U}(\bar{Q})$ and the property \eqref{bilinearbar} of the bilinear form defined on $\bar\fg\oplus\C d^\rs=\fg(\bar A)$. For the case when $j\in J_+\setminus\bar{J}_+$, since
\[
\left(d^\rs\mid e^{\ad_{\bar{U}(\bar{Q})}}{\Ld}_j\right)=\left(e^{-\ad_{\bar{U}(\bar{Q})}}d^\rs\mid \Ld_j\right),
\]
where $e^{-\ad_{\bar{U}(\bar{Q})}}d^\rs$ lies in $\bar\fg\oplus\C d^\rs$, we only need to show that the relation
\begin{equation}\label{Ldgbar}
(X\mid\Ld_j)=0
\end{equation}
holds true for any $X\in \bar\fg\oplus\C d^\rs$ and  $j\in J_+\setminus\bar{J}_+$.
In fact, it is easy to see that
$(d^\rs\mid\Ld_j)=0$, and that $(X\mid\Ld_j)=0$ for any $X$ lying in $[\Ld, \bar{\fg}]$ or in $\bar{\mathcal{H}}$ (we note that $1\notin J_+\setminus\bar{J}_+$). Thus \eqref{Ldgbar} follows from the decomposition \eqref{gbardec} of $\bar\fg$.
The lemma is proved.
\end{prf}

\begin{lem}\label{thm-ham}
The Hamiltonian densities
\[
h_j(Q)=-\left(\Ld_j\mid H(Q)\right), \quad \bar{h}_j(\bar Q)=-\left(\bar\Ld_j\mid \bar{H}(\bar Q)\right)^-
\]
of the Drinfeld-Sokolov hierarchies associated to $(\fg, \rs, \mathds{1})$ and to $(\bar\fg, \bar\rs, \mathds{1})$ satisfy the following relations:
\begin{equation}\label{}
h_j(\bar{Q})=\left\{\begin{array}{cl}
                      \dfrac{1}{\kappa\sqrt{\mu}}\bar{h}_j(\bar{Q}), & j\in \bar{J}_+; \\
                      0, & j\in  J_+\setminus\bar{J}_+.
                    \end{array}
 \right.
\end{equation}
\end{lem}
\begin{prf}
For $j\in\bar{J}_+$, by using \eqref{Ldjbar} and  \eqref{bilinearbar},  we obtain
\begin{equation}\label{hhbar}
\bar{h}_j(\bar{Q})=\left(-\bar{\Ld}_j\mid\bar{H}(\bar{Q})\right)^-=\kappa\sqrt{\mu}\left(-\Ld_j\mid
H(\bar{Q}) \right)=\kappa\sqrt{\mu}\,h_j(\bar{Q}).
\end{equation}
For $j\in  J_+\setminus\bar{J}_+$, Lemma~\ref{thm-red} implies that $H(\bar{Q})$ belongs to $\bar\fg$, then $h_j(\bar{Q})$ vanishes due to \eqref{Ldgbar}. Thus the lemma is proved.
\end{prf}

\vskip 2ex

\begin{prfof}{Theorem~\ref{thm-qred}}
As it was done in the previous subsection, we take a $\sg$-invariant complementary subspace $\mathcal{V}\subset\mathcal{B}$.
Note that $Q^\mV$ is a gauge of $Q\in C^\infty(\R, \mathcal{B})$, then $\bar{Q}^{\bar{\mV}}=\left.Q^\mV\right|_{\bar{\mathcal{B}}}$ taking value in $\bar\mV=\mV\cap\bar\fg$ is a gauge of $\bar Q=\left.Q\right|_{\bar{\mathcal{B}}}$. Let $\sL^\mV$ and ${\bar{\sL}}^{\bar{\mV}}$ be the corresponding operators
defined in \eqref{QV} and \eqref{sLbar}.

By using Lemma~\ref{thm-red}, together with the decompositions \eqref{BV} and \eqref{BbarBhat} of the Borel subalgebra $\mathcal{B}$, the equations \eqref{Lt3} are recast to
\begin{align}\label{}
\frac{\pd\bar{\sL}^{\bar\mV}}{\pd\bar{t}_j} =&\left[-(e^{\ad_{\bar{U}(\bar{Q}^{\bar{\mV}}) }
}\bar{\Ld}_j)_{\ge0}+{R}(\bar{Q}^{\bar{\mV}}, \bar{\Ld}_j), {\bar\sL}^{\bar{\mV}} \right] \nn\\
=&\sqrt{\mu} \left[-(e^{\ad_{{U}(\bar{Q}^{\bar{\mV}}) }
}{\Ld}_j)_{\ge0}+{R}(\bar{Q}^{\bar{\mV}}, \Ld_j), {\bar\sL}^{\bar{\mV}} \right] \nn\\
=&\sqrt{\mu} \left[-(e^{\ad_{{U}(\left.Q^\mV\right|_{\bar{\mathcal{B}}} ) }
}{\Ld}_j)_{\ge0}+{R}(\left.Q^\mV\right|_{\bar{\mathcal{B}}}, \Ld_j), \frac{\od}{\od x}+\Ld+\left.Q^\mV\right|_{\bar{\mathcal{B}}}\right] \nn\\
=&
 \sqrt{\mu}\left.\frac{\pd\sL^\mV}{\pd
t_j}\right|_{\bar{\mathcal{R}}}, \quad j\in\bar{J}_+.
\end{align}
So the first assertion of Theorem~\ref{thm-qred} is proved.

Now let us proceed to show the second assertion of
Theorem~\ref{thm-qred}.  We represent the coordinates of $\mathcal{V}$
as $\mathbf{u}=(\bar{\mathbf{u}},\hat{\mathbf{u}})$, where
\begin{equation}\label{udec}
\bar{\mathbf{u}}=\{u_i\mid \sg^*
u_i=u_i\}, \quad \hat{\mathbf{u}}=\{u_i\mid \sg^* u_i\ne u_i\}.
\end{equation}
Recall the Hamiltonian representation \eqref{utham} of the flows of the Drinfeld-Sokolov hierarchy.
For any $j\in J^\sg_+\setminus\bar{J}_+$, according to Proposition~\ref{thm-sg} and Lemma~\ref{thm-ham}, the Hamiltonian densities $h_j=h_j(Q^\mV)\in\mR$ satisfy the
relations
\[
\sg^*h_j=h_j, \quad h_j|_{\bar\mR}=0.
\]
They imply that each monomial term of
the differential polynomial $h_j$ contain factors in
$\hat{\mathbf{u}}$ and their $x$-derivatives of total degree no less than $2$,
hence the variational derivatives of $\mathscr{H}_j=\int h_j \od x$ satisfy
\[
\left.\frac{\dt \mathscr{H}_j}{\dt
u_m}\right|_{\bar\mR}=\sum_{k\ge0}\left(-\frac{\pd}{\pd x}\right)^k\left.\frac{\pd h_j}{\pd
u_m^{(k)}}\right|_{\bar\mR}=0, \quad m=1, 2, \dots, \ell.
\]
Thus we have
$\left({\pd\mathbf{u}}/{\pd{t}_j}\right)|_{\bar\mR}=0$ for $j\in J^\sg_+\setminus\bar{J}_+$.
The theorem is proved.
\end{prfof}

With the notations introduced in \eqref{udec}, the Drinfeld-Sokolov hierarchy \eqref{zh-9}
associated to $(\fg, \rs, \mathds{1})$ can be represented as:
\begin{equation}\label{vt}
\frac{\pd\bar{\mathbf{u}}}{\pd t_j}=\bar{\mathbf{X}}(\bar{\mathbf{u}},\hat{\mathbf{u}}), \quad
\frac{\pd\hat{\mathbf{u}}}{\pd t_j}=\hat{\mathbf{X}}(\bar{\mathbf{u}},\hat{\mathbf{u}}), \qquad
j\in J_+,
\end{equation}
where  $\bar{\mathbf{X}}, \hat{\mathbf{X}}\in \mR$.
It follows from the $\Gm$-reduction theorem (Theorem~\ref{thm-qred}) that
\begin{equation}\label{ubart}
\left.\frac{\pd\bar{\mathbf{u}}}{\pd
{\bar{t}}_j}\right|_{\bar\mR}=\sqrt{\mu}\, \bar{\mathbf{X}}(\bar{\mathbf{u}},0), \quad \left.\frac{\pd\hat{\mathbf{u}}}{\pd
{\bar{t}}_j}\right|_{\bar\mR}=\sqrt{\mu}\,\hat{\mathbf{X}}(\bar{\mathbf{u}},0)=0, \quad j\in\bar{J}_+,
\end{equation}
and that the reduced flows form the Drinfeld-Sokolov hierarchy associated to $(\bar\fg, \bar\rs, \mathds{1})$.

\subsection{Reduction properties of the tau functions}

We proceed to study properties of solutions of the Drinfeld-Sokolov hierarchies, and in particular we will concentrate on solutions of formal series in the time variables. As before we fix a $\sg$-invariant complementary subspace $\mV$ in \eqref{BV}.
Given any solution $\mathbf{u}(\mathbf{t})=(u_1(\bt), u_2(\bt),\dots, u_\ell(\bt))$ of the Drinfeld-Sokolov hierarchy \eqref{LVt} (or \eqref{zh-9} ) associated to $(\fg, \rs, \mathds{1})$ such that each $u_i(\bt)\in \mathcal{S}:=\C[[\mathbf{t}]]$ with  $\mathbf{t}=\{t_j\}_{j\in J_+}$, there is an action on $\mathcal{S}$ induced by \eqref{sgv} such that
\begin{equation}\label{sgut}
\sg^*: ~\mathbf{u}(\mathbf{t}) \mapsto (\ep_1 u_1(\bt), \dots, \ep_\ell u_\ell(\bt)).
\end{equation}
On the other hand, the assertion (iv) of Proposition \ref{thm-sg} leads us to
define the map on the space of the time variables \footnote{In general, we can define the map on the space of time variables as  $\sg^* t_j=\zeta_j t_j+c_j$ for some shift constants $c_j$ with $j\in J_+$. Without lose of generality, we choose all such shift constants to be zero, for the reason that in this case the solution $\tilde{\mathbf{u}}(\mathbf{t})$ of the Drinfeld-Sokolov hierarchy define by \eqref{tildeut} still belongs to $\mathcal{S}$.}
\[
\sg^* t_j=\zeta_j t_j, \quad j\in J_+,
\]
then one sees that
\begin{equation}\label{tildeut}
\tilde{\mathbf{u}}(\bt):=\sg^*\left(\, \mathbf{u}\left( (\sg^*)^{-1}\mathbf{t}\right) \, \right)
\end{equation}
is also a solution of the Drinfeld-Sokolov hierarchy. Indeed, by using Proposition~\ref{thm-sg} one has, for any $j\in J_+$,
\begin{align}\label{}
  \frac{\p\tilde{u}_i(\bt)}{\p t_j}=& \frac{\p }{\p t_j} \sg^*\left(\, \mathbf{u}\left( (\sg^*)^{-1}\mathbf{t}\right) \, \right) =
   \frac{\ep_i}{\zeta_j}\left.\frac{\p u_i(\bt)}{\p t_j}\right|_{\bt\mapsto(\sg^*)^{-1}\mathbf{t} } \nn\\
   =&  \frac{\ep_i}{\zeta_j}\left. X_j^i(\mathbf{u}(\mathbf{t}))\right|_{\bt\mapsto(\sg^*)^{-1}\mathbf{t} }
   = \left.X_j^i(\sg^*\mathbf{u}(\mathbf{t}))\right|_{\bt\mapsto(\sg^*)^{-1}\mathbf{t} } \nn\\
 =&X_j^i\left(\,\sg^*\left(\bu( (\sg^*)^{-1}\bt) \right)\,\right)=X_j^i(\tilde{\mathbf{u}}(\bt) ).
\end{align}
For the solution $\tilde{\bu}(\bt)$, let $\tilde{\tau}^\rs(\bt)$ be the tau function defined by \eqref{tauOm}.
\begin{prp}\label{thm-sgtau}
The following equality holds true (up to the addition of a linear function in $\bt$):
\begin{equation}\label{sgtau}
\log\tilde\tau^\rs(\bt)=\log\tau^\rs\left( (\sg^*)^{-1}\bt\right).
\end{equation}
\end{prp}
\begin{prf}
We only need to compare
the second-order derivatives of both sides of \eqref{sgtau}. In
fact, by using Proposition~\ref{thm-sg}, for any $j,k\in J_+$ we have
\begin{align*}
\frac{\pd^2\log\tilde{\tau}^\rs(\bt)}{\pd t_j\pd
t_k}=& \left.\Om_{j k}^\rs(\mathbf{u})\right|_{\mathbf{u}=\sg^*(\,\mathbf{u}( (\sg^*)^{-1}\bt)\,)}= \left.\Om_{j k}^\rs(\sg^*(\mathbf{u}(\bt) ))\right|_{\bt\mapsto (\sg^*)^{-1}\bt } \\
=& \sg^*\left.\Om_{j k}^\rs( \mathbf{u}(\bt) )\right|_{\bt\mapsto (\sg^*)^{-1}\bt } =  \zeta_j^{-1}\zeta_k^{-1}\left.\Om_{j k}^\rs( \mathbf{u}(\bt) )\right|_{\bt\mapsto (\sg^*)^{-1}\bt } \\
=&  \zeta_j^{-1}\zeta_k^{-1}\left.
  \frac{\pd^2 \log\tau^\rs(\bt)}{\pd  t_j \pd  t_k}\right|_{\bt\mapsto (\sg^*)^{-1}\bt } =\frac{\pd^2 \log\tau^\rs\left((\sg^*)^{-1}\bt\right)}{\pd t_j \pd t_k}
\end{align*}
Thus the proposition is proved.
\end{prf}

The proposition implies that, up to the addition of a linear function of $\bt$,  $\log\tilde{\tau}^\rs(\bt)=\log\tau^\rs(\bt)$ if and only if
\begin{equation}\label{tausgt}
\log\tau^\rs(\sg^*\bt)=\log\tau^\rs(\bt).
\end{equation}
\begin{defn} \label{def-Gmtau}
The tau function $\tau^\rs(\bt)$ for a certain solution $\bu(\bt)$ of the Drinfeld-Sokolov
hierarchy \eqref{LVt} associated to $(\fg, \rs, \mathds{1})$ is called $\Gm$-invariant if it satisfies \eqref{tausgt}.
\end{defn}

Now let us consider the tau functions $\bar{\tau}^{\bar\rs}$ of the Drinfeld-Sokolov
hierarchy \eqref{Lt3} associated to $(\bar\fg, \bar\rs, \mathds{1})$ which is defined by
the formulae
\begin{equation}\label{bartau}
\frac{\pd^2\log\bar\tau^{\bar\rs}}{\pd \bar{t}_i \pd
\bar{t}_j}=\frac{1}{\bar{h}^{\bar{\rs}} }
\left(d^{\bar\rs}\mid
\left[\left( e^{\ad_{ \bar{U}(\bar{Q}^{\bar\mV}) } }\bar{\Ld}_i\right)_{\ge0}, e^{\ad_{\bar{U}(\bar{Q}^{\bar\mV}) } }{\bar\Ld}_j\right]\right)^-, \quad i,j\in \bar{J}_+
\end{equation}
with
\begin{equation}\label{}
\bar{h}^{\bar\rs}=(d^{\bar\rs}\mid\bar{c})^-=\kappa(d^\rs\mid \mu{c})=\kappa\mu
h^\rs.
\end{equation}
For the purpose of clarifying the relation between $\bar{\tau}^{\bar{\rs}}$ and $\tau^\rs$, let us introduce the following notations:
\begin{equation}\label{tthat}
\tilde{\bt}=\{t_j\}_{j\in\bar{J}_+}, \quad \hat{\bt}=\{t_j\}_{j\in J_+\setminus\bar{J}_+}, \quad \bar{\bt}=\{\bar{t}_j\}_{j\in\bar{J}_+}.
\end{equation}
\begin{thm}\label{thm-taured}
Let the affine Kac-Moody algebras $\fg$ and $\bar\fg$ be given as in Tables~\ref{table-diag1} and \ref{table-diag2}. Suppose a tau function $\tau^\rs(\bt)$, satisfying the condition $\log\tau^\rs(\bt)\in \mathcal{S}$,  of the Drinfeld-Sokolov hierarchy \eqref{LVt} associated to $(\fg, \rs, \mathds{1})$
 is $\Gm$-invariant, then the function $\bar{\tau}^{\bar\rs}(\bar\bt)$ defined by
\begin{equation}\label{tautaubar}
\log\bar{\tau}^{\bar\rs}(\bar\bt)=\left.\frac{1}{\mu}\log\tau^\rs(\bt)
\right|_{\hat{\bt}=0; \,\tilde{\bt}=\sqrt{\mu}\,\bar{\bt}}
\end{equation}
is a tau function of the Drinfeld-Sokolov hierarchy \eqref{Lt3} associated to $(\bar\fg, \bar\rs, \mathds{1})$.
\end{thm}

\begin{prf}
Since $\tau^\rs(\bt)$ is $\Gm$-invariant, it follows from \eqref{tauh} that the Hamiltonian densities $h_j$ with  $j\in J_+$ satisfy
\begin{align}\label{sgh}
\left.h_j(\,\bu(\bt)\,)\right|_{\hat{\bt}=0}=&\frac{h}{j}\left.\frac{\p^2\log\tau^\rs(\bt)}{\p x\p t_j}\right|_{\hat{\bt}=0}=\frac{h}{j}\left.\frac{\p^2\log\tilde{\tau}^\rs(\bt)}{\p x\p t_j}\right|_{\hat{\bt}=0} \nn\\
=&\left.h_j(\tilde{\bu}(\bt))\right|_{\hat{\bt}=0} =\left.\sg^*h_j(\,\bu((\sg^*)^{-1}\bt)\,)\right|_{\hat{\bt}=0}  \nn\\
=&\frac{1}{\zeta_j}\left.h_j(\,\bu((\sg^*)^{-1}\bt)\,)\right|_{\hat{\bt}=0}
=\frac{1}{\zeta_j}\left.h_j(\,\bu(\bt)\,)\right|_{\hat{\bt}=0},
\end{align}
where the last two equalities follow from the assertion (i) of Proposition~\ref{thm-sg} and the fact that $\sg^*\bar{\bt}=\bar{\bt}$. For any case listed in Tables~\ref{table-diag1} and \ref{table-diag2}, we have $\zeta_j\ne1$ whenever  $j\in J_+\setminus\bar{J}_+$, hence
\begin{equation}\label{hutred}
h_j(\,\bu(\bt)\,)|_{\hat{\bt}=0}=0, \quad j\in J_+\setminus\bar{J}_+.
\end{equation}
Let us recall the fact, as we explained in Remark~\ref{rmk-hj},
that the Hamiltonian densities
$\bh=(h_{m_1}, h_{m_2}, \dots, h_{m_\ell})$, where $m_1, m_2, \dots, m_\ell$ are the first $\ell$ exponents in $J_+$, are related to $\bu=(u_1, u_2, \dots, u_\ell)$ by a Miura-type transformation. Hence, we can represent $\bu$ in terms of $\bh$ as
follows:
\begin{equation}
u_i=g_i(\bh)\in C^\infty(\bh)[[\bh', \bh'', \dots]],\quad i=1,\dots, \ell.
\end{equation}
Moreover, such a representation must be consistent with the action of $\sg^*$, namely, we have
\[\sg^* u_i=g_i(\sg^*\bh).\]
By using \eqref{sgut},  \eqref{hutred} and that $\sg^*h_j=h_j$ for $j\in \bar{J}_+$,
we have
\begin{align}
\ep_i u_i(\bt)|_{\hat{\bt}=0}=\sg^*(u_i(\bt))|_{\hat{\bt}=0}=g_i(\, \sg^*\bh( \bu(\bt) )\,)|_{\hat{\bt}=0}=g_i(\, \bh( \bu(\bt) )\,)|_{\hat{\bt}=0}=u_i(\bt)|_{\hat{\bt}=0}.
\end{align}
It follows from \eqref{udec} that $\hat{\mathbf{u}}|_{\hat{\bt}=0}=0$, hence  $\bar{\mathbf{u}}|_{\hat{\bt}=0}$ satisfies the equations given in \eqref{ubart}, which means that $\bar{\mathbf{u}}|_{\hat{\bt}=0; \,\tilde{\bt}=\sqrt{\mu}\,\bar{\bt}}$
is a solution of the
Drinfeld-Sokolov hierarchy associated to $(\bar\fg, \bar\rs, \mathds{1})$.
In other words, these solutions of the Drinfeld-Sokolov
hierarchies associated to $(\bar\fg, \bar\rs, \mathds{1})$ and to $(\fg, \rs, \mathds{1})$ are related by
 $\bar{Q}^{\bar{\mV}}=\left.Q^\mV\right|_{\hat{\bt}=0; \,
\tilde{\bt}=\sqrt{\mu}\,\bar{\bt}}$.
Consequently, by using Lemma~\ref{thm-red}, \eqref{Ldjbar},\eqref{bilinearbar} and \eqref{bartau} we have, for $i,\, j\in\bar{J}_+$, that
\begin{align*}
\frac{\pd^2\log\bar{\tau}^{\bar\rs}}{\pd\bar{t}_i\pd\bar{t}_j}
=&\frac{\kappa\mu}{\kappa\mu h^\rs}\left(d^\rs\mid
\left[\left(e^{\ad_{{U}(\bar{Q}^{\bar\mV})}}{\Ld}_i\right)_{\ge0},
e^{\ad_{{U}(\bar{Q}^{\bar\mV})}}{\Ld}_j\right]\right) \\
=&\left.\frac{1}{ h^\rs}\left(d^\rs\mid
\left[\left(e^{\ad_{{U}(Q^\mV)}}{\Ld}_i\right)_{\ge0},
e^{\ad_{{U}(Q^\mV)}}{\Ld}_j\right]\right)\right|_{\hat{\bt}=0; \,\tilde{\bt}=\sqrt{\mu}\,\bar{\bt}} \\
=&\left.\frac{\pd^2\log{\tau^\rs}}{\pd {t}_i\pd
{t}_j}\right|_{\hat{\bt}=0; \,\tilde{\bt}=\sqrt{\mu}\,\bar{\bt}}.
\end{align*}
Therefore, in consideration of $\pd/\pd\bar{t}_j=\sqrt{\mu}\,\pd/\pd {t}_j$, we complete
the proof of the theorem.
\end{prf}

As a conclusion, from Theorems~\ref{thm-qred} and \ref{thm-taured} we arrive at Theorem~\ref{thm-main}.

\begin{exa}
Let us compute some explicit solutions of the Drinfeld-Sokolov hierarchy associated to $(\fg=A_2^{(1)}, \rs=(1, 0, 1), \mathds{1})$. The first few flows are given in  Example~\ref{exa-A2sg} with unknown functions $u(\bt), v(\bt)$.
Note that every solution is determined by its initial value $(u(\bt), v(\bt))|_{t_{>1}=0}=(u^0(x), v^0(x))$. We take $u^0(x)=a x$ and $v^0(x)=b$ with arbitrary constants $a$ and $b$, and we arrive at the following result:
\begin{align*}
  &u(\bt)|_{t_{i >6}=0}\\
  =&a x+2 a b t_{2}+\frac{4}{3} t_{4} \left(3 a^2 b x-a b^3\right)-\frac{5}{9} t_{5} \left(a^3 x^2-18 a^2 b^2 x-3 a
   b^4\right)-\frac{1}{3} a^2
   t_{2}^2\\
   &-\frac{4}{27} t_{4}^2 \left(5 a^4 x^2-126 a^3
   b^2 x+45 a^2 b^4\right)+\frac{25}{81} t_{5}^2 \left(2 a^5 x^3-114
   a^4 b^2 x^2+a^4+246 a^3 b^4 x+42 a^2 b^6\right)\\
   &-\frac{80}{9} t_{5} t_{2} \left(a^3 b x-2 a^2
   b^3\right)-\frac{4}{3} t_{4} t_{2} \left(a^3 x-7 a^2
   b^2\right)-\frac{20}{27} t_{4} t_{5} \left(19 a^4 b x^2-122 a^3 b^3
   x+3 a^2 b^5\right)+ \mathrm{h.o.t.}, \\
   &\\
  &v(\bt)|_{t_{i>6}=0} \\
  =&b-\frac{1}{3} a t_{2}-\frac{2}{9} t_{4}
   \left(a^2 x-3 a b^2\right)  -\frac{10}{9} t_{5} \left(a^2 b x+a b^3\right)
   -\frac{32}{27} t_{4}^2 \left(a^3 b x-a^2
   b^3\right)\\
   &+\frac{50}{81} t_{5}^2 \left(3 a^4 b x^2-10 a^3 b^3 x+3
   a^2 b^5\right) -\frac{8}{9} a^2 b t_{2}
   t_{4}+\frac{10}{27} t_{2} t_{5} \left(a^3 x-3 a^2
   b^2\right)\\
   &+\frac{10}{27} t_{4} t_{5} \left(a^4 x^2-22 a^3 b^2
   x-7 a^2 b^4\right)+\mathrm{h.o.t.},
\end{align*}
where `$\mathrm{h.o.t.}$' means higher order terms in $t_2,\ t_4,\ t_5$.
The tau function has the expression
{\small
\begin{align*}
&\log\tau^\rs(\bt)|_{t_{i>6}=0}\\
=&-\frac{b^2 x^2}{2}+\frac{a x^3}{18}
+t_{2} \left(\frac{2}{3} a b x^2+\frac{4 b^3
   x}{3}\right)
+t_{4} \left(\frac{8}{27} a^2 b x^3-\frac{8}{9} a b^3 x^2-\frac{8 b^5
   x}{3}\right) \\
   & +t_{5}
   \left(-\frac{5 a^3 x^4}{324}+\frac{25}{27} a^2 b^2
   x^3+\frac{25}{18} a b^4 x^2+\frac{35 b^6 x}{9}\right)
   +\frac{1}{2} t_{2}^2 \left(-\frac{2 a^2
   x^2}{9}+\frac{4}{3} a b^2 x-2 b^4\right) \\
   &
   +\frac{1}{2} t_{4}^2 \left(-\frac{4 a^4 x^4}{81}+x
   \left(-\frac{4 a^3}{81}-\frac{176 a b^6}{27}\right)+\frac{80}{27}
   a^3 b^2 x^3-\frac{136}{27} a^2 b^4 x^2+\frac{20 a^2
   b^2}{27}-\frac{100 b^8}{9}\right)
   \\
   &+\frac{1}{2} t_{5}^2
   \left(\frac{5 a^5 x^5}{243}-\frac{25}{9} a^4 b^2
   x^4+\frac{950}{81} a^3 b^4 x^3+x \left(-\frac{350 a^3
   b^2}{81}-\frac{575 a b^8}{27}\right)+\frac{175 a^2 b^4}{81}\right. \\
   &\left.+x^2
   \left(\frac{25 a^4}{243}-\frac{50 a^2 b^6}{81}\right)-\frac{245
   b^{10}}{9}\right)
   +t_{2} t_{5}
   \left(-\frac{20}{27} a^3 b x^3+\frac{100}{27} a^2 b^3
   x^2-\frac{10 a^2 b}{27}-\frac{20}{9} a b^5 x-\frac{20
   b^7}{3}\right)
   \\ &
   +t_{2} t_{4} \left(-\frac{8 a^3 x^3}{81}+\frac{8}{3}
   a^2 b^2 x^2+\frac{8}{9} a b^4 x+\frac{40 b^6}{9}\right)
   +t_{4} t_{5}
   \left(-\frac{40}{81} a^4 b x^4+x \left(\frac{320 a
   b^7}{27}-\frac{20 a^3 b}{27}\right) \right. \\
   &\left.+\frac{640}{81} a^3 b^3
   x^3+\frac{80}{27} a^2 b^5 x^2+\frac{20 a^2 b^3}{9}+\frac{1400
   b^9}{81}\right)
    +\mathrm{h.o.t.}
\end{align*}
}
Note that
\[
\sg^*(u(\bt),v(\bt))=(u(\bt),-v(\bt)), \quad \sg^*t_j=(-1)^j t_j \hbox{ for } j\in  3\Z_++\{1, 2\}.
\]
It can be seen that
\[
(\tilde{u}(\bt), \tilde{v}(\bt))=\left(u( (\sg^*)^{-1}\bt),-v((\sg^*)^{-1}\bt) \right)
\]
is just the solution with initial data $(\tilde{u}(\bt), \tilde{v}(\bt))|_{t_j=x \dt_{j,1}}=(a x, -b)$,
and the corresponding tau function satisfies the relation
\[
\log{\tilde\tau}^\rs(\bt)=\log\tau^\rs(\bt)|_{b\mapsto-b;~ \bt\mapsto (\sg^*)^{-1}\bt}.
\]
This observation agrees with Proposition~\ref{thm-sgtau}.
One also observes that the tau function $\tau^\rs(\bt)$ is $\Gm$-invariant if and only if $b=0$.

In this case, the tau function of the Drinfeld-Sokolov hierarchy associated to
$(\bar\fg=A_2^{(2)}, \bar\rs=(1,0), \mathds{1})$  is given by \eqref{tautaubar}  with $\mu=2$. Namely, the tau function satisfies
\begin{align}
\log\bar{\tau}^{\bar{\rs}}(\bar{\bt})|_{\bar{t}_{>6}=0}=&\frac{1}{2}\log\tau^\rs(\bt)|_{b=0,\, x=\sqrt{2}\,\bar{t}_1, \, t_5=\sqrt{2}\,\bar{t}_5,\, t_2=t_4=0,\, t_{>6}=0} \nn\\
=& \frac{a}{9\sqrt{2}} \bar{t}_1^3  -\frac{5 a^3 }{81 \sqrt{2}}\bar{t}_1^5\bar{t}_{5}+
   \left(\frac{10\sqrt{2}\, a^5 }{243}\bar{t}_1^5+\frac{25 a^4
   }{243}\bar{t}_1^2\right)\bar{t}_{5}^2+\mathrm{h.o.t.} \label{A2taured}
\end{align}
\end{exa}

\begin{rmk}\label{rmk-tauTab3}
Theorem~\ref{thm-taured} generalizes the second part of
Theorem~\ref{thm-LRZ} to the affine Kac-Moody algebras listed in Table~\ref{table-diag2}, however, such a result does not hold for the cases listed in Table~\ref{table-diag3}. The reason is that, in the latter case, the equality  \eqref{sgh} is not sufficient to derive $h_{m_i}|_{\hat{\bt}=0}=0$ when $m_i\in J^\sg_+\setminus\bar{J}_+$. For instance, considering the Drinfeld-Sokolov hierarchy given in Example~\ref{exa-C1tau}, let us compute its solution $(u,v,w)$ with the initial value $(u,v,w)|_{t_{i>1}=0}=(a x, b, c)$, where $a,\ b,\ c$ are constants. We have
\begin{align*}
  &u(\bt)|_{t_{i>6}=0} \\
  =&a
   x+  t_{5} \left(-\frac{5 a^3 x^2}{9}-\frac{10}{3} a^2 b^2 x+5 a
   b^4+\frac{5 a c^2}{6}\right)+\frac{1}{2} t_{5}^2
   \left(\frac{100 a^5 x^3}{81}-\frac{100}{27} a^4 b^2 x^2+\frac{50
   a^4}{81} \right. \\
   &\left. +\frac{500}{9} a^3 b^4 x+\frac{25}{18} a^3 b
   c-\frac{100}{3} a^2 b^6+\frac{50}{9} a^2 b^2 c^2\right)+ t_{3} t_{5} \left(-\frac{20}{3} a^3
   b^2 x+20 a^2 b^4-\frac{5 a^2 c^2}{3}\right)+ \mathrm{h.o.t.}, \\
   &\\
 &v(\bt)|_{t_{>6}=0} \\
  =&b+a b t_{3}+\frac{10}{9} a^2 b
   x t_{5}+\frac{1}{2} a^2 b t_{3}^2+\frac{1}{2} t_{5}^2
   \left(-\frac{200}{27} a^3 b^3 x+\frac{25 a^3 c}{108}+\frac{100
   a^2 b^5}{9}-\frac{25}{27} a^2 b c^2\right) \\
   &+
    \frac{10}{9} a^3 b x t_{3} t_{5}+\mathrm{h.o.t.}, \\
    &\\
    &w(\bt)|_{t_{>6}=0} \\
  =&c-a c t_{3}+  t_{5}
   \left(\frac{5 a^2 b}{9}+\frac{5}{9} a^2 c x-\frac{5}{3} a b^2
   c\right)+\frac{1}{2} a^2 c t_{3}^2+\frac{1}{2} t_{5}^2 \left(-\frac{25}{81} a^4 c
   x^2-\frac{275 a^3 b^3}{27}-\frac{25}{9} a^2 b^4 c \right. \\
   &\left. +\frac{25 a^2
   c^3}{27}+x \left(-\frac{25 a^4 b}{27}-\frac{50}{27} a^3 b^2
   c\right)\right) +t_{5} t_{3} \left(\frac{5 a^3
   b}{9}-\frac{5}{9} a^3 c x-\frac{5}{3} a^2 b^2
   c\right)+\mathrm{h.o.t.};\\
&\\
&\log\tau^\rs(\bt)|_{t_{>6}=0}\\
=&-\frac{b^2 x^2}{2}+\frac{a
   x^3}{18}
   +t_{3}
   \left(-a b^2 x^2-\frac{c^2 x}{2}\right)
   +t_{5} \left(-\frac{5}{324} a^3 x^4-\frac{5}{9} a^2 b^2
   x^3+x^2 \left(\frac{5 a b^4}{6}+\frac{5 a c^2}{36}\right) \right. \\
   &\left. +x
   \left(\frac{5 a b c}{3}-\frac{5 b^6}{3}+\frac{5 b^2
   c^2}{2}\right)\right)+\frac{1}{2} t_{3}^2 \left(-2 a^2 b^2 x^2+3 a b c+a c^2 x-6 b^6+3 b^2
   c^2\right)
    \\
    & +\frac{1}{2} t_{5}^2 \left(\frac{5
   a^5 x^5}{243}-\frac{25}{81} a^4 b^2 x^4+\frac{50}{9} a^3 b^4
   x^3+x \left(\frac{25 a^3 b^2}{81}+\frac{25 a b^8}{3}-\frac{50}{9}
   a b^4 c^2+\frac{25 a c^4}{108}\right) \right. \\
   &\left. -\frac{125 a^2 c^2}{648}+x^2
   \left(\frac{25 a^4}{243}-\frac{50}{3} a^2 b^6+\frac{50}{27} a^2
   b^2 c^2\right)-5 b^{10}-\frac{25 b^2 c^4}{36}\right)
   \\
   &+t_{3} t_{5} \left(-\frac{10}{9} a^3 b^2
   x^3+x^2 \left(\frac{10 a^2 b^4}{3}-\frac{5 a^2
   c^2}{18}\right)-\frac{5 a^2 b^2}{6}-10 a b^6 x-\frac{5 b^4
   c^2}{2}-\frac{5 c^4}{24}\right)  +\mathrm{h.o.t.}
\end{align*}
Note that $\sg^*t_j=t_j$ for all $j$,  so $\tau^\rs(\bt)$ is automatically $\Gm$-invariant. However, the formula \eqref{tautaubar} does not give a tau function of the hierarchy associated to  $(\bar\fg=A_2^{(2)}, \bar\rs=(1, 0), \mathds{1})$, since the reduced tau function and the associated function $u$ still depends on the parameters $b$ and $c$. In fact, if we take $b=c=0$, then $v(\bt)=w(\bt)=0$,
hence the tau function $\bar{\tau}^{\bar{\rs}}$ of the Drinfeld-Sokolov hierarchy associated to $(\bar\fg=A_2^{(2)}, \bar\rs=(1, 0), \mathds{1})$ satisfies
\begin{align*}
\log\bar{\tau}^{\bar{\rs}}(\bar{\bt})|_{\bar{t}_{>6}=0}=&\frac{1}{2}\log\tau^\rs(\bt)|_{b=c=0,\, x=\sqrt{2}\,\bar{t}_1, \, t_5=\sqrt{2}\,\bar{t}_5,\, t_3=0,\, t_{>6}=0} \nn\\
=& \frac{a}{9\sqrt{2}} \bar{t}_1^3  -\frac{5 a^3 }{81 \sqrt{2}}\bar{t}_1^5\bar{t}_{5}+
   \left(\frac{10\sqrt{2}\, a^5 }{243}\bar{t}_1^5+\frac{25 a^4
   }{243}\bar{t}_1^2\right)\bar{t}_{5}^2+\mathrm{h.o.t.}.
\end{align*}
which coincides with \eqref{A2taured}.
\end{rmk}

\section{Concluding remarks}
In this paper, we prove a $\Gm$-reduction theorem for the Drinfeld-Sokolov hierarchies associated to
affine Kac-Moody algebras which admit diagram automorphisms. We expect that this result is useful
to the study of the FJRW theory.

In \cite{LRZ}, it was shown that the partition function of the FJRW theory of BCFG types can be obtained from
that of the FJRW theory of ADE types, by using the fact that the associated Drinfeld-Sokolov
hierarchies and the string equations uniquely determine the topological
solutions, which are exactly the partition functions of the corresponding cohomological field theories. The topological solutions of the Drinfeld-Sokolov hierarchies of ADE and BCFG types can be computed explicitly, for instance, via an algorithm proposed in \cite{CW2}.
However, the Drinfeld-Sokolov hierarchy associated to a twisted affine Kac-Moody algebra does not have an analogue of the string equation,
so one can not use it to pick up a particular solution. Actually, there is
no definition for the \emph{topological solution} in the twisted cases.

For the untwsited cases, the string equation $L_{-1}\tau=0$ is the first one of the Virasoro constraints.
In the twisted cases, the Virasoro constraints
\[L_m \tau=0, \quad m\ge 0\]
start from the zeroth one.
One can show that these constraints together with the integrable hierarchy itself determine a collection of
solutions, which are parametrized by a finite-dimensional space. We observe that the ODEs that determine the initial conditions of these solutions possess the
Painlev\'e property, and that there exist affine Weyl group actions on the solution space of these ODEs.
We hope that these observations will be useful to the construction of the FJRW theory corresponding to the twisted affine
Kac-Moody algebras. We will study this problem in a separate publication.

\vskip 0.5truecm \noindent{\bf Acknowledgments.}
The authors thank J\"urgen Fuchs and Yongbin Ruan for useful discussions.
The work is partially supported by NSFC  No.\,11771238, No.\,11771461 and No.\,11831017.
The author S.-Q. Liu is supported by the National Science Fund for Distinguished Young Scholars No.\,11725104.

\begin{appendices}
\renewcommand{\thesection}{A}
\renewcommand{\theequation}{A.\arabic{equation}}

\section{Proof of the Serre relations for $\bar{\g}$}\label{zh-3}

Let $\g$ be the derived algebra of the Kac-Moody algebra associated to an affine generalized Cartan matrix $\left(a_{ij}\right)_{i, j=0, \dots, \ell}$.
Suppose \[\{e_i,\ f_i,\ \al^\vee_i\mid i=0, \dots, \ell\}\] is a collection of Chevalley generators of $\g$, then they satisfy the following
Serre relations
\begin{align*}
& [\al^\vee_i, \al^\vee_j] =0, \quad
[e_i, f_j] =\delta_{ij}\al^\vee_i, \quad
[\al^\vee_i, e_j] =a_{ij}e_j, \quad
[\al^\vee_i, f_j] =-a_{ij}f_j, \\
& \left(\ad_{e_i}\right)^{1-a_{ij}}\left(e_j\right)=0, \quad
\left(\ad_{f_i}\right)^{1-a_{ij}}\left(f_j\right)=0.
\end{align*}

Suppose $\bsigma$ is a diagram automorphism of $\left(a_{ij}\right)_{i, j=0, \dots, \ell}$, i.e. a bijection \[\bsigma:\{0, \dots, \ell\} \to \{0, \dots, \ell\}\] such that
\[a_{ij}=a_{\bsigma(i)\bsigma(j)},\]
then one can define a Lie algebra automorphism $\sigma:\g \to \g$ such that
\[\sigma\left(e_i\right)=e_{\bsigma(i)}, \quad \sigma\left(f_i\right)=f_{\bsigma(i)}, \quad \sigma\left(\al^\vee_i\right)=\al^\vee_{\bsigma(i)}.\]

For an index $i\in \{0, \dots, \ell\}$, we denote its $\bsigma$-orbit  by $\ang{i}$, and $N_i=\#\ang{i}$, then define
\[b_i=3-\sum_{j\in\ang{i}}a_{ji}.\]
If we choose another representative $i'=\bsigma^m(i)\in\ang{i}$, then
\[b_{i'}=3-\sum_{j\in\ang{i'}}a_{ji'}=3-\sum_{j\in\ang{\bsigma^m(i)}}a_{j\bsigma^m(i)}=3-\sum_{\bsigma^{-m}(j)\in\ang{i}}a_{\bsigma^{-m}(j)i}=b_i,\]
so $b_i$ is independent of the choice of the representative $i\in\ang{i}$.

It is easy to see that $b_i$'s are integers satisfying $b_i\ge 1$. Following \cite{FSS}, we say that $\bsigma$ satisfies \emph{the linking condition} if
\[b_i\le 2, \quad \forall i \in \{0, \dots, \ell\}.\]
Note that in affine cases, all diagram automorphisms satisfy this condition, except the diagram automorphism of $A_\ell^{(1)}$ with $\bsigma(i)\equiv i+1$
$\left(\mathrm{mod}\ \ell+1\right)$ or its inverse. We assume that $\bsigma$ satisfies the linking condition from now on.

For each orbit $\ang{i}$, define
\begin{equation}\label{chevalleygbar}
E_i=\sum_{k\in\ang{i}}e_k, \quad F_i=b_i\sum_{k\in\ang{i}}f_k, \quad H_i=b_i\sum_{k\in\ang{i}}\al^\vee_k,
\end{equation}
and define
\begin{equation}\label{Abar}
A_{ij}=b_i\sum_{k\in\ang{i}}a_{kj}
\end{equation}
for two orbits $\ang{i}$, $\ang{j}$.
Then it is easy to see that these elements of $\g$ and integers are also independent of the choice of the representative $i\in\ang{i}$ or $j\in\ang{j}$.

Let $I$ be the set of all orbits $I=\{\ang{i}\mid i\in \{0, \dots, \ell\}\}$.
It is shown in \cite{FSS} that $\left(A_{ij}\right)_{i,j \in I}$ is
also an affine generalized Cartan matrix\footnote{The definition of Catan matrix used in \cite{FSS} is different from
the one we used in the present paper, i.e. the Cartan matrix defined in \cite{FSS} is the transposed matrix of the one
we used. However, this difference doesn't affect the applicability of \cite{FSS}'s result used here.}. In this appendix,
we show that the elements $\{E_i,\ F_i,\ H_i\}_{i\in I}$ of $\g$ satisfy the Serre relations associated to this
generalized Cartan matrix.
\begin{lem}
The elements $\{E_i,\ F_i,\ H_i\}_{i\in I}$ of $\g$ satisfy
\begin{align*}
& [H_i, H_j] =0, \\
& [E_i, F_j] =\delta_{ij}H_i, \\
& [H_i, E_j] =A_{ij}E_j, \\
& [H_i, F_j] =-A_{ij}F_j.
\end{align*}
\end{lem}
\begin{prf}
We only prove the third one:
\begin{align*}
[H_i,\ E_j] =& \left[b_i \sum_{k\in\ang{i}}\al^\vee_k,\ \sum_{l\in\ang{j}}e_l\right] = b_i  \sum_{k\in\ang{i}}\sum_{l\in\ang{j}} a_{kl}e_l \\
=& b_i  \sum_{\alpha=0}^{N_i-1}\sum_{\beta=0}^{N_j-1} a_{\bsigma^\alpha(i)\bsigma^\beta(j)}e_{\bsigma^\beta(j)}
= \sum_{\beta=0}^{N_j-1}\left(b_i\sum_{\alpha=0}^{N_i-1}a_{\bsigma^{\alpha-\beta}(i)j}\right)e_{\bsigma^\beta(j)} \\
=& A_{ij} \sum_{\beta=0}^{N_j-1}e_{\bsigma^\beta(j)}=A_{ij}E_j.
\end{align*}
Proofs for the other three identities are similar, so we omit them here.
\end{prf}

Up to now we have verified that $\bar\fg$ (recall that the center is assumed to be trivial in the current proof) is in fact a reduced contragrediant Lie algebra with the Cartan matrix $(A_{i j})_{i, j\in I}$ of affine type. Thus, according to Proposition~13 of \cite{Kac68}, we conclude that $\bar\fg$ is the affine Kac-Moody algebra for the folded Cartan matrix  $(A_{i j})_{i, j\in I}$.

To make this paper self-contained, let us give a straightforward proof of the rest of the Serre relations.
\begin{lem}
If $b_i=1$, then we have
\[\left(\ad_{E_i}\right)^{1-A_{ij}}\left(E_j\right)=0, \quad
\left(\ad_{F_i}\right)^{1-A_{ij}}\left(F_j\right)=0.\]
\end{lem}
\begin{prf}
We only prove the first one, since the second one is similar. The condition $b_i=1$ means that $a_{ij}=0$ for all $j\in\ang{i}$ and $j\ne i$,
then it is easy to see that $[e_j, e_k]=0$ and $[\ad_{e_j}, \ad_{e_k}]=0$ for all $j, k\in \ang{i}$.

For an arbitrary positive integer $m$, we have
\begin{align}
\left(\ad_{E_i}\right)^m=& \left(\sum_{k\in\ang{i}}\ad_{e_k}\right)^m
= \left(\sum_{\alpha=0}^{N_i-1}\ad_{e_{\bsigma^{\alpha}(i)}}\right)^m \nn \\
=& \sum_{\left(d_1, \dots, d_{N_i}\right)}\frac{m!}{d_1!\cdots d_{N_i}!}\left(\ad_{e_{\bsigma^0(i)}}\right)^{d_1}\cdots
\left(\ad_{e_{\bsigma^{N_i-1}(i)}}\right)^{d_{N_i}}, \label{multinomial}
\end{align}
where the summation is taken over integer tuples $\left(d_1, \dots, d_{N_i}\right)$ satisfying
\[d_1, \dots, d_{N_i}\ge 0, \quad d_1+\cdots+d_{N_i}=m.\]
On the other hand,
\[
\left(\ad_{E_i}\right)^m\left(E_j\right)=\left(\ad_{E_i}\right)^m\left(\sum_{l\in\ang{j}}e_l\right)
=\sum_{\beta=0}^{N_j-1}\sigma^\beta\left(\left(\ad_{E_i}\right)^m\left(e_j\right)\right),
\]
so we only need to show that $\left(\ad_{E_i}\right)^m\left(e_j\right)=0$.

If $d_k\ (k=1, \dots, N_i)$ satisfies $d_k\ge 1-a_{\bsigma^{k-1}(i)j}$, then
\[\left(\ad_{e_{\bsigma^{k-1}(i)}}\right)^{d_k}\left(e_j\right)=0,\]
so we do not need to consider the tuples $\left(d_1, \dots, d_{N_i}\right)$ with such kind of $d_k$'s in the summation \eqref{multinomial}
for $\left(\ad_{E_i}\right)^m\left(e_j\right)$.

If all $d_k$'s satisfy $d_k\le -a_{\bsigma^{k-1}(i)j}$, then
\[m=d_1+\cdots+d_{N_i}\le -\sum_{\alpha=0}^{N_i-1}a_{\bsigma^{\alpha}(i)j}=-A_{ij}.\]
So, when $m\ge 1-A_{ij}$, we have $\left(\ad_{E_i}\right)^m\left(e_j\right)=0$. The lemma is proved.
\end{prf}

\vskip 1em

If $b_i=2$, there is exactly one index $k\in \ang{i}$ such that $a_{i k}=-1$. Actually, in all the affine cases satisfying the linking condition,
the automorphism $\bsigma$ with some $b_i=2$ must be an involution, i.e. $\ang{i}=\{i, \bsigma(i)\}$. We denote
\[X=\ad_{e_i}, \quad Y=\ad_{e_{\bsigma(i)}}, \quad Z=[X, Y],\]
then $[X, Z]=[Y, Z]=0$. Denote by
\[e_j=v, \quad p=-a_{ij},\quad q=-a_{\bsigma(i)j},\]
then we have $X^{p+1}(v)=0$, $Y^{q+1}(v)=0$. The Serre relation \[\left(\ad_{E_i}\right)^{1-A_{ij}}\left(E_j\right)=0\] is equivalent to
\[\left(X+Y\right)^{2p+2q+1}(v)=0.\]
So the problem is reduced to the following conjecture:
\begin{con} \label{linear-conj}
Let $V$ be a linear space over a field $\mathbbm{k}$ with
$\mathrm{char}\mathbbm{k}=0$.
Suppose $X, Y, Z\in \mathrm{End}\left(V\right)$ satisfy
\[Z=[X, Y], \quad [X, Z]=[Y, Z]=0,\]
and $v\in V$, $p, q\in\mathbb{N} $ satisfy
\[X^{p+1}(v)=0, \quad Y^{q+1}(v)=0,\]
then we have
\[\left(X+Y\right)^{2p+2q+1}(v)=0.\]
\end{con}

\begin{lem}
Conjecture \ref{linear-conj} is correct when $p=0$ or $q=0$.
\end{lem}
\begin{prf}
Assume that $q=0$, i.e. $Y(v)=0$.
The Baker-Campbell-Hausdorff formula implies that
\[e^{t\left(X+Y\right)}(v)=e^{\frac{t^2}{2}Z}e^{t X}(v).\]
We need to prove that the left hand side is actually a polynomial in $t$ with degree $\le 2p$.
The right hand side reads
\[e^{\frac{t^2}{2}Z}e^{t X}(v)=\sum_{k, l\ge0}\frac{t^{2k+l}}{k!\, l!\, 2^k}Z^k\left(X^l(v)\right).\]
We are to show that, when $2k+l>2p$, $Z^k\left(X^l(v)\right)=0$.

By using the identity $[X^l, Y]=l X^{l-1}Z$, we have
\begin{align*}
Z^k X^l (v) =& X^l Z^{k-1}(XY-YX)(v)=-X^l Y X Z^{k-1}(v) \\
=& -l Z^k X^l(v)-YX^{l+1}Z^{k-1}(v),
\end{align*}
which implies that
\[Z^k X^l (v)=-\frac{1}{l+1}YX^{l+1}Z^{k-1}(v)=\cdots=\frac{(-1)^k}{(l+1)\cdots(l+k)}Y^kX^{k+l}(v).\]
When $2k+l>2p$, we have $k+l>p$, so $X^{k+l}(v)=0$. The lemma is proved.
\end{prf}

\begin{lem}
Conjecture \ref{linear-conj} is correct when $p=q=1$.
\end{lem}
\begin{prf}
First, we have
\[(X+Y)^2(v)=(X^2+XY+YX+Y^2)(v)=(XY+YX)(v),\]
then
\begin{align*}
& (X+Y)^3(v) = (X^2Y+XYX+YXY+Y^2X)(v)\\
=& \left((YX^2+2ZX)+(YX+Z)X+(XY-Z)Y+(XY^2-2ZY)\right)(v)\\
=&3Z(X-Y)(v).
\end{align*}
Next
\[(X+Y)^4(v)=3Z(X^2+YX-XY-Y^2)(z)=-3Z^2(v),\]
and
\[(X+Y)^5(v)=-3Z^2(X+Y)(v).\]

Note that
\begin{align*}
Z^2X(v) =& \frac{1}{2}Z[X^2, Y](v) = \frac{1}{2}Z X^2 Y(v) = \frac{1}{4}X[X^2, Y]Y(v)\\
=& -\frac{1}{4}XYX^2Y(v) = -\frac{1}{4} XY(X^2Y-YX^2)(v)\\
=& -\frac{1}{2}XYXZ(v) = -\frac{1}{2}Z(XY-YX)X(v)=-\frac{1}{2}Z^2X(v),
\end{align*}
so we have $Z^2X(v)=0$. Similarly, $Z^2Y(v)=0$. The lemma is proved.
\end{prf}

In all the affine cases satisfying the linking condition, either $p=0$, $q=0$, or $p=q=1$ (since $|A_{i j}|\le 4$), so the Serre relation for the cases with $b_i=2$ is proved.

Finally, when $\fg$ is an affine Kac-Moody algebra of type $X_{\ell'}^{(r)}$ given in Tables~\ref{table-diag1}--\ref{table-diag3}, then one can check  case by case  that the generalized Cartan matrix $(A_{i j})_{i,j \in I}$ is of type $X_{\bar{\ell}'}^{(\bar r)}$.

\renewcommand{\thesection}{B}
\renewcommand{\theequation}{B.\arabic{equation}}
\section{List of Affine Kac-Moody algebras and their subalgebras}\label{zh-2}

In this appendix we use the same notations as those appeared in Sections~\ref{sec-g} and \ref{sec-gsg}.
Let $\fg$ be the derived subalgebra of an affine Kac-Moody algebra  $\fg(A)$ of type $X_{\ell'}^{(r)}$
(we write $\fg=X_{\ell'}^{(r)}$ for short)
listed in Tables~\ref{table-diag1}--\ref{table-diag3}. For $\fg(A)=\fg\oplus \C d$, let $J$ be the set of
exponents,  $h$ be the Coxeter number, $c$ be the canonical central
element, and $(\cdot\mid\cdot)$ be the standard invariant bilinear form.
Similar notations for the subalgebras
$\bar{\fg}, \, \fg^\sg\subset\fg$ introduced in Subsection~\ref{sec-gsg} are used. In particular, $\bar\fg$ is the derived subalgebra of the Affine Kac-Moody algebras
$\mathfrak{g}(\bar A)$ of type $X_{\bar{\ell}'}^{(\bar r)}$ ($\bar\fg=X_{\bar{\ell}'}^{(\bar r)}$ for short), whose  canonical central element and standard invariant bilinear form are given by
\[ \bar{c}=\mu c, \quad (X\mid Y)^-=\kappa
(X\mid Y) \hbox{ for } X, Y \in  \fg(\bar A)
\]
with constants $\mu$ and $\kappa$.
Such data are listed in what follows.
\begin{enumerate} \renewcommand{\labelenumi}{(\alph{enumi})}

\item[(a1)]
\begin{itemize}
\item $\fg=A_{2n-1}^{(1)}$ ~ $(n\ge2)$,
\item[] $J=\{1,2,3, \dots, 2n-1\}+h\Z$, \quad    $h=2n$,
\item $\bar\sg(0)=0, \qquad \bar\sg(i)=2n-i$, \quad $i=1, 2, \dots, 2n-1$,
\item $\bar{\fg}=C_{n}^{(1)}$,
\item[] $\bar{J}=\{1, 3, 5, \dots,
2n-1\}+\bar{h}\Z$, \quad $\bar{h}=2n$,
\item[] $\mu=1, \quad \kappa=1$,
\item $\fg^\sg=\bar\fg, \quad J^\sg=\bar{J}$
\end{itemize}

\item[(a2)]
\begin{itemize}
\item $\fg=A_{2n}^{(1)}$ ~ $(n\ge1)$,
\item[] $J=\{1,2,3, \dots, 2n\}+h\Z$, \quad    $h=2n+1$,
\item  $\bar\sg(i)=2n-i$, \quad $i=0,1, 2, \dots, 2n$,
\item $\bar{\fg}=A_{2n}^{(2)}$,
\item[]  $\bar{J}=\{1, 3, \dots, 2n-1, 2n+3, 2n+5, \dots,
4n+1\}+2\bar{h}\Z$, \quad $\bar{h}=2n+1$,
\item[] $\mu=2, \quad \kappa=\frac{1}{2}$,
\item $\fg^\sg=\bar\fg, \quad J^\sg=\bar{J}$
\end{itemize}

\item[(a3)]
\begin{itemize}
\item $\fg=A_{2n+1}^{(1)}$ ~ $(n\ge2)$,
\item[] $J=\{1,2,3, \dots, 2n+1\}+h\Z$, \quad    $h=2n+2$,
\item  $\bar\sg(i)=2n-i+1$, \quad $i=0,1, 2, \dots, 2n+1$,
\item $\bar{\fg}=D_{n+1}^{(2)}$,
\item[] $\bar{J}=\{1, 3, 5, \dots,
2n+1\}+2\bar{h}\Z$, \quad $\bar{h}=n+1$,
\item[] $\mu=2, \quad \kappa=\frac{1}{4}$,
\item $\fg^\sg=\bar\fg, \quad J^\sg=\bar{J}$
\end{itemize}

\item[(a4)]
\begin{itemize}
\item $\fg=A_{2(n+1)-1}^{(2)}$ ~ $(n\ge2)$,
\item[] $J=\{1,3,5, \dots, 4n+1\}+2h\Z$, \quad    $h=2n+1$,
\item $\bar\sg(0)=1, \quad \bar\sg(1)=0, \quad \bar\sg(i)=i$, \quad $i=2,3, \dots, n+1$,
\item $\bar{\fg}=A_{2n}^{(2)}$,
\item[] $\bar{J}=\{1, 3, 5, \dots, 2n-1, 2n+3, 2n+3,\dots
4n+1\}+2\bar{h}\Z$, \quad $\bar{h}=2n+1$,
\item[] $\mu=1, \quad \kappa=1$,
\item  $\fg^\sg\ne\bar\fg, \quad J^\sg=\bar{J}\cup (2n+1)\Z^{\mathrm{odd}}$
\end{itemize}

\item[(b)]
\begin{itemize}
\item $\fg=B_{n+1}^{(1)}$ ~ $(n\ge2)$,
\item[] $J=\{1,3,5, \dots, 2n+1\}+h\Z$, \quad    $h=2n+2$,
\item  $\bar\sg(0)=1, \quad \bar\sg(1)=0, \quad \bar\sg(i)=i$, \quad $i=2,3, \dots, n+1$,
\item $\bar{\fg}=D_{n+1}^{(2)}$,
\item[] $\bar{J}=\{1, 3, 5, \dots, 2n+1\}+2\bar{h}\Z$, \quad
$\bar{h}=n+1$,
\item[] $\mu=1, \quad \kappa=\frac{1}{2}$,
\item $\fg^\sg=\bar\fg, \quad J^\sg=\bar{J}$
\end{itemize}

\item[(c1)]
\begin{itemize}
\item $\fg=C_{2n}^{(1)}$ ~ $(n\ge2)$,
\item[] $J=\{1, 3, 5, \dots, 4n+1\}+h\Z$, \quad    $h=4n$,
\item  $\bar\sg(i)=2n-i$, \quad $i=0,1,2, \dots, 2n$,
\item $\bar{\fg}=C_{n}^{(1)}$,
\item[] $\bar{J}=\{1, 3, 5, \dots,
2n-1\}+\bar{h}\Z$, \quad $\bar{h}=2n$,
\item[] $\mu=1, \quad \kappa=\frac{1}{2}$,
\item $\fg^\sg=\bar\fg, \quad J^\sg=\bar{J}$
\end{itemize}

\item[(c2)]
\begin{itemize}
\item $\fg=C_{2n+1}^{(1)}$ ~ $(n\ge1)$,
\item[] $J=\{1, 3, 5, \dots, 4n-1\}+h\Z$, \quad    $h=4n+2$,
\item  $\bar\sg(i)=2n-i+1$, \quad $i=0,1,2, \dots, 2n+1$,
\item $\bar{\fg}=A_{2n}^{(2)}$,
\item[] $\bar{J}=\{1, 3, 5, \dots, 2n-1, 2n+3, 2n+3,\dots
4n+1\}+2\bar{h}\Z$, \quad $\bar{h}=2n+1$,
\item[] $\mu=2, \quad \kappa=\frac{1}{4}$,
\item $\fg^\sg\ne\bar\fg, \quad J^\sg=\bar{J}\cup (2n+1)\Z^{\mathrm{odd}}$
\end{itemize}

\item[(d1)]
\begin{itemize}
\item $\fg=D_{n+1}^{(1)}$ ~ $(n\ge3)$,
\item[] $J=\{1, 3, 5, \dots, 2n-3, 2n-1, n\}+h\Z$, \quad    $h=2n$,
\item $\bar\sg(n)=n+1, \quad \bar\sg(n+1)=n$, \quad  $\bar\sg(i)=i$, \quad $i=0,1,2, \dots, n-1$,
\item $\bar{\fg}=B_{n}^{(1)}$,
\item[] $\bar{J}=\{1, 3, 5, \dots,
2n-1\}+\bar{h}\Z$, \quad $\bar{h}=2n$,
\item[] $\mu=1, \quad \kappa=1$,
\item $\fg^\sg=\bar\fg, \quad J^\sg=\bar{J}$
\end{itemize}
\setcounter{enumi}{8}

\item[(d2)]
\begin{itemize}
\item $\fg=D_{4}^{(1)}$,
\item[] $J=\{1, 3, 3, 5\}+h\Z$, \quad    $h=6$,
\item $\bar\sg(0,1,3,4,2)=(0, 3,4,1,2)$,
\item $\bar{\fg}=G_{2}^{(1)}$,
\item[] $\bar{J}=\{1,  5\}+6\Z$, \quad $\bar{h}=6$,
\item[] $\mu=1, \quad \kappa=1$,
\item $\fg^\sg=\bar\fg, \quad J^\sg=\bar{J}$
\end{itemize}

\item[(d3)]
\begin{itemize}
\item $\fg=D_{2n+1}^{(1)}$ ~ $(n\ge2)$,
\item[] $J=\{1, 3, 5, \dots, 4n-3, 4n-1, 2n\}+h\Z$, \quad    $h=4n$,
\item $\bar\sg(i)=2n-i+1$, \quad $i=0,1,2, \dots, 2n+1$,
\item $\bar{\fg}=B_{n}^{(1)}$,
\item[] $\bar{J}=\{1, 3, 5, \dots,
2n-1\}+\bar{h}\Z$, \quad $\bar{h}=2n$,
\item[] $\mu=1, \quad \kappa=\frac{1}{2}$,
\item $\fg^\sg=\bar\fg, \quad J^\sg=\bar{J}$
\end{itemize}

\item[(d4)]
\begin{itemize}
\item $\fg=D_{2n}^{(1)}$ ~ $(n\ge2)$,
\item[] $J=\{1, 3, 5, \dots, 4n-5, 4n-3, 2n-1\}+h\Z$ ~~($(2n-1)\Z^{\mathrm{odd}}$ are double exponents), \quad    $h=4n-2$,
\item $\bar\sg(i)=2n-i$, \quad $i=0,1,2, \dots, 2n$,
\item $\bar{\fg}=A_{2n-1}^{(2)}$,
\item[] $\bar{J}=\{1, 3, 5, \dots,
4n-3\}+2\bar{h}\Z$, \quad $\bar{h}=2n-1$,
\item[] $\mu=1, \quad \kappa=\frac{1}{2}$,
\item $\fg^\sg\ne\bar\fg, \quad J^\sg=\bar{J}\coprod
(2n-1)\Z^{\mathrm{odd}}$
\end{itemize}

\item[(d5)]
\begin{itemize}
\item $\fg=D_{4}^{(1)}$,
\item[] $J=\{1, 3, 3, 5\}+h\Z$, \quad    $h=6$,
\item $\bar\sg(0,1,3,4,2)=(1,3,4,0,2)$,
\item $\bar{\fg}=A_{2}^{(2)}$,
\item[] $\bar{J}=\{1,  5\}+6\Z$, \quad $\bar{h}=3$,
\item[] $\mu=1, \quad \kappa=\frac{1}{2}$,
\item $\fg^\sg\ne\bar\fg, \quad J^\sg=\bar{J}\cup 3\Z^{\mathrm{odd}}$
\end{itemize}

\item[(d6)]
\begin{itemize}
\item $\fg=D_{(2n+1)+1}^{(2)} \quad (n\ge1)$,
\item[] $J=\{1, 3, 5, \dots, 4n+3\}+2h\Z$, \quad    $h=2n+2$,
\item $\bar\sg(i)=2n-i+1$, \quad $i=0,1,2, \dots, 2n+1$,
\item $\bar{\fg}=D_{n+1}^{(2)}$,
\item[] $\bar{J}=\{1, 3, 5, \dots, 2n+1\}+2\bar{h}\Z$, \quad
$\bar{h}=n+1$,
\item[] $\mu=1, \quad \kappa=\frac{1}{2}$,
\item $\fg^\sg=\bar\fg, \quad J^\sg=\bar{J}$
\end{itemize}

\item[(d7)]
\begin{itemize}
\item $\fg=D_{2n+1}^{(2)} \quad (n\ge1)$,
\item[] $J=\{1, 3, 5, \dots, 4n+1\}+2h\Z$, \quad    $h=2n+1$,
\item $\bar\sg(i)=2n-i$, \quad $i=0,1,2, \dots, 2n$,
\item $\bar{\fg}=A_{2n}^{(2)}$,
\item[] $\bar{J}=\{1, 3, 5, \dots, 2n-1, 2n+3, 2n+5,\dots,
4n+1\}+2\bar{h}\Z$, \quad $\bar{h}=2n+1$,
\item[] $\mu=1, \quad \kappa=1$,
\item $\fg^\sg\ne\bar\fg, \quad J^\sg=\bar{J}\cup (2n+1)\Z^{\mathrm{odd}}$
\end{itemize}

\item[(e1)]
\begin{itemize}
\item $\fg=E_{6}^{(1)}$,
\item[] $J=\{1, 4, 5, 7, 8,11\}+h\Z$, \quad    $h=12$,
\item $\bar\sg(0,1,2,3,4,5,6)=(0,5,4,3,2,1,6)$,
\item $\bar{\fg}=F_{4}^{(1)}$,
\item[] $\bar{J}=\{1, 5,7,11\}+12\Z$, \quad $\bar{h}=12$,
\item[] $\mu=1, \quad \kappa=1$,
\item $\fg^\sg=\bar\fg, \quad J^\sg=\bar J$
\end{itemize}

\item[(e2)]
\begin{itemize}
\item $\fg=E_{6}^{(1)}$,
\item[] $J=\{1, 4, 5, 7, 8,11\}+h\Z$, \quad    $h=12$,
\item $\bar\sg(0,1,5,2,4,6,3)=(1,5,0,4,6,2,3)$,
\item $\bar{\fg}=D_{4}^{(3)}$,
\item[] $\bar{J}=\{1, 5,7,11\}+12\Z$, \quad $\bar{h}=4$,
\item[] $\mu=1, \quad \kappa=\frac{1}{3}$,
\item $\fg^\sg\ne\bar\fg, \quad J^\sg=J$
\end{itemize}

\item[(e3)]
\begin{itemize}
\item $\fg=E_{7}^{(1)}$,
\item[] $J=\{1, 5, 7, 9,11,13,17\}+h\Z$, \quad    $h=18$,
\item $\bar\sg(7)=7$, \quad $\bar\sg(i)=6-i, \quad i=0,1,2, 3, 4,5, 6$,
\item $\bar{\fg}=E_{6}^{(2)}$,
\item[] $\bar{J}=\{1, 5, 7, 11,13,17\}+18\Z$, \quad $\bar{h}=9$,
\item[] $\mu=1, \quad \kappa=\frac{1}{2}$,
\item $\fg^\sg\ne\bar\fg, \quad J^\sg=J$
\end{itemize}

\end{enumerate}

\end{appendices}

\end{document}